\documentclass[reprint, aps, pra, footinbib, letterpaper, superscriptaddress]{revtex4-2}
\usepackage[T1]{fontenc} % Use modern font encodings

\usepackage{amsmath,color, tikz}
\usetikzlibrary{matrix}
\usepackage{amssymb}
\usepackage{amsfonts}
\usepackage{amsthm}
\usepackage{mathtools}
\usepackage{comment}
\usepackage{hyperref}
\usepackage{cleveref}
\usepackage{algorithm}
\usepackage{algpseudocode}
\usepackage{dsfont}

\usepackage{bbold}
\usepackage{chemformula}
\usepackage{siunitx}

\DeclarePairedDelimiterX\braket[2]{\langle}{\rangle}{#1 \delimsize\vert #2}
\DeclarePairedDelimiterX\braket3[3]{\langle}{\rangle}{#1 \delimsize\vert #2 \delimsize\vert #3}

\usepackage{accents}

\usepackage{fancyvrb}

\usepackage{pdfpages}

\newcommand{\red}[1]{{\color{black} #1}}

	\makeatletter
	\AtBeginDocument{\let\LS@rot\@undefined}
	\makeatother

\begin{document}

\title{Selective Excitation of IR-Inactive Modes via Vibrational Polaritons: Insights from Atomistic Simulations}

        \author{Xinwei Ji}
        \affiliation{Department of Physics and Astronomy, University of Delaware, Newark, Delaware 19716, USA}
	
	\author{Tao E. Li}%
	\email{taoeli@udel.edu}
	\affiliation{Department of Physics and Astronomy, University of Delaware, Newark, Delaware 19716, USA}

    %\begin{tocentry}
        %\includegraphics[width=1.0\linewidth]{toc_ch4.png}
    %\end{tocentry}

        \begin{abstract}
        Vibrational polaritons, hybrid light-matter states formed between molecular vibrations and infrared (IR) cavity modes, provide a novel approach for modifying chemical reaction pathways and energy transfer processes. For vibrational polaritons involving condensed-phase molecules, the short polariton lifetime raises debate over whether pumping  polaritons may produce different effects on molecules compared to directly exciting the molecules in free space or under weak coupling. Here,  for liquid methane under vibrational strong coupling,  classical cavity molecular dynamics simulations show that pumping the upper polariton (UP) formed by the asymmetric bending mode of methane can sometimes selectively excite the IR-inactive symmetric bending mode. This finding is validated when the molecular system is described using  both empirical force fields and machine-learning potentials\red{, also in qualitative agreement with analytical theory of polariton energy transfer rates based on Fermi's golden rule calculations.} Additionally, \red{our study suggests} that  polariton-induced energy transfer \red{to IR-inactive modes} reaches maximal efficiency when the UP has significant  contributions from both photons and molecules, underscoring the importance of light-matter hybridization.  As IR-inactive vibrational modes are generally inaccessible to direct IR excitation, our study  highlights the unique role of polariton formation in selectively controlling IR-inactive vibrations. \red{Since this polariton-induced process occurs after the polariton decays, it} may  impact IR photochemistry on a timescale longer than the polariton lifetime, as observed in experiments.
	\end{abstract}

	\maketitle

    \section{Introduction} 
    
    Exploring novel methods to modify chemical reaction and energy transfer pathways is a crucial objective in the field of chemistry. Over the past decade, experimental studies have demonstrated that these pathways can be efficiently controlled by the formation of vibrational polaritons \cite{Shalabney2015,Long2015,Thomas2016, Thomas2019_science,Xiang2020Science,Chen2022,Ahn2023Science}. These quasiparticles stem from strong light-matter interactions between molecular vibrational transitions and infrared (IR) cavity photon modes, which are frequently prepared by confining  a macroscopic layer of condensed-phase molecules within a planar Fabry--P\'erot microcavity \cite{Ribeiro2018,Herrera2019,Li2022Review,Fregoni2022,Simpkins2023,Mandal2023ChemRev,Ruggenthaler2023,Xiang2024}.  
    
    In this vibrational strong coupling (VSC) regime, pump-probe and two-dimensional IR (2D-IR) spectroscopies demonstrate that vibrational polaritons typically decay within a few ps after external excitation \cite{Dunkelberger2016,Xiang2018,Grafton2020,Xiang2020Science,Chen2022,Pyles2024}. During the fast relaxation of polaritons, the polariton energy can be transferred to other molecular excited-state degrees of freedom, creating  incoherent reservoir excitations of molecules. The dynamical response of these reservoir excitations has been shown to be indistinguishable from the molecular dynamics outside the cavity \cite{Grafton2020,Pyles2024,Xiang2021JCP}. This experimental evidence supports the perspective that polaritons function as optical filters \cite{Schwennicke2024}. In other words, pumping vibrational polaritons would not create a molecular vibrational excited-state manifold inaccessible by pumping molecules under weak coupling or outside the cavity \cite{Schwennicke2024}. Viewing polaritons as optical filters, however, may not fully align with recent experimental observations, which suggest that exciting polaritons can significantly influence molecular energy transfer and reaction pathways on a time scale much longer than the polariton lifetime \cite{Xiang2020Science,Chen2022}.

    Here, we explore how pumping vibrational polaritons may nontrivially transfer energy to other molecular degrees of freedom through numerical simulations. Specifically, we examine whether exciting vibrational polaritons with an IR pulse may directly accumulate energy in IR-inactive vibrational states of molecules \cite{Hirschmann2024}. This proposed mechanism showcases the advantage of polariton pumping, as the IR light alone cannot efficiently excite IR-inactive vibrational modes in ambient conditions.
    
    Our simulation approach employs the classical cavity molecular dynamics (CavMD) scheme \cite{Li2020Water,Li2020Nonlinear,Li2021Solute,Li2021Relaxation}. Although several numerical and analytical methods have been proposed in recent years to study VSC  \cite{Galego2019,Hernandez2019,Campos-Gonzalez-Angulo2019,Hoffmann2020,Botzung2020,LiHuo2021,Fischer2021,YangCao2021,Wang2022JPCL,Flick2017,Riso2022,Schafer2021,Bonini2021,Yang2021,Rosenzweig2022,Triana2020Shape,Haugland2020,Philbin2022,Poh2023,Suyabatmaz2023,Yu2024}, the CavMD approach offers a distinct advantage in describing nonequilibrium polariton relaxation dynamics in realistic condensed-phase  molecular systems. For instance, this approach reveals that pumping the lower polariton (LP) can directly transfer energy to  highly excited vibrational states of molecules due to the energy match between twice the LP energy and the $0\rightarrow 2$ vibrational transition of molecules \cite{Li2020Nonlinear}. This prediction is in qualitative agreement with experiments \cite{Xiang2019State,Xiang2021JCP} and analytical theory \cite{Ribeiro2020}.

    \begin{figure*}
	    \centering
	    \includegraphics[width=0.7\linewidth]{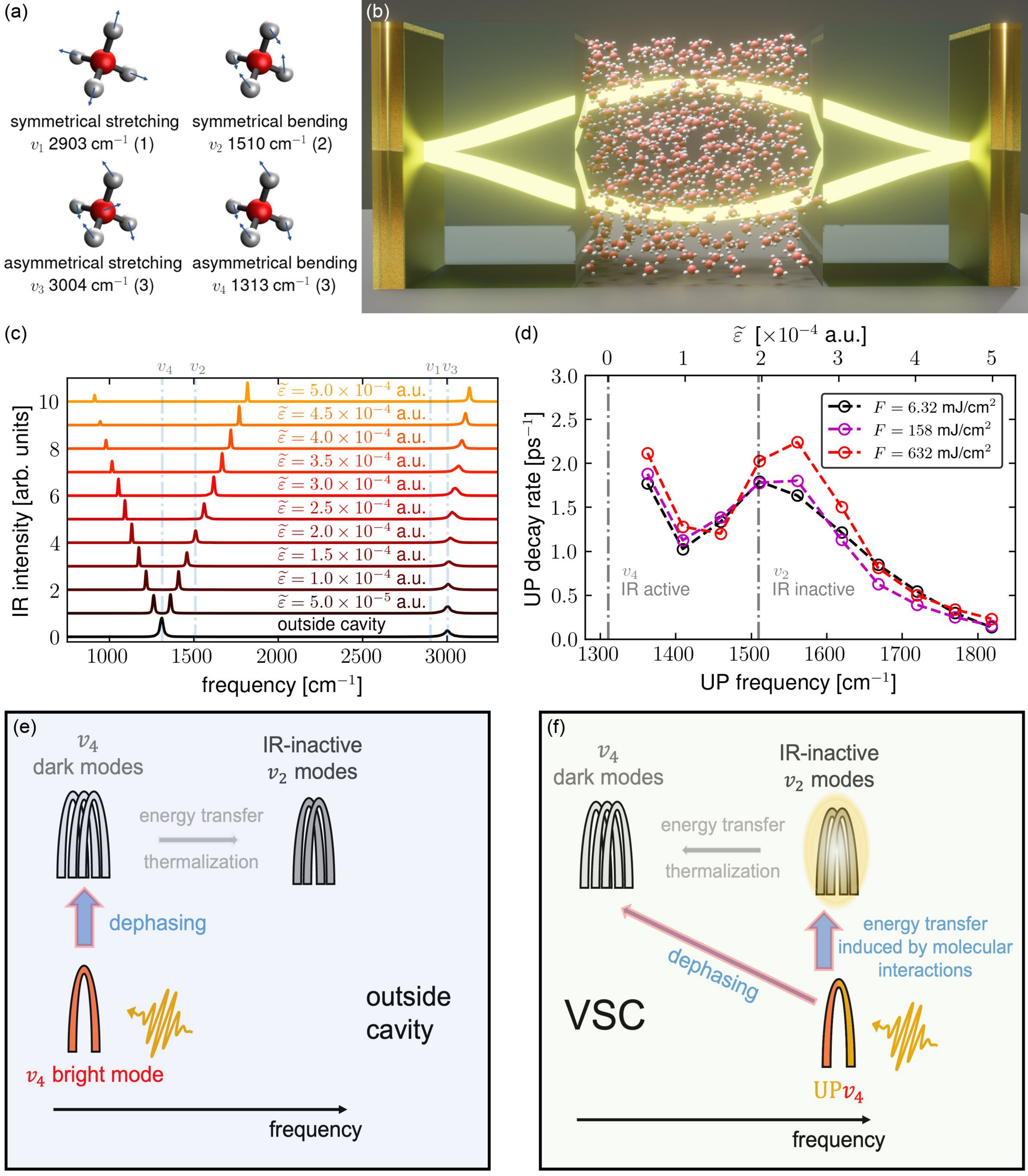}
	    \caption{Liquid-phase methane under VSC. (a) Schematic representation of the four unique vibrational modes in \ch{CH4}. The corresponding vibrational frequencies are obtained from liquid-phase molecular dynamics simulations at 110 K, with the degeneracy of each mode labeled in parentheses. (b) The CavMD simulation setup consists of $N_{\rm {simu}} = 400$ \ch{CH4} molecules coupled to a single cavity mode polarized along both the $x$- and $y$-directions. (c) Simulated linear IR spectra of the molecular system under increased effective light-matter coupling strengths $\widetilde{\varepsilon}$ (bottom to top). The cavity frequency is set to $\omega_{\rm c} = 1311$ cm$^{-1}$, at resonance with the $v_4$ asymmetric bending mode. The locations of the $v_1$-$v_4$ vibrational modes are marked by vertical gray lines. (d) Simulated UP$_{v_4}$ decay rates corresponding to part (c), ranked in ascending order of the polariton frequency. The corresponding  $\widetilde{\varepsilon}$ values for the polaritons are also labeled on the top $x$-axis. The polariton decay rates are fitted from nonequilibrium photonic energy dynamics when a cw pulse is used to resonantly excite the polaritons.  Three different pulse fluences are considered: $F = 6.32$ mJ/cm$^{2}$ (black), 158 mJ/cm$^{2}$ (magenta), and 632 mJ/cm$^{2}$ (red). (e,f) Simplified energy transfer mechanism under external pumping of (e) the $v_4$ mode outside the cavity or (f) the UP$_{v_4}$ under VSC. The thickness of the arrows represents the relative rates of energy transfer. }
	    \label{fig:setup_polariton_rate}
    \end{figure*}

    To investigate the role of IR-inactive vibrational modes during polariton pumping, we study liquid methane (\ch{CH4}) under VSC. This strong coupling system bridges two types of experimentally reported VSC setups, gas-phase methane \cite{Wright2023} and \ch{W(CO)6} or \ch{Fe(CO)5} molecules in the liquid phase \cite{Dunkelberger2016,Xiang2018,Xiang2020Science,Chen2022}. On the one hand, studying the liquid phase allows us to avoid the complexity associated with the vast number of rovibrational  transitions as in gas-phase methane \cite{Wright2023}. On the other hand, using methane instead of \ch{W(CO)6} or \ch{Fe(CO)5} molecules makes it more feasible to analyze the nonequilibrium dynamics of all vibrational normal modes.

    \section{Results} 
    
    As depicted in Fig. \ref{fig:setup_polariton_rate}a, molecular dynamics simulations reveal that due to the tetrahedral ($T_d$) symmetry, liquid \ch{CH4} contains both a triply degenerate, IR-active $v_4$ asymmetric bending mode near 1300 cm$^{-1}$, and a doubly degenerate, IR-inactive symmetric bending mode near 1500 cm$^{-1}$ (see also SI Fig. S1), in good agreement with experimental observations \cite{Crawford1952,Chapados1972}.  We focus on studying the dynamics of the upper polariton (UP) formed by the $v_4$ asymmetric bending mode, which we will refer to as UP$_{v_4}$ henceforth.

    For studying VSC, the CavMD simulation setup, shown in Fig. \ref{fig:setup_polariton_rate}b, consists of $N_{\text{simu}}=400$ liquid-phase methane molecules coupled to a single cavity mode at 110 K. The optical cavity is assumed to be placed along the $z$-direction, with the cavity mode polarized along both the $x$- and $y$-directions. For most calculations presented in this manuscript, the methane molecules are modeled using the standard COMPASS force field \cite{Sun1998} under periodic boundary conditions. This force field is optimized for condensed-phase applications and has been validated against the structures and vibrational frequencies of common organic liquids, including methane \cite{Sun1998}. While the COMPASS force field provides a highly accurate potential energy surface for \ch{CH4} dimers, its description of condensed-phase intermolecular interactions is less satisfactory \cite{Veit2019}.  To address this limitation, we also apply a machine-learning  potential for liquid \ch{CH4} based on the Gaussian Approximation Potential (GAP) framework \cite{Veit2019,Deringer2021}, which improves the modeling of intermolecular interactions.  Nuclear forces are evaluated using LAMMPS \cite{Thompson2022}, and the CavMD simulations are performed using the modified i-PI package \cite{Litman2024,Li2020Water}. Additional simulation details can be  found in SI Secs. II and III. 
    
    When the $v_4$ asymmetric bending  mode of liquid methane is resonantly coupled to a lossless cavity mode at  $\omega_{\rm c} = 1311$ cm$^{-1}$, the linear-response IR spectrum of methane, evaluated from the dipole autocorrection function \cite{McQuarrie1976,Gaigeot2003,Habershon2008,Li2020Nonlinear}, is plotted in Fig. \ref{fig:setup_polariton_rate}c. \red{Outside the cavity (bottom line), apart from the strong $v_4$ band, a very weak peak at the $v_2$ location (approximately 2\% of the $v_4$ intensity) also emerges in the spectrum, which arises due to intramolecular couplings between the $v_2$ and $v_4$ transitions \cite{Childs1939,Robiette1979,Tipping2001}.}  As the effective light-matter coupling per molecule, $\widetilde{\varepsilon}$, is gradually amplified from zero to $5\times 10^{-4}$ a.u. in increments of $5\times 10^{-5}$ a.u. (from bottom to top), a pair of LP and UP states emerges in the spectrum, with the associated Rabi splitting gradually increasing. \red{Very  interestingly, when $\widetilde{\varepsilon}$ reaches $2\times 10^{-5} \sim 3 \times 10^{-5}$ a.u., i.e., when the UP frequency is close to the $v_2$ IR-inactive mode, the UP linewidth becomes broadened. As discussed in the literature\cite{ying2024theory}, such UP linewidth broadening might indicate the UP energy transfer to specific dark states, known as inter-branch scattering.}

    \subsection{UP$_{v_4}$ decay rates} 
    
    We investigate the nonequilibrium  relaxation dynamics of UP$_{v_4}$ by exciting the molecular system with a continuous-wave (cw) pulse. The pulse is defined as $\mathbf{E}(t)=E_0\cos(\omega t)\mathbf{e}_x$ and is applied over the time window $0.1 \text{\ ps} < t < 0.6$ ps, where $E_0$ denotes the pulse amplitude and $\mathbf{e}_x$ represents the unit vector along the $x$-direction. Following the resonant excitation of each UP$_{v_4}$ in Fig. \ref{fig:setup_polariton_rate}c, an exponential fit of the nonequilibrium photon energy dynamics (SI Fig. S2) yields the corresponding UP$_{v_4}$ decay rates,  as shown in Fig. \ref{fig:setup_polariton_rate}d. These decay rates are ordered in the ascending frequency of the UP$_{v_4}$ defined in the linear spectra in Fig. \ref{fig:setup_polariton_rate}c. The corresponding light-matter coupling strength $\widetilde{\varepsilon}$ for each UP$_{v_4}$ frequency is also indicated on the top $x$-axis. 
    
    Overall, Fig. \ref{fig:setup_polariton_rate}d demonstrates that the polariton decay rates are on the order of ps$^{-1}$. Since the cavity loss is turned off during the simulations, this rapid polariton relaxation can only be attributed to the energy transfer to other vibrationally excited-state manifolds of the molecules, such as the dark modes or asymmetric combinations of the $v_4$ vibrations.  Importantly, for three vastly different pulse fluences [$F = 6.32$ mJ/cm$^{2}$ (black), 158 mJ/cm$^{2}$ (magenta), and 632 mJ/cm$^{2}$ (red)], the UP$_{v_4}$ decay rates consistently exhibit a double-peak behavior as the polariton frequency increases. Such behavior cannot be solely explained by the polariton energy transfer to the $v_4$ dark modes. According to this mechanism \cite{Groenhof2019,Li2021Relaxation,Chng2024}, the polariton decay rate should monotonically decrease  as the energy gap between the UP$_{v_4}$ and the original $v_4$ lineshape increases. However,  since the second peak appears around the $v_2$ symmetric bending mode at 1510 cm$^{-1}$, Fig. \ref{fig:setup_polariton_rate}d strongly suggests the involvement of the IR-inactive $v_2$ vibrations during the UP$_{v_4}$ relaxation. \red{Notably, the enhanced UP decay rate when the UP frequency is near the $v_2$ transition is strongly correlated to the UP linewidth broadening in Fig. \ref{fig:setup_polariton_rate}c at  the same frequency range.}

    \begin{figure*}
	    \centering
	    \includegraphics[width=1.0\linewidth]{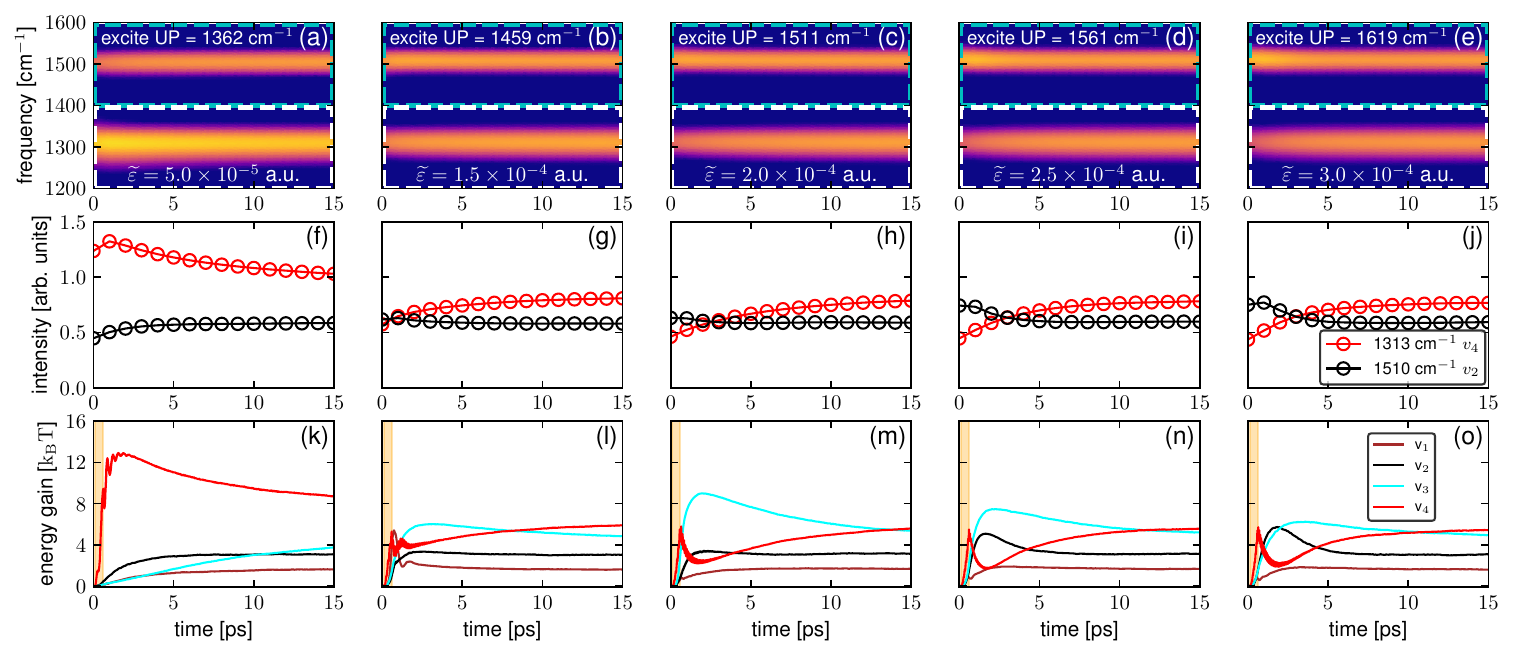}
	    \caption{Average vibrational energy dynamics per molecule following the UP$_{v_4}$ pumping under the pulse fluence of $F = 632$ mJ/cm$^{2}$. (a-e) Average time-resolved spectra of individual \ch{CH4} bending angles under different light-matter coupling strengths $\widetilde{\varepsilon}$. The associated UP frequencies are labeled in each graph. %At each time snapshot $T_i$, the transient spectrum is obtained by Fourier transforming the angle autocorrelation function over the time period $[T_i, T_i + \Delta T_i]$, where $\Delta T_i = 5$ ps. 
        (f-j) Integrated peak intensity dynamics corresponding to the top panel. The integration over the frequency range highlighted in the cyan box (top panel) represents the $v_2$ mode (black dots), whereas that in the white box (top panel) corresponds to the $v_4$ mode (red dots). (k-o) Corresponding average vibrational energy dynamics per molecule using symmetry coordinates. The energy distribution among four vibrational modes is shown: $v_1$ (brown), $v_2$ (black), $v_3$ (cyan), and $v_4$ (red). The orange region in each part represents the time window during which the cw pulse is applied (0.1 ps < $t$ < 0.6 ps). The total vibrational energy in the $v_1$-$v_4$ modes, rather than the vibrational energy divided by the corresponding mode degeneracy, is shown. Under the UP$_{v_4}$ pumping, the IR-inactive $v_2$ mode can sometimes be transiently excited more strongly than the $v_4$ mode within 5 ps. }
	    \label{fig:angle_autocorrelation&symmetry_coordinate_6e-3}
    \end{figure*}

    \subsection{Time-resolved bending dynamics of individual \ch{CH4} molecules}
    
    Because both the $v_4$ and $v_2$ modes correspond to the bending motions of methane, their population dynamics during the UP$_{v_4}$ relaxation can be captured by directly analyzing the motions of individual \ch{H-C-H} bending angles. Figs. \ref{fig:angle_autocorrelation&symmetry_coordinate_6e-3}a-e present the average time-resolved spectra of individual \ch{CH4} bending angles   after resonantly exciting the UP at a pulse fluence of $F = 632$ mJ/cm$^{2}$. Each subplot, from left to right, represents the strong coupling system under an increased light-matter coupling strength $\widetilde{\varepsilon}$. The value of $\widetilde{\varepsilon}$ and the corresponding UP frequency are labeled in each graph, respectively, with the associated linear polariton spectrum provided in Fig. \ref{fig:setup_polariton_rate}c. At each time snapshot $T_i$, the transient bending-angle spectrum  is computed by taking the Fourier transform of the angle autocorrelation function  over the time interval $[T_i, T_i + \Delta T_i]$ with $\Delta T_i = 5$ ps; see SI Sec. III for details. Although the time resolution is low (5 ps), Figs. \ref{fig:angle_autocorrelation&symmetry_coordinate_6e-3}a-e  demonstrate that both the $v_4$ and $v_2$ vibrations, peaking around 1300 cm$^{-1}$ and 1500 cm$^{-1}$, respectively, are excited following the UP relaxation. As clearly shown in the integrated intensity dynamics of the two peaks (Figs. \ref{fig:angle_autocorrelation&symmetry_coordinate_6e-3}f-j), once the UP frequency exceeds approximately 1500 cm$^{-1}$,  the UP$_{v_4}$ pumping leads to a stronger excitation of the IR-inactive $v_2$ bending mode (black dots) compared to the IR-active $v_4$ mode (red dots).

     \subsection{UP-induced energy transfer using symmetry coordinates}
     
    In an effort to understand the UP$_{v_4}$ energy transfer with a higher time resolution, we further analyze the population dynamics of the four symmetry coordinates of \ch{CH4} \cite{Lazzeretti1987,Wang2003} during the UP$_{v_4}$ relaxation. Due to the $T_d$ symmetry of \ch{CH4} molecules, the four symmetry coordinates, which are linear combinations of \ch{H-C-H} angles or \ch{C-H} bond lengths, separate the $v_1$-$v_4$ vibrational modes  (See SI Sec. III for details) \cite{Lazzeretti1987,Wang2003}. With these symmetry coordinates, Figs. \ref{fig:angle_autocorrelation&symmetry_coordinate_6e-3}k-o present the average vibrational population dynamics of the $v_1$-$v_4$ modes per molecule corresponding to  Figs. \ref{fig:angle_autocorrelation&symmetry_coordinate_6e-3}f-j with a time resolution of 0.5 fs (the CavMD simulation time step). Overall, the vibrational dynamics of $v_2$ (black) and $v_4$ (red) modes confirm that, once the UP$_{v_4}$ frequency exceeds approximately 1500 cm$^{-1}$, the  transient $v_2$ excitation  becomes more prominent than that of $v_4$ within 5 ps. This trend is in qualitative agreement with the findings in Figs. \ref{fig:angle_autocorrelation&symmetry_coordinate_6e-3}f-j.  
    
    Moreover, with a high time resolution, the $v_4$ dynamics in Figs. \ref{fig:angle_autocorrelation&symmetry_coordinate_6e-3}m-o show an initial increase in population during the pulse excitation in the time window $0.1 \mathrm{\ ps} < t < 0.6$ ps (orange region). This occurs because the bright mode of $v_4$ vibrations constitutes approximately half of the UP. Consequently, exciting the UP leads to an increase in the $v_4$ population. At later times ($t \lesssim 2$ ps),  the $v_4$ population undergoes a rapid decrease during the UP$_{v_4}$ relaxation, indicating energy transfer from the $v_4$ bright mode to other vibrational modes rather than to the $v_4$ dark modes. After the UP relaxation ($t \gtrsim 2$ ps), the $v_4$ population is accumulated again, as the excess vibrational energy in the other modes gradually transfers back to $v_4$ \red{due to intramolecular vibrational energy redistribution}. The fast $v_4$ population relaxation when $t \lesssim 2$ ps is not clearly observed in Figs. \ref{fig:angle_autocorrelation&symmetry_coordinate_6e-3}k and \ref{fig:angle_autocorrelation&symmetry_coordinate_6e-3}l. This suggests that, at small Rabi splitting values, most the UP energy is transferred (or dephased) to the $v_4$ dark modes. After all, the bright mode and dark modes of $v_4$ vibrations cannot be distinguished from the average $v_4$ vibrational energy per molecule. The $v_2$ and $v_4$ dynamics presented in Figs. \ref{fig:angle_autocorrelation&symmetry_coordinate_6e-3}e,j,o are further validated when liquid \ch{CH4} is modeled using the GAP machine-learning potential trained on first-principles potential energy surfaces \cite{Veit2019}; see SI Fig. S16 for details.

    A perhaps surprising finding in Figs. \ref{fig:angle_autocorrelation&symmetry_coordinate_6e-3}k-o is the strong transient excitation of the $v_3$ asymmetric stretching mode at 3004 cm$^{-1}$ (cyan) when the UP frequency is close to 1500 cm$^{-1}$  (Fig. \ref{fig:angle_autocorrelation&symmetry_coordinate_6e-3}m). This behavior cannot be explained by direct energy transfer from the UP to $v_3$ vibrations due to their spectral overlap. If this were the underlying mechanism, one would expect an even stronger excitation of $v_3$ with a further increase in the UP frequency, which contradicts the weaker $v_3$ excitation shown in Figs. \ref{fig:angle_autocorrelation&symmetry_coordinate_6e-3}n and \ref{fig:angle_autocorrelation&symmetry_coordinate_6e-3}o. We postulate that the strong transient excitation of $v_3$  originates from the UP$_{v_4}$ $+\  v_2 \rightarrow v_3$ energy transfer pathway, which reaches resonance when  the UP frequency is close to 1500 cm$^{-1}$. Since  this side mechanism depletes the $v_2$ population,  it explains the relatively weak energy gain in the IR-inactive $v_2$ mode when  UP$_{v_4}$ is near 1500 cm$^{-1}$ (Fig. \ref{fig:angle_autocorrelation&symmetry_coordinate_6e-3}m), despite  UP$_{v_4}$ $\rightarrow v_2$ also reaching resonance at this frequency. 
    
    When the UP frequency exceeds 1500 cm$^{-1}$ (Figs. \ref{fig:angle_autocorrelation&symmetry_coordinate_6e-3}n and \ref{fig:angle_autocorrelation&symmetry_coordinate_6e-3}o),  both UP$_{v_4}$ $\rightarrow v_2$  and UP$_{v_4}$ $+\  v_2 \rightarrow v_3$ should slow down due to the reduced spectral overlap. However, since UP$_{v_4}$ $+\  v_2 \rightarrow v_3$ is a second-order process which depends on the generation rate of $v_2$, at large Rabi splitting values, the UP$_{v_4}$ $+\  v_2 \rightarrow v_3$ pathway is expected to slow down more drastically than the first-order process, UP$_{v_4}$ $\rightarrow v_2$. Due to this competing behavior, the transient $v_2$ population can be  more effectively preserved at large Rabi splitting values, as observed in Figs. \ref{fig:angle_autocorrelation&symmetry_coordinate_6e-3}n and \ref{fig:angle_autocorrelation&symmetry_coordinate_6e-3}o. 
    
    Further analysis of vibrational energy transfer under different pulse fluences suggests a weak nonlinear pathway, 2 UP$_v{_{4}} \rightarrow v_1$, when UP$_v{_{4}}$ = 1459 cm$^{-1}$ (SI Fig. S3). Because $v_1$ is IR-inactive, this nonlinear pathway differs from conventional multiphoton processes.  This analysis also suggests the possible existence of a similar nonlinear pathway, 2 UP$_v{_{4}} \rightarrow v_3$. However, SI Fig. S3 indicates that even if this nonlinear pathway exists, this pathway would be significantly weaker than the UP$_v{_{4}} \rightarrow v_2$ and UP$_{v_4}$ $+\  v_2 \rightarrow v_3$ pathways, all of which are at resonance when UP$_v{_{4}}$  is near 1500 cm$^{-1}$. \red{Overall, while the linear IR spectrum in Fig. \ref{fig:setup_polariton_rate}c suggests the UP$_{v_4}$ linewidth broadening around 1500 cm$^{-1}$, which correlates with the enhanced UP$_{v_4}$ decay rate, the above analysis using symmetry coordinates reveal multiple competing UP$_{v_4}$ energy transfer pathways around 1500 cm$^{-1}$, highlighting the unique role of realistic simulations. }

    \subsection{Understanding UP$_{v_4}$ decay with Fermi's golden rule}
    
    Guided by the CavMD results on UP$_{v_4}$ energy transfer to $v_2$ and $v_4$ vibrational modes, we use the Fermi's golden rule to evaluate the decay rate of UP$_{v_4}$, denoted as $\gamma_{{\mathrm{UP}_{v_4}}}$. Because the cavity loss is turned off in the simulations, the  UP$_{v_4}$ decay rate can be expressed as  $\gamma_{{\mathrm{UP}_{v_4}}} = \gamma_{{\mathrm{UP}_{v_4} \rightarrow \mathrm{D}_{v_4}}} + \gamma_{{\mathrm{UP}_{v_4} \rightarrow {v_2}}}$. The first term, $\gamma_{{\mathrm{UP}_{v_4} \rightarrow \mathrm{D}_{v_4}}}$, represents the polariton dephasing rate to the dark modes of $v_4$ vibrations, while the second term, $\gamma_{{\mathrm{UP}_{v_4} \rightarrow {v_2}}}$, accounts for the polariton energy transfer to the IR-inactive $v_2$ vibrations. As the second-order UP$_{v_4}$ $+\  v_2 \rightarrow v_3$ pathway depends on the generation rate of $v_2$ vibrations, the rate of this side mechanism is proportional to $\gamma_{{\mathrm{UP}_{v_4} \rightarrow {v_2}}}$,  thus effectively rescaling the value of $\gamma_{{\mathrm{UP}_{v_4} \rightarrow {v_2}}}$. For simplicity, we do not explicitly include the UP$_{v_4}$ $+\  v_2 \rightarrow v_3$ pathway in analytical derivations.

    In the harmonic limit, UP$_{v_4}$ is decoupled from the dark modes of $v_4$ vibrations and the IR-inactive $v_2$ vibrations. By introducing various inter- or intramolecular interactions perturbatively, following Ref. \citenum{Li2021Relaxation}, we obtain the analytical decay rates as follows (see SI Sec. I for detailed derivations):
    \begin{subequations}\label{eq:UP_decay_rate}
        \begin{align}
            \gamma_{{\mathrm{UP}_{v_4} \rightarrow \mathrm{D}_{v_4}}} &= 2\pi (\Delta_{\mathrm{dd}}^2 + \Xi_{44}^2) \vert X_{+}^{(\mathrm{B})}\vert ^2 J_{\mathrm{UP}_{v_4}, {v_4}} \label{eq:UP_decay_rate_v4} ,\\
            \gamma_{{\mathrm{UP}_{v_4} \rightarrow {v_2}}} &= 2\pi (\Xi_{24}^2 + Z_{24}^2) \vert X_{+}^{(\mathrm{B})}\vert ^2 J_{\mathrm{UP}_{v_4}, {v_2}} \label{eq:UP_decay_rate_v2} .
        \end{align}
    \end{subequations}
    Here, $\Delta_{\mathrm{dd}}$ denotes the average intermolecular dipole-dipole coupling between the IR-active $v_4$ vibrations; $\Xi_{44}$ and $\Xi_{24}$ represent the average intramolecular anharmonic coupling within the triply degenerate $v_4$ vibrations and that between $v_2$ and $v_4$ vibrations, respectively; $Z_{24}$ denotes the intramolecular Coriolis interaction between  $v_2$ and $v_4$ due to rovibrational coupling \cite{Childs1939,Robiette1979,Tipping2001}. $\vert X_{+}^{(\mathrm{B})}\vert ^2$ represents the weight of the $v_4$ bright mode in UP$_{v_4}$, which is approximately $1/2$ at resonance strong coupling. $J_{\mathrm{UP}_{v_4}, {v_4}}$ and $J_{\mathrm{UP}_{v_4}, {v_2}}$ are the overlap integrals between the UP$_{v_4}$ and the vibrational density of states of $v_4$ and $v_2$ modes, respectively. \red{Due to the important role of the overlap integral, it is clear that the LP formed by the $v_4$ mode cannot efficiently transfer energy to the $v_2$ state.} 

    \begin{figure*}
	    \centering
	    \includegraphics[width=0.7\linewidth]{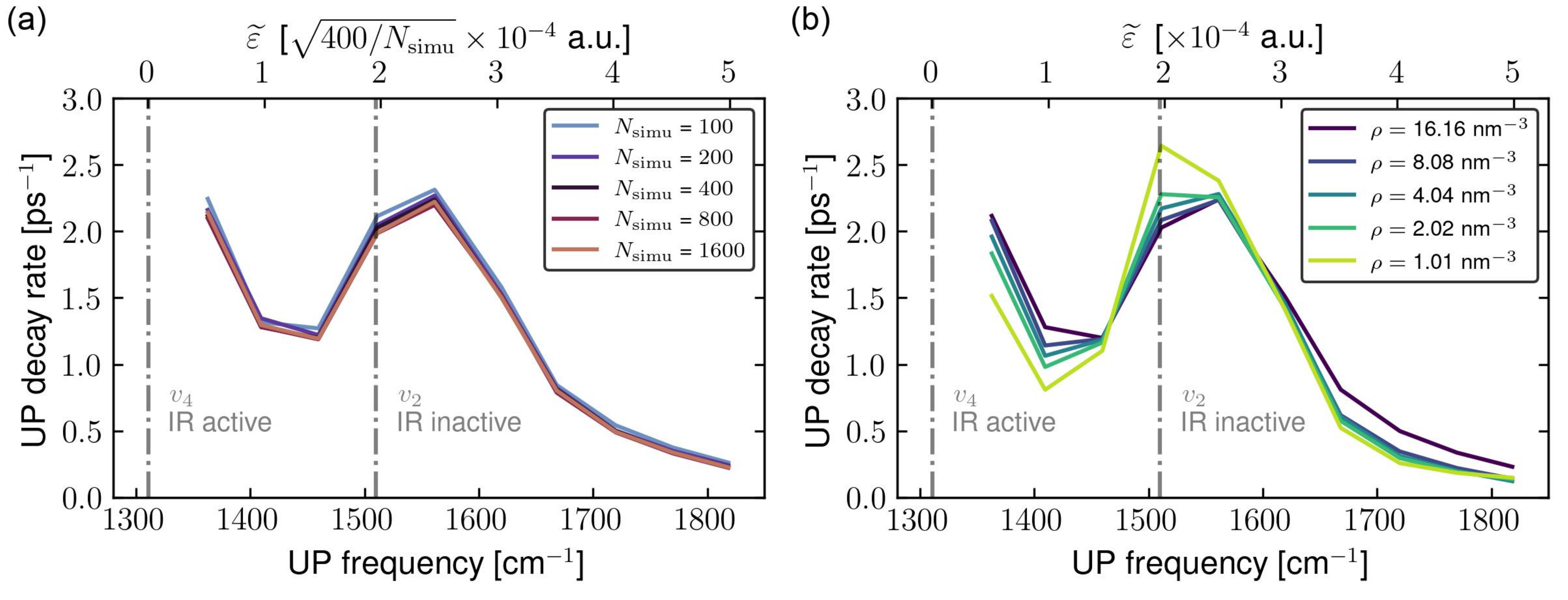}
	    \caption{Parameter dependence on the fitted UP$_{v_4}$ decay rates analogous to Fig. \ref{fig:setup_polariton_rate}d. (a) Simulations conducted at the constant molecular density. The number of simulated molecules, $N_{\rm{simu}}$, is varied as 100 (blue), 200 (purple), 400 (black), 800 (brown), and 1600 (magenta), while $\widetilde{\varepsilon} \propto 1/\sqrt{N_{\rm{simu}}}$ is adjusted to maintain a fixed polariton frequency as $N_{\rm{simu}}$ changes. (b) Simulations conducted at the constant molecular number and varied molecular number densities: $\rho = $ 16.16, 8.08, 4.04, 2.02, and 1.01 nm$^{-3}$ (from black to light green). The highest density corresponds to the liquid phase. The molecular density is reduced by increasing the simulation cell size. The parameter dependence of  the fitted UP$_{v_4}$ decay rates qualitatively agrees with the analytical results in Eq. \eqref{eq:UP_decay_rate}.}
	    \label{fig:parameter_dependence}
    \end{figure*}

    \subsection{Examining the   golden rule UP$_{v_4}$ decay rate with simulations}
    
    Eq. \eqref{eq:UP_decay_rate} indicates that the  $\mathrm{UP}_{v_4}$ decay rate is independent of the total number of molecules, provided that the molecular interactions, the molecular weight $\vert X_{+}^{(\mathrm{B})}\vert ^2$,  and spectral overlaps remain fixed. Our simulations can test this predicted invariance with respect to the simulated molecular number $N_{\rm{simu}}$ under the following constraints: (i) the fixed molecular interactions are ensured by simulating a molecular system with constant density and temperature; (ii) the Rabi splitting is maintained by reducing the effective light-matter coupling strength $\widetilde{\varepsilon} \propto 1/\sqrt{N_{\rm{simu}}}$ as $N_{\rm{simu}}$ increases \cite{Li2020Nonlinear}; and (iii) given an unchanged Rabi splitting, $\vert X_{+}^{(\mathrm{B})}\vert ^2$ remains constant by fixing the cavity frequency at $\omega_{\rm c} = 1311$ cm$^{-1}$. 
    
    With these constraints, Fig. \ref{fig:parameter_dependence}a demonstrates the UP$_{v_4}$ decay rate as a function of its frequency under various values of $N_{\rm{simu}}$. Notably, for $N_{\rm{simu}}\geq 200$ (all lines except the light blue one), the simulated UP$_{v_4}$ decay rates remain unchanged. This  independence from $N_{\rm{simu}}$ is further confirmed by directly comparing the symmetry coordinate dynamics of molecules (SI Fig. S4). Overall, the invariance with respect to $N_{\rm{simu}}$  validates that our microscopic simulations  can be directly compared to VSC experiments involving a macroscopic number of molecules. 

    More interestingly, for $N_{\rm{simu}} = 400$, if the molecular density is gradually reduced by increasing the volume of the simulation cell, the simulated UP$_{v_4}$ decay rate in Fig. \ref{fig:parameter_dependence}b decreases when the UP$_{v_4}$ frequency is close to the $v_4$ vibrations, whereas the rate increases when the frequency is near the IR-inactive $v_2$ vibrations. Since all the other simulation conditions remain the same as those in Fig. \ref{fig:parameter_dependence}a, these density-dependent UP$_{v_4}$ decay rates can be attributed to molecular interactions driving the UP$_{v_4}$ energy transfer. 
    
    When the UP$_{v_4}$ is near the $v_4$ vibrations, the ${{\mathrm{UP}_{v_4} \rightarrow \mathrm{D}_{v_4}}}$ pathway dominates the UP$_{v_4}$ relaxation. Because the intermolecular $v_4$ dipole-dipole coupling $\Delta_{\mathrm{dd}}$ is greatly weakened at lower molecular densities, the corresponding UP$_{v_4}$ decay rate, as described in Eq. \eqref{eq:UP_decay_rate_v4}, decreases, consistent with previous calculations \cite{Li2021Relaxation}.  By contrast, when the UP$_{v_4}$ frequency is close to the $v_2$ lineshape, the ${{\mathrm{UP}_{v_4} \rightarrow {v_2}}}$ pathway dominates the UP$_{v_4}$ relaxation. According to Eq. \eqref{eq:UP_decay_rate_v2}, as polariton energy transfer in this case involves only intramolecular interactions, the UP$_{v_4}$ decay rate cannot be reduced when the molecular density decreases. Instead, as reducing the molecular density librates the molecular rotations, it enhances the $v_2$-$v_4$  intramolecular Coriolis interaction $Z_{24}$, thereby significantly amplifying the UP$_{v_4}$ decay rate. The nonequilibrium symmetry coordinate dynamics corresponding to Fig. \ref{fig:parameter_dependence}b are also plotted in SI Fig. S5. Overall, Fig. \ref{fig:parameter_dependence} provides numerical results which are  in qualitative agreement with the analytical UP$_{v_4}$ decay rate presented in Eq. \eqref{eq:UP_decay_rate}.

    \section{Discussion}
    
    Our simulations demonstrate transient, strong energy accumulation in IR-inactive $v_2$ vibrations following the $\mathrm{UP}_{v_4}$ excitation under suitable conditions. However, our simulations above have three major limitations: (i) the cw pulse, which is broad in the frequency domain, may not selectively excite the $\mathrm{UP}_{v_4}$; (ii) the pulse is assumed to excite the molecular subsystem, whereas in realistic cavities, the cavity modes --- rather than the molecules --- exhibit strong optical absorption \cite{Carusotto2013}; and (iii) the cavity is assumed to be lossless. While applying these three approximations simplifies the numerical fitting of the polariton relaxation rates, reduces the number of required parameters, and facilitates the comparison with analytical theory, it also limits the transferability of our conclusions and hinder the direct comparison with experiments.

    To better account for realistic experimental conditions, following Ref. \citenum{Li2021Solute}, we will instead apply a Gaussian pulse to excite the cavity mode. The cavity loss will be incorporated by coupling the cavity mode to a Langevin thermostat. The cavity lifetime will be set to 0.75 ps, a value that balances with  the cavity transition dipole moment in accordance with the input-output theory \cite{Li2021Solute,Carusotto2013}; see SI Sec. III for simulation details.

    \subsection{More realistic simulations} 
    
    In Fig. \ref{fig:figure4}a, we present additional symmetry coordinate dynamics analogous to those in Fig. \ref{fig:angle_autocorrelation&symmetry_coordinate_6e-3}m when  a Gaussian pulse is applied to excite the lossy cavity mode.  Overall, using these more realistic parameters reproduces symmetry coordinate dynamics that are highly similar to those in Fig. \ref{fig:angle_autocorrelation&symmetry_coordinate_6e-3}m, providing validation of our simulations above. The symmetry coordinate and photonic dynamics at different UP frequencies are also included in SI Figs. S6 and S7. 

    \subsection{The role of polariton formation}
    
\begin{figure*}
	    \centering
	    \includegraphics[width=1.0\linewidth]{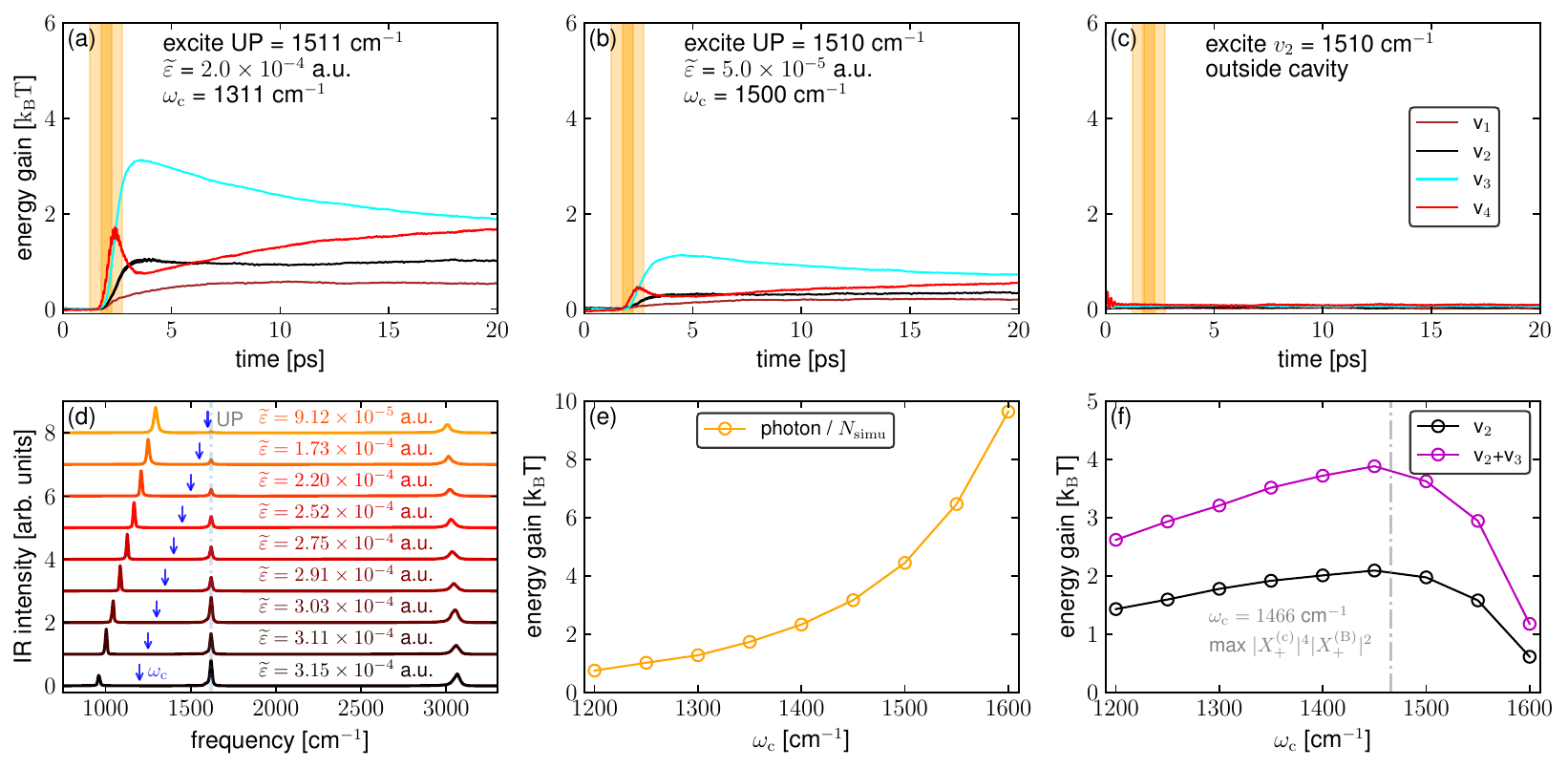}
	    \caption{UP$_{v_4}$ pumping simulations with a Gaussian pulse exciting a lossy cavity mode. (a) Average vibrational energy dynamics per molecule analogous to Fig.  \ref{fig:angle_autocorrelation&symmetry_coordinate_6e-3}m, with $\omega_{\rm c} = 1311$ cm$^{-1}$, $\widetilde{\varepsilon} = 2\times 10^{-4}$ a.u., and the UP frequency 1511 cm$^{-1}$. (b) Analogous vibrational energy dynamics when $\omega_{\rm c} = 1500$ cm$^{-1}$ and $\widetilde{\varepsilon} = 5\times 10^{-5}$ a.u., yielding nearly the same UP frequency as in part (a). (c) Analogous vibrational energy dynamics when the Gaussian pulse excites the molecules outside the cavity at frequency 1510 cm$^{-1}$.  (d) Linear polariton spectra calculated under different $\{\omega_{\rm c}, \widetilde{\varepsilon}\}$ values. These two parameters (labeled on each lineshape) are properly balanced to maintain the fixed UP frequency at 1619 cm$^{-1}$ (vertical gray line). (e,f) Maximum transient energy gain during the  Gaussian pulse excitation of each UP labeled in part (d).  Three degrees of freedom are shown: (e)  cavity mode energy normalized  by the number of molecules (orange dots);  (f) vibrational energy per molecule for $v_2$ (black dots) and $v_2 + v_3$ (magenta dots).   }
	    \label{fig:figure4}
    \end{figure*}

    If a blue-shifted cavity mode at $\omega_{\rm c} = 1500$ cm$^{-1}$ is coupled to the molecular system with an effective coupling strength of $\widetilde{\varepsilon} = 5\times 10^{-5}$ a.u. (corresponding to a Rabi splitting of approximately 100 cm$^{-1} \sim 0.33$ ps), the resulting UP peaks at 1510 cm$^{-1}$, nearly identical to the UP frequency under resonance strong coupling in Fig. \ref{fig:figure4}a. When a Gaussian pulse resonantly excites the UP, the resulting symmetry coordinate dynamics, as shown in Fig. \ref{fig:figure4}b, are significantly suppressed. Since the cavity lifetime is set to 0.75 ps (corresponding to a cavity decay rate approximately half the Rabi splitting when $\widetilde{\varepsilon} = 5\times 10^{-5}$ a.u.), this simulation lies at the interface between strong coupling and weak coupling. More comprehensive polariton-induced energy transfer dynamics when $\omega_{\rm c} = 1500$ cm$^{-1}$ are also included in SI Figs. S8-S10.
    
    Moreover, outside the cavity, when the same Gaussian pulse is  used to directly excite the $v_2$ vibrations at 1510 cm$^{-1}$, almost no molecular response is observed. Comparing the resonance strong coupling result in Fig. \ref{fig:figure4}a with Figs. \ref{fig:figure4}b and \ref{fig:figure4}c,  where the UP and (or) pulse frequency remain mostly unchanged, it is evident that strong coupling plays a crucial role in achieving significant transient energy accumulation in IR-inactive $v_2$ vibrations and other vibrationally excited-state manifolds.

    To further investigate the role of polariton formation in promoting the IR-inactive $v_2$ excitation, we systematically compare a set of strong coupling systems under different combinations of $\omega_{\rm{c}}$ and $\widetilde{\varepsilon}$; see Fig. \ref{fig:figure4}d for the corresponding linear polariton spectra and the associated parameter values. The combinations of $\omega_{\rm{c}}$ and $\widetilde{\varepsilon}$ are carefully balanced to maintain the UP frequencies fixed at 1619 cm$^{-1}$, the same as in Fig. \ref{fig:angle_autocorrelation&symmetry_coordinate_6e-3}o. Since the cavity lifetime is 0.75 ps and  $\widetilde{\varepsilon} \gtrsim 10^{-4}$ a.u. (corresponding to the Rabi splitting of $\gtrsim$ 200 cm$^{-1} \sim 0.17$ ps), the strong coupling condition is satisfied throughout Fig. \ref{fig:figure4}d.

    When a Gaussian pulse resonantly excites each UP$_{v_4}$ shown  in Fig. \ref{fig:figure4}d, SI Figs. S11 and S12 demonstrate the time-resolved polariton-induced molecular dynamics. Fig. \ref{fig:figure4}e illustrates the maximum transient photonic energy (orange dots) during the UP$_{v_4}$ excitation as a function of the corresponding cavity mode frequency $\omega_{\rm c}$. The transient photon energy monotonically increases as $\omega_{\rm c}$ blue-shifts. This occurs because, for the fixed UP frequency, an increase in $\omega_{\rm c}$ results in  a larger photonic weight in the corresponding UP. As the cavity photon exclusively responds to the external field, increasing the photonic weight in the UP leads to  stronger excitation of the photonic degrees of freedom. 
    
    By contrast, as shown in Fig. \ref{fig:figure4}f, the maximum transient $v_2$ energy (black dots, corresponding to the peak value in the symmetry coordinate dynamics) exhibits a turnover behavior around $\omega_{\rm c} \sim 1450$ cm$^{-1}$. The same turnover behavior is also observed in the maximum energy gain of $v_2 + v_3$ (magenta dots). According to the UP$_{v_4}$ $+\  v_2 \rightarrow v_3$ pathway, the generation of $v_3$ requires the consumption of $v_2$ excitations, so the maximum energy gain in $v_2 + v_3$ provides an alternative  measure of the magnitude of $v_2$ excitations. 
    
    We now explore the underlying mechanism of this turnover behavior. The magnitude of polaritonic energy absorption due to external pulse excitation is proportional to $|\mu_{\rm UP}E_0|^2$, where $\mu_{\rm UP}$ represents the transition dipole moment of the UP. Denoting the photonic weight in the UP$_{v_4}$ as $|X_{+}^{(\rm{c})}|^2$, $\mu_{\rm UP}$ can be expressed as $|X_{+}^{(\rm{c})}|^2\mu_{\rm c}$, where $\mu_{\rm c}$ represents the transition dipole moment of the cavity mode. Consequently, (i) the magnitude of polaritonic energy absorption is proportional to $E_0^2 |X_{+}^{(\rm{c})}|^4$. According to the analytical UP energy transfer rate in Eq. \eqref{eq:UP_decay_rate}, (ii) the $v_2$ energy gain is proportional to the molecular weight $|X_{+}^{(\rm{B})}|^2$. Combining these two considerations, we can qualitatively express the magnitude of $v_2$ excitation under external pulse pumping as
    \begin{equation}\label{eq:v2_efficiency}
        E_{v2} \propto E_0^2|X_{+}^{(\rm{c})}|^4 |X_{+}^{(\rm{B})}|^2 .
    \end{equation}
    Since $|X_{+}^{(\rm{c})}|^2 + |X_{+}^{(\rm{B})}|^2 =1$, the maximum $v_2$ excitation should occur at $|X_{+}^{(\rm{c})}|^2 = 2/3$, corresponding to a cavity frequency of $\omega_{\rm c} = 1466$ cm$^{-1}$ based on our parameters; see SI Sec. III for detailed calculations. This peak $\omega_{\rm c}$ value agrees with the $v_2$ energy gain trend in Fig. \ref{fig:figure4}f. 
    
    Overall, Eq. \eqref{eq:v2_efficiency} suggests that the maximum $\mathrm{UP}_{v_4}\rightarrow v_2$ excitation requires meaningful contributions from both photons and the molecules. Therefore,  polariton formation --- i.e., the hybridization of light and matter states --- is crucial for promoting this energy transfer. Eq. \eqref{eq:v2_efficiency} also indicates that when the strong coupling system deviates from the optimal light-matter hybridization (i.e., $|X_{+}^{(\rm{c})}|^2 = 2/3$), the same magnitude of transient $v_2$ excitation can still be achieved by increasing the pulse amplitude $E_0$. This observation suggests that, given the fixed pulse pumping intensity, the optimal light-matter hybridization (i.e., when $|X_{+}^{(\rm{c})}|^2 = 2/3$) provides the most  efficient means of accumulating energy in IR-inactive $v_2$ vibrations. This finding suggests that strong coupling might provide an advantageous platform  for improving the energy conversion efficiency in, e.g., solar cells.

    Our simulations reveal that polariton pumping can selectively excite IR-inactive vibrational modes, but rapid energy redistribution in liquids limits the practical applicability of this mechanism. SI Fig. S13 shows that lowering molecular density can prolong the selective excitation of IR inactive modes. \red{This observation also suggests that our finding is robust with respect to the solvent environment, as intramolecular interactions are important for inducing this polariton energy transfer [Eq. \eqref{eq:UP_decay_rate}].} Additionally, control isotope simulations for liquid \ch{CD4} under VSC (SI Fig. S15) confirm the universality of the mechanism and highlight that a substantial frequency gap between IR-active and IR-inactive modes is crucial for effective energy transfer to IR-inactive modes. \red{ Lastly, while the self-consistency between classical CavMD simulations and analytical golden rule rates suggests the robustness of our findings, future work should address how including nuclear and photonic quantum effects may impact polariton-induced energy transfer.}

    \section{Conclusion}
    
    The schematic in Fig. \ref{fig:setup_polariton_rate}f provides a simplified summary of how exciting the $\mathrm{UP}_{v_4}$ may yield a transient, strong energy accumulation in IR-inactive $v_2$ vibrations via polariton pumping. Outside the cavity, this energy accumulation cannot occur when an IR pulse is used to excite the $v_4$ vibrations directly (Fig. \ref{fig:setup_polariton_rate}e). Both numerical simulations and analytical derivations highlight that this polariton-induced energy accumulation in IR-inactive modes requires the polariton to simultaneously contain significant photonic and molecular components. Therefore, reaching the optimal light-matter hybridization (i.e., when $|X_{+}^{(\rm{c})}|^2 = 2/3$) provides the most efficient means for observing this polariton-induced process. 

    Based on this study, a mechanistic understanding of existing ultrafast VSC experiments, despite their involvement of more complex molecular systems with high symmetry \cite{Xiang2020Science,Chen2022}, is within reach. Particularly, although the lifetime of vibrational polaritons is very short, our simulations highlight that exciting polaritons can indeed create a unique molecular excited-state distribution in the dark states. This process could ultimately alter IR photochemistry on a timescale much longer than the polariton lifetime as observed in experiments \cite{Chen2022}.    Beyond VSC, our simulations may also provide  insights on understanding energy transfer between exciton-polaritons and triplet states \cite{Bhuyan2023}, a process for which first-principles simulations remain highly challenging. Overall, numerical simulations offer a unique perspective on understanding how  polariton formation can be leveraged to control symmetry-protected molecular excitations in the dark, a mechanism which may impact a broad range of light-induced chemical processes.

      \section{Acknowledgments}
      \red{This material is based upon work supported by the U.S. National Science Foundation under Grant No. CHE-2502758.}
      This research is also supported in part through the use of Information Technologies (IT) resources at the University of Delaware, specifically the high-performance computing resources. We thank Prof. Marissa Weichman, Prof. Wei Xiong, Dr. Jeff Owrutsky, Dr. Blake Simpkins, and Dr. Michael Michon for insightful discussions.

      \section{Data Availability Statement}
      The code and input files are available on Github at \url{https://github.com/TaoELi/cavity-md-ipi}. The SI contains analytical derivation of the UP$_{v_4}$ decay rate, methods and simulation details, and supplementary simulation data.

    %\bibliography{references.bib}

\begin{thebibliography}{70}%
\makeatletter
\providecommand \@ifxundefined [1]{%
 \@ifx{#1\undefined}
}%
\providecommand \@ifnum [1]{%
 \ifnum #1\expandafter \@firstoftwo
 \else \expandafter \@secondoftwo
 \fi
}%
\providecommand \@ifx [1]{%
 \ifx #1\expandafter \@firstoftwo
 \else \expandafter \@secondoftwo
 \fi
}%
\providecommand \natexlab [1]{#1}%
\providecommand \enquote  [1]{``#1''}%
\providecommand \bibnamefont  [1]{#1}%
\providecommand \bibfnamefont [1]{#1}%
\providecommand \citenamefont [1]{#1}%
\providecommand \href@noop [0]{\@secondoftwo}%
\providecommand \href [0]{\begingroup \@sanitize@url \@href}%
\providecommand \@href[1]{\@@startlink{#1}\@@href}%
\providecommand \@@href[1]{\endgroup#1\@@endlink}%
\providecommand \@sanitize@url [0]{\catcode `\\12\catcode `\$12\catcode `\&12\catcode `\#12\catcode `\^12\catcode `\_12\catcode `\%12\relax}%
\providecommand \@@startlink[1]{}%
\providecommand \@@endlink[0]{}%
\providecommand \url  [0]{\begingroup\@sanitize@url \@url }%
\providecommand \@url [1]{\endgroup\@href {#1}{\urlprefix }}%
\providecommand \urlprefix  [0]{URL }%
\providecommand \Eprint [0]{\href }%
\providecommand \doibase [0]{https://doi.org/}%
\providecommand \selectlanguage [0]{\@gobble}%
\providecommand \bibinfo  [0]{\@secondoftwo}%
\providecommand \bibfield  [0]{\@secondoftwo}%
\providecommand \translation [1]{[#1]}%
\providecommand \BibitemOpen [0]{}%
\providecommand \bibitemStop [0]{}%
\providecommand \bibitemNoStop [0]{.\EOS\space}%
\providecommand \EOS [0]{\spacefactor3000\relax}%
\providecommand \BibitemShut  [1]{\csname bibitem#1\endcsname}%
\let\auto@bib@innerbib\@empty
%</preamble>
\bibitem [{\citenamefont {Shalabney}\ \emph {et~al.}(2015)\citenamefont {Shalabney}, \citenamefont {George}, \citenamefont {Hutchison}, \citenamefont {Pupillo}, \citenamefont {Genet},\ and\ \citenamefont {Ebbesen}}]{Shalabney2015}%
  \BibitemOpen
  \bibfield  {author} {\bibinfo {author} {\bibfnamefont {A.}~\bibnamefont {Shalabney}}, \bibinfo {author} {\bibfnamefont {J.}~\bibnamefont {George}}, \bibinfo {author} {\bibfnamefont {J.}~\bibnamefont {Hutchison}}, \bibinfo {author} {\bibfnamefont {G.}~\bibnamefont {Pupillo}}, \bibinfo {author} {\bibfnamefont {C.}~\bibnamefont {Genet}},\ and\ \bibinfo {author} {\bibfnamefont {T.~W.}\ \bibnamefont {Ebbesen}},\ }\bibfield  {title} {\bibinfo {title} {{Coherent Coupling of Molecular Resonators with a Microcavity Mode}},\ }\href {https://doi.org/10.1038/ncomms6981} {\bibfield  {journal} {\bibinfo  {journal} {Nat. Commun.}\ }\textbf {\bibinfo {volume} {6}},\ \bibinfo {pages} {5981} (\bibinfo {year} {2015})}\BibitemShut {NoStop}%
\bibitem [{\citenamefont {Long}\ and\ \citenamefont {Simpkins}(2015)}]{Long2015}%
  \BibitemOpen
  \bibfield  {author} {\bibinfo {author} {\bibfnamefont {J.~P.}\ \bibnamefont {Long}}\ and\ \bibinfo {author} {\bibfnamefont {B.~S.}\ \bibnamefont {Simpkins}},\ }\bibfield  {title} {\bibinfo {title} {{Coherent Coupling between a Molecular Vibration and Fabry–Perot Optical Cavity to Give Hybridized States in the Strong Coupling Limit}},\ }\href {https://doi.org/10.1021/ph5003347} {\bibfield  {journal} {\bibinfo  {journal} {ACS Photonics}\ }\textbf {\bibinfo {volume} {2}},\ \bibinfo {pages} {130} (\bibinfo {year} {2015})}\BibitemShut {NoStop}%
\bibitem [{\citenamefont {Thomas}\ \emph {et~al.}(2016)\citenamefont {Thomas}, \citenamefont {George}, \citenamefont {Shalabney}, \citenamefont {Dryzhakov}, \citenamefont {Varma}, \citenamefont {Moran}, \citenamefont {Chervy}, \citenamefont {Zhong}, \citenamefont {Devaux}, \citenamefont {Genet}, \citenamefont {Hutchison},\ and\ \citenamefont {Ebbesen}}]{Thomas2016}%
  \BibitemOpen
  \bibfield  {author} {\bibinfo {author} {\bibfnamefont {A.}~\bibnamefont {Thomas}}, \bibinfo {author} {\bibfnamefont {J.}~\bibnamefont {George}}, \bibinfo {author} {\bibfnamefont {A.}~\bibnamefont {Shalabney}}, \bibinfo {author} {\bibfnamefont {M.}~\bibnamefont {Dryzhakov}}, \bibinfo {author} {\bibfnamefont {S.~J.}\ \bibnamefont {Varma}}, \bibinfo {author} {\bibfnamefont {J.}~\bibnamefont {Moran}}, \bibinfo {author} {\bibfnamefont {T.}~\bibnamefont {Chervy}}, \bibinfo {author} {\bibfnamefont {X.}~\bibnamefont {Zhong}}, \bibinfo {author} {\bibfnamefont {E.}~\bibnamefont {Devaux}}, \bibinfo {author} {\bibfnamefont {C.}~\bibnamefont {Genet}}, \bibinfo {author} {\bibfnamefont {J.~A.}\ \bibnamefont {Hutchison}},\ and\ \bibinfo {author} {\bibfnamefont {T.~W.}\ \bibnamefont {Ebbesen}},\ }\bibfield  {title} {\bibinfo {title} {{Ground-State Chemical Reactivity under Vibrational Coupling to the Vacuum Electromagnetic Field}},\ }\href {https://doi.org/10.1002/anie.201605504} {\bibfield  {journal} {\bibinfo  {journal}
  {Angew. Chemie Int. Ed.}\ }\textbf {\bibinfo {volume} {55}},\ \bibinfo {pages} {11462} (\bibinfo {year} {2016})}\BibitemShut {NoStop}%
\bibitem [{\citenamefont {Thomas}\ \emph {et~al.}(2019)\citenamefont {Thomas}, \citenamefont {Lethuillier-Karl}, \citenamefont {Nagarajan}, \citenamefont {Vergauwe}, \citenamefont {George}, \citenamefont {Chervy}, \citenamefont {Shalabney}, \citenamefont {Devaux}, \citenamefont {Genet}, \citenamefont {Moran},\ and\ \citenamefont {Ebbesen}}]{Thomas2019_science}%
  \BibitemOpen
  \bibfield  {author} {\bibinfo {author} {\bibfnamefont {A.}~\bibnamefont {Thomas}}, \bibinfo {author} {\bibfnamefont {L.}~\bibnamefont {Lethuillier-Karl}}, \bibinfo {author} {\bibfnamefont {K.}~\bibnamefont {Nagarajan}}, \bibinfo {author} {\bibfnamefont {R.~M.~A.}\ \bibnamefont {Vergauwe}}, \bibinfo {author} {\bibfnamefont {J.}~\bibnamefont {George}}, \bibinfo {author} {\bibfnamefont {T.}~\bibnamefont {Chervy}}, \bibinfo {author} {\bibfnamefont {A.}~\bibnamefont {Shalabney}}, \bibinfo {author} {\bibfnamefont {E.}~\bibnamefont {Devaux}}, \bibinfo {author} {\bibfnamefont {C.}~\bibnamefont {Genet}}, \bibinfo {author} {\bibfnamefont {J.}~\bibnamefont {Moran}},\ and\ \bibinfo {author} {\bibfnamefont {T.~W.}\ \bibnamefont {Ebbesen}},\ }\bibfield  {title} {\bibinfo {title} {{Tilting a Ground-State Reactivity Landscape by Vibrational Strong Coupling}},\ }\href {https://doi.org/10.1126/science.aau7742} {\bibfield  {journal} {\bibinfo  {journal} {Science}\ }\textbf {\bibinfo {volume} {363}},\ \bibinfo {pages} {615}
  (\bibinfo {year} {2019})}\BibitemShut {NoStop}%
\bibitem [{\citenamefont {Xiang}\ \emph {et~al.}(2020)\citenamefont {Xiang}, \citenamefont {Ribeiro}, \citenamefont {Du}, \citenamefont {Chen}, \citenamefont {Yang}, \citenamefont {Wang}, \citenamefont {Yuen-Zhou},\ and\ \citenamefont {Xiong}}]{Xiang2020Science}%
  \BibitemOpen
  \bibfield  {author} {\bibinfo {author} {\bibfnamefont {B.}~\bibnamefont {Xiang}}, \bibinfo {author} {\bibfnamefont {R.~F.}\ \bibnamefont {Ribeiro}}, \bibinfo {author} {\bibfnamefont {M.}~\bibnamefont {Du}}, \bibinfo {author} {\bibfnamefont {L.}~\bibnamefont {Chen}}, \bibinfo {author} {\bibfnamefont {Z.}~\bibnamefont {Yang}}, \bibinfo {author} {\bibfnamefont {J.}~\bibnamefont {Wang}}, \bibinfo {author} {\bibfnamefont {J.}~\bibnamefont {Yuen-Zhou}},\ and\ \bibinfo {author} {\bibfnamefont {W.}~\bibnamefont {Xiong}},\ }\bibfield  {title} {\bibinfo {title} {{Intermolecular Vibrational Energy Transfer Enabled by Microcavity Strong Light--Matter Coupling}},\ }\href {https://doi.org/10.1126/science.aba3544} {\bibfield  {journal} {\bibinfo  {journal} {Science}\ }\textbf {\bibinfo {volume} {368}},\ \bibinfo {pages} {665} (\bibinfo {year} {2020})}\BibitemShut {NoStop}%
\bibitem [{\citenamefont {Chen}\ \emph {et~al.}(2022)\citenamefont {Chen}, \citenamefont {Du}, \citenamefont {Yang}, \citenamefont {Yuen-Zhou},\ and\ \citenamefont {Xiong}}]{Chen2022}%
  \BibitemOpen
  \bibfield  {author} {\bibinfo {author} {\bibfnamefont {T.-T.}\ \bibnamefont {Chen}}, \bibinfo {author} {\bibfnamefont {M.}~\bibnamefont {Du}}, \bibinfo {author} {\bibfnamefont {Z.}~\bibnamefont {Yang}}, \bibinfo {author} {\bibfnamefont {J.}~\bibnamefont {Yuen-Zhou}},\ and\ \bibinfo {author} {\bibfnamefont {W.}~\bibnamefont {Xiong}},\ }\bibfield  {title} {\bibinfo {title} {{Cavity-enabled Enhancement of Ultrafast Intramolecular Vibrational Redistribution over Pseudorotation}},\ }\href {https://doi.org/10.1126/science.add0276} {\bibfield  {journal} {\bibinfo  {journal} {Science}\ }\textbf {\bibinfo {volume} {378}},\ \bibinfo {pages} {790} (\bibinfo {year} {2022})}\BibitemShut {NoStop}%
\bibitem [{\citenamefont {Ahn}\ \emph {et~al.}(2023)\citenamefont {Ahn}, \citenamefont {Triana}, \citenamefont {Recabal}, \citenamefont {Herrera},\ and\ \citenamefont {Simpkins}}]{Ahn2023Science}%
  \BibitemOpen
  \bibfield  {author} {\bibinfo {author} {\bibfnamefont {W.}~\bibnamefont {Ahn}}, \bibinfo {author} {\bibfnamefont {J.~F.}\ \bibnamefont {Triana}}, \bibinfo {author} {\bibfnamefont {F.}~\bibnamefont {Recabal}}, \bibinfo {author} {\bibfnamefont {F.}~\bibnamefont {Herrera}},\ and\ \bibinfo {author} {\bibfnamefont {B.~S.}\ \bibnamefont {Simpkins}},\ }\bibfield  {title} {\bibinfo {title} {{Modification of Ground-State Chemical Reactivity via Light–Matter Coherence in Infrared Cavities}},\ }\href {https://doi.org/10.1126/science.ade7147} {\bibfield  {journal} {\bibinfo  {journal} {Science}\ }\textbf {\bibinfo {volume} {380}},\ \bibinfo {pages} {1165} (\bibinfo {year} {2023})}\BibitemShut {NoStop}%
\bibitem [{\citenamefont {Ribeiro}\ \emph {et~al.}(2018)\citenamefont {Ribeiro}, \citenamefont {Mart{\'{i}}nez-Mart{\'{i}}nez}, \citenamefont {Du}, \citenamefont {Campos-Gonzalez-Angulo},\ and\ \citenamefont {Yuen-Zhou}}]{Ribeiro2018}%
  \BibitemOpen
  \bibfield  {author} {\bibinfo {author} {\bibfnamefont {R.~F.}\ \bibnamefont {Ribeiro}}, \bibinfo {author} {\bibfnamefont {L.~A.}\ \bibnamefont {Mart{\'{i}}nez-Mart{\'{i}}nez}}, \bibinfo {author} {\bibfnamefont {M.}~\bibnamefont {Du}}, \bibinfo {author} {\bibfnamefont {J.}~\bibnamefont {Campos-Gonzalez-Angulo}},\ and\ \bibinfo {author} {\bibfnamefont {J.}~\bibnamefont {Yuen-Zhou}},\ }\bibfield  {title} {\bibinfo {title} {{Polariton Chemistry: Controlling Molecular Dynamics with Optical Cavities}},\ }\href {https://doi.org/10.1039/C8SC01043A} {\bibfield  {journal} {\bibinfo  {journal} {Chem. Sci.}\ }\textbf {\bibinfo {volume} {9}},\ \bibinfo {pages} {6325} (\bibinfo {year} {2018})}\BibitemShut {NoStop}%
\bibitem [{\citenamefont {Herrera}\ and\ \citenamefont {Owrutsky}(2020)}]{Herrera2019}%
  \BibitemOpen
  \bibfield  {author} {\bibinfo {author} {\bibfnamefont {F.}~\bibnamefont {Herrera}}\ and\ \bibinfo {author} {\bibfnamefont {J.}~\bibnamefont {Owrutsky}},\ }\bibfield  {title} {\bibinfo {title} {{Molecular Polaritons for Controlling Chemistry with Quantum Optics}},\ }\href {https://doi.org/10.1063/1.5136320} {\bibfield  {journal} {\bibinfo  {journal} {J. Chem. Phys.}\ }\textbf {\bibinfo {volume} {152}},\ \bibinfo {pages} {100902} (\bibinfo {year} {2020})}\BibitemShut {NoStop}%
\bibitem [{\citenamefont {Li}\ \emph {et~al.}(2022{\natexlab{a}})\citenamefont {Li}, \citenamefont {Cui}, \citenamefont {Subotnik},\ and\ \citenamefont {Nitzan}}]{Li2022Review}%
  \BibitemOpen
  \bibfield  {author} {\bibinfo {author} {\bibfnamefont {T.~E.}\ \bibnamefont {Li}}, \bibinfo {author} {\bibfnamefont {B.}~\bibnamefont {Cui}}, \bibinfo {author} {\bibfnamefont {J.~E.}\ \bibnamefont {Subotnik}},\ and\ \bibinfo {author} {\bibfnamefont {A.}~\bibnamefont {Nitzan}},\ }\bibfield  {title} {\bibinfo {title} {{Molecular Polaritonics: Chemical Dynamics Under Strong Light–Matter Coupling}},\ }\href {https://doi.org/10.1146/annurev-physchem-090519-042621} {\bibfield  {journal} {\bibinfo  {journal} {Annu. Rev. Phys. Chem.}\ }\textbf {\bibinfo {volume} {73}},\ \bibinfo {pages} {43} (\bibinfo {year} {2022}{\natexlab{a}})}\BibitemShut {NoStop}%
\bibitem [{\citenamefont {Fregoni}\ \emph {et~al.}(2022)\citenamefont {Fregoni}, \citenamefont {Garcia-Vidal},\ and\ \citenamefont {Feist}}]{Fregoni2022}%
  \BibitemOpen
  \bibfield  {author} {\bibinfo {author} {\bibfnamefont {J.}~\bibnamefont {Fregoni}}, \bibinfo {author} {\bibfnamefont {F.~J.}\ \bibnamefont {Garcia-Vidal}},\ and\ \bibinfo {author} {\bibfnamefont {J.}~\bibnamefont {Feist}},\ }\bibfield  {title} {\bibinfo {title} {{Theoretical Challenges in Polaritonic Chemistry}},\ }\href {https://doi.org/10.1021/acsphotonics.1c01749} {\bibfield  {journal} {\bibinfo  {journal} {ACS Photonics}\ }\textbf {\bibinfo {volume} {9}},\ \bibinfo {pages} {1096} (\bibinfo {year} {2022})}\BibitemShut {NoStop}%
\bibitem [{\citenamefont {Simpkins}\ \emph {et~al.}(2023)\citenamefont {Simpkins}, \citenamefont {Dunkelberger},\ and\ \citenamefont {Vurgaftman}}]{Simpkins2023}%
  \BibitemOpen
  \bibfield  {author} {\bibinfo {author} {\bibfnamefont {B.~S.}\ \bibnamefont {Simpkins}}, \bibinfo {author} {\bibfnamefont {A.~D.}\ \bibnamefont {Dunkelberger}},\ and\ \bibinfo {author} {\bibfnamefont {I.}~\bibnamefont {Vurgaftman}},\ }\bibfield  {title} {\bibinfo {title} {{Control, Modulation, and Analytical Descriptions of Vibrational Strong Coupling}},\ }\href {https://doi.org/10.1021/acs.chemrev.2c00774} {\bibfield  {journal} {\bibinfo  {journal} {Chem. Rev.}\ }\textbf {\bibinfo {volume} {123}},\ \bibinfo {pages} {5020} (\bibinfo {year} {2023})}\BibitemShut {NoStop}%
\bibitem [{\citenamefont {Mandal}\ \emph {et~al.}(2023)\citenamefont {Mandal}, \citenamefont {Taylor}, \citenamefont {Weight}, \citenamefont {Koessler}, \citenamefont {Li},\ and\ \citenamefont {Huo}}]{Mandal2023ChemRev}%
  \BibitemOpen
  \bibfield  {author} {\bibinfo {author} {\bibfnamefont {A.}~\bibnamefont {Mandal}}, \bibinfo {author} {\bibfnamefont {M.~A.}\ \bibnamefont {Taylor}}, \bibinfo {author} {\bibfnamefont {B.~M.}\ \bibnamefont {Weight}}, \bibinfo {author} {\bibfnamefont {E.~R.}\ \bibnamefont {Koessler}}, \bibinfo {author} {\bibfnamefont {X.}~\bibnamefont {Li}},\ and\ \bibinfo {author} {\bibfnamefont {P.}~\bibnamefont {Huo}},\ }\bibfield  {title} {\bibinfo {title} {{Theoretical Advances in Polariton Chemistry and Molecular Cavity Quantum Electrodynamics}},\ }\href {https://doi.org/10.1021/acs.chemrev.2c00855} {\bibfield  {journal} {\bibinfo  {journal} {Chem. Rev.}\ }\textbf {\bibinfo {volume} {123}},\ \bibinfo {pages} {9786} (\bibinfo {year} {2023})}\BibitemShut {NoStop}%
\bibitem [{\citenamefont {Ruggenthaler}\ \emph {et~al.}(2023)\citenamefont {Ruggenthaler}, \citenamefont {Sidler},\ and\ \citenamefont {Rubio}}]{Ruggenthaler2023}%
  \BibitemOpen
  \bibfield  {author} {\bibinfo {author} {\bibfnamefont {M.}~\bibnamefont {Ruggenthaler}}, \bibinfo {author} {\bibfnamefont {D.}~\bibnamefont {Sidler}},\ and\ \bibinfo {author} {\bibfnamefont {A.}~\bibnamefont {Rubio}},\ }\bibfield  {title} {\bibinfo {title} {{Understanding Polaritonic Chemistry from Ab Initio Quantum Electrodynamics}},\ }\href {https://doi.org/10.1021/acs.chemrev.2c00788} {\bibfield  {journal} {\bibinfo  {journal} {Chem. Rev.}\ }\textbf {\bibinfo {volume} {123}},\ \bibinfo {pages} {11191} (\bibinfo {year} {2023})}\BibitemShut {NoStop}%
\bibitem [{\citenamefont {Xiang}\ and\ \citenamefont {Xiong}(2024)}]{Xiang2024}%
  \BibitemOpen
  \bibfield  {author} {\bibinfo {author} {\bibfnamefont {B.}~\bibnamefont {Xiang}}\ and\ \bibinfo {author} {\bibfnamefont {W.}~\bibnamefont {Xiong}},\ }\bibfield  {title} {\bibinfo {title} {{Molecular Polaritons for Chemistry, Photonics and Quantum Technologies}},\ }\href {https://doi.org/10.1021/ACS.CHEMREV.3C00662/ASSET/IMAGES/LARGE/CR3C00662_0036.JPEG} {\bibfield  {journal} {\bibinfo  {journal} {Chem. Rev.}\ }\textbf {\bibinfo {volume} {124}},\ \bibinfo {pages} {2512} (\bibinfo {year} {2024})}\BibitemShut {NoStop}%
\bibitem [{\citenamefont {Dunkelberger}\ \emph {et~al.}(2016)\citenamefont {Dunkelberger}, \citenamefont {Spann}, \citenamefont {Fears}, \citenamefont {Simpkins},\ and\ \citenamefont {Owrutsky}}]{Dunkelberger2016}%
  \BibitemOpen
  \bibfield  {author} {\bibinfo {author} {\bibfnamefont {A.~D.}\ \bibnamefont {Dunkelberger}}, \bibinfo {author} {\bibfnamefont {B.~T.}\ \bibnamefont {Spann}}, \bibinfo {author} {\bibfnamefont {K.~P.}\ \bibnamefont {Fears}}, \bibinfo {author} {\bibfnamefont {B.~S.}\ \bibnamefont {Simpkins}},\ and\ \bibinfo {author} {\bibfnamefont {J.~C.}\ \bibnamefont {Owrutsky}},\ }\bibfield  {title} {\bibinfo {title} {{Modified Relaxation Dynamics and Coherent Energy Exchange in Coupled Vibration-Cavity Polaritons}},\ }\href {https://doi.org/10.1038/ncomms13504} {\bibfield  {journal} {\bibinfo  {journal} {Nat. Commun.}\ }\textbf {\bibinfo {volume} {7}},\ \bibinfo {pages} {1} (\bibinfo {year} {2016})}\BibitemShut {NoStop}%
\bibitem [{\citenamefont {Xiang}\ \emph {et~al.}(2018)\citenamefont {Xiang}, \citenamefont {Ribeiro}, \citenamefont {Dunkelberger}, \citenamefont {Wang}, \citenamefont {Li}, \citenamefont {Simpkins}, \citenamefont {Owrutsky}, \citenamefont {Yuen-Zhou},\ and\ \citenamefont {Xiong}}]{Xiang2018}%
  \BibitemOpen
  \bibfield  {author} {\bibinfo {author} {\bibfnamefont {B.}~\bibnamefont {Xiang}}, \bibinfo {author} {\bibfnamefont {R.~F.}\ \bibnamefont {Ribeiro}}, \bibinfo {author} {\bibfnamefont {A.~D.}\ \bibnamefont {Dunkelberger}}, \bibinfo {author} {\bibfnamefont {J.}~\bibnamefont {Wang}}, \bibinfo {author} {\bibfnamefont {Y.}~\bibnamefont {Li}}, \bibinfo {author} {\bibfnamefont {B.~S.}\ \bibnamefont {Simpkins}}, \bibinfo {author} {\bibfnamefont {J.~C.}\ \bibnamefont {Owrutsky}}, \bibinfo {author} {\bibfnamefont {J.}~\bibnamefont {Yuen-Zhou}},\ and\ \bibinfo {author} {\bibfnamefont {W.}~\bibnamefont {Xiong}},\ }\bibfield  {title} {\bibinfo {title} {{Two-dimensional Infrared Spectroscopy of Vibrational Polaritons}},\ }\href {https://doi.org/10.1073/pnas.1722063115} {\bibfield  {journal} {\bibinfo  {journal} {Proc. Natl. Acad. Sci.}\ }\textbf {\bibinfo {volume} {115}},\ \bibinfo {pages} {4845} (\bibinfo {year} {2018})}\BibitemShut {NoStop}%
\bibitem [{\citenamefont {Grafton}\ \emph {et~al.}(2021)\citenamefont {Grafton}, \citenamefont {Dunkelberger}, \citenamefont {Simpkins}, \citenamefont {Triana}, \citenamefont {Hern{\'{a}}ndez}, \citenamefont {Herrera},\ and\ \citenamefont {Owrutsky}}]{Grafton2020}%
  \BibitemOpen
  \bibfield  {author} {\bibinfo {author} {\bibfnamefont {A.~B.}\ \bibnamefont {Grafton}}, \bibinfo {author} {\bibfnamefont {A.~D.}\ \bibnamefont {Dunkelberger}}, \bibinfo {author} {\bibfnamefont {B.~S.}\ \bibnamefont {Simpkins}}, \bibinfo {author} {\bibfnamefont {J.~F.}\ \bibnamefont {Triana}}, \bibinfo {author} {\bibfnamefont {F.~J.}\ \bibnamefont {Hern{\'{a}}ndez}}, \bibinfo {author} {\bibfnamefont {F.}~\bibnamefont {Herrera}},\ and\ \bibinfo {author} {\bibfnamefont {J.~C.}\ \bibnamefont {Owrutsky}},\ }\bibfield  {title} {\bibinfo {title} {{Excited-State Vibration-Polariton Transitions and Dynamics in Nitroprusside}},\ }\href {https://doi.org/10.1038/s41467-020-20535-z} {\bibfield  {journal} {\bibinfo  {journal} {Nat. Commun.}\ }\textbf {\bibinfo {volume} {12}},\ \bibinfo {pages} {214} (\bibinfo {year} {2021})}\BibitemShut {NoStop}%
\bibitem [{\citenamefont {Pyles}\ \emph {et~al.}(2024)\citenamefont {Pyles}, \citenamefont {Simpkins}, \citenamefont {Vurgaftman}, \citenamefont {Owrutsky},\ and\ \citenamefont {Dunkelberger}}]{Pyles2024}%
  \BibitemOpen
  \bibfield  {author} {\bibinfo {author} {\bibfnamefont {C.~G.}\ \bibnamefont {Pyles}}, \bibinfo {author} {\bibfnamefont {B.~S.}\ \bibnamefont {Simpkins}}, \bibinfo {author} {\bibfnamefont {I.}~\bibnamefont {Vurgaftman}}, \bibinfo {author} {\bibfnamefont {J.~C.}\ \bibnamefont {Owrutsky}},\ and\ \bibinfo {author} {\bibfnamefont {A.~D.}\ \bibnamefont {Dunkelberger}},\ }\bibfield  {title} {\bibinfo {title} {{Revisiting Cavity-coupled 2DIR: A Classical Approach Implicates Reservoir Modes}},\ }\href {https://doi.org/10.1063/5.0239301/3326988} {\bibfield  {journal} {\bibinfo  {journal} {J. Chem. Phys.}\ }\textbf {\bibinfo {volume} {161}},\ \bibinfo {pages} {234202} (\bibinfo {year} {2024})}\BibitemShut {NoStop}%
\bibitem [{\citenamefont {Xiang}\ and\ \citenamefont {Xiong}(2021)}]{Xiang2021JCP}%
  \BibitemOpen
  \bibfield  {author} {\bibinfo {author} {\bibfnamefont {B.}~\bibnamefont {Xiang}}\ and\ \bibinfo {author} {\bibfnamefont {W.}~\bibnamefont {Xiong}},\ }\bibfield  {title} {\bibinfo {title} {{Molecular Vibrational Polariton: Its Dynamics and Potentials in Novel Chemistry and Quantum Technology}},\ }\href {https://doi.org/10.1063/5.0054896} {\bibfield  {journal} {\bibinfo  {journal} {J. Chem. Phys.}\ }\textbf {\bibinfo {volume} {155}},\ \bibinfo {pages} {050901} (\bibinfo {year} {2021})}\BibitemShut {NoStop}%
\bibitem [{\citenamefont {Schwennicke}\ \emph {et~al.}(2024)\citenamefont {Schwennicke}, \citenamefont {Koner}, \citenamefont {P{\'{e}}rez-S{\'{a}}nchez}, \citenamefont {Xiong}, \citenamefont {Giebink}, \citenamefont {Weichman},\ and\ \citenamefont {Yuen-Zhou}}]{Schwennicke2024}%
  \BibitemOpen
  \bibfield  {author} {\bibinfo {author} {\bibfnamefont {K.}~\bibnamefont {Schwennicke}}, \bibinfo {author} {\bibfnamefont {A.}~\bibnamefont {Koner}}, \bibinfo {author} {\bibfnamefont {J.~B.}\ \bibnamefont {P{\'{e}}rez-S{\'{a}}nchez}}, \bibinfo {author} {\bibfnamefont {W.}~\bibnamefont {Xiong}}, \bibinfo {author} {\bibfnamefont {N.~C.}\ \bibnamefont {Giebink}}, \bibinfo {author} {\bibfnamefont {M.~L.}\ \bibnamefont {Weichman}},\ and\ \bibinfo {author} {\bibfnamefont {J.}~\bibnamefont {Yuen-Zhou}},\ }\bibfield  {title} {\bibinfo {title} {{When Do Molecular Polaritons Behave like Optical Filters?}},\ }\href@noop {} {\bibfield  {journal} {\bibinfo  {journal} {arXiv}\ ,\ \bibinfo {pages} {DOI: 10.48550/arXiv.2408.05036 (accessed 2024}} (\bibinfo {year} {2024})}\BibitemShut {NoStop}%
\bibitem [{\citenamefont {Hirschmann}\ \emph {et~al.}(2024)\citenamefont {Hirschmann}, \citenamefont {Bhakta},\ and\ \citenamefont {Xiong}}]{Hirschmann2024}%
  \BibitemOpen
  \bibfield  {author} {\bibinfo {author} {\bibfnamefont {O.}~\bibnamefont {Hirschmann}}, \bibinfo {author} {\bibfnamefont {H.~H.}\ \bibnamefont {Bhakta}},\ and\ \bibinfo {author} {\bibfnamefont {W.}~\bibnamefont {Xiong}},\ }\bibfield  {title} {\bibinfo {title} {{The role of IR inactive mode in W(CO)6 polariton relaxation process}},\ }\href {https://doi.org/10.1515/NANOPH-2023-0589/DOWNLOADASSET/SUPPL/J_NANOPH-2023-0589_SUPPL_001.DOCX} {\bibfield  {journal} {\bibinfo  {journal} {Nanophotonics}\ }\textbf {\bibinfo {volume} {13}},\ \bibinfo {pages} {2029} (\bibinfo {year} {2024})}\BibitemShut {NoStop}%
\bibitem [{\citenamefont {Li}\ \emph {et~al.}(2020)\citenamefont {Li}, \citenamefont {Subotnik},\ and\ \citenamefont {Nitzan}}]{Li2020Water}%
  \BibitemOpen
  \bibfield  {author} {\bibinfo {author} {\bibfnamefont {T.~E.}\ \bibnamefont {Li}}, \bibinfo {author} {\bibfnamefont {J.~E.}\ \bibnamefont {Subotnik}},\ and\ \bibinfo {author} {\bibfnamefont {A.}~\bibnamefont {Nitzan}},\ }\bibfield  {title} {\bibinfo {title} {{Cavity Molecular Dynamics Simulations of Liquid Water under Vibrational Ultrastrong Coupling}},\ }\href {https://doi.org/10.1073/pnas.2009272117} {\bibfield  {journal} {\bibinfo  {journal} {Proc. Natl. Acad. Sci.}\ }\textbf {\bibinfo {volume} {117}},\ \bibinfo {pages} {18324} (\bibinfo {year} {2020})}\BibitemShut {NoStop}%
\bibitem [{\citenamefont {Li}\ \emph {et~al.}(2021{\natexlab{a}})\citenamefont {Li}, \citenamefont {Nitzan},\ and\ \citenamefont {Subotnik}}]{Li2020Nonlinear}%
  \BibitemOpen
  \bibfield  {author} {\bibinfo {author} {\bibfnamefont {T.~E.}\ \bibnamefont {Li}}, \bibinfo {author} {\bibfnamefont {A.}~\bibnamefont {Nitzan}},\ and\ \bibinfo {author} {\bibfnamefont {J.~E.}\ \bibnamefont {Subotnik}},\ }\bibfield  {title} {\bibinfo {title} {{Cavity Molecular Dynamics Simulations of Vibrational Polariton-Enhanced Molecular Nonlinear Absorption}},\ }\href {https://doi.org/10.1063/5.0037623} {\bibfield  {journal} {\bibinfo  {journal} {J. Chem. Phys.}\ }\textbf {\bibinfo {volume} {154}},\ \bibinfo {pages} {094124} (\bibinfo {year} {2021}{\natexlab{a}})}\BibitemShut {NoStop}%
\bibitem [{\citenamefont {Li}\ \emph {et~al.}(2022{\natexlab{b}})\citenamefont {Li}, \citenamefont {Nitzan},\ and\ \citenamefont {Subotnik}}]{Li2021Solute}%
  \BibitemOpen
  \bibfield  {author} {\bibinfo {author} {\bibfnamefont {T.~E.}\ \bibnamefont {Li}}, \bibinfo {author} {\bibfnamefont {A.}~\bibnamefont {Nitzan}},\ and\ \bibinfo {author} {\bibfnamefont {J.~E.}\ \bibnamefont {Subotnik}},\ }\bibfield  {title} {\bibinfo {title} {{Energy-Efficient Pathway for Selectively Exciting Solute Molecules to High Vibrational States via Solvent Vibration-Polariton Pumping}},\ }\href {https://doi.org/10.1038/s41467-022-31703-8} {\bibfield  {journal} {\bibinfo  {journal} {Nat. Commun.}\ }\textbf {\bibinfo {volume} {13}},\ \bibinfo {pages} {4203} (\bibinfo {year} {2022}{\natexlab{b}})}\BibitemShut {NoStop}%
\bibitem [{\citenamefont {Li}\ \emph {et~al.}(2022{\natexlab{c}})\citenamefont {Li}, \citenamefont {Nitzan},\ and\ \citenamefont {Subotnik}}]{Li2021Relaxation}%
  \BibitemOpen
  \bibfield  {author} {\bibinfo {author} {\bibfnamefont {T.~E.}\ \bibnamefont {Li}}, \bibinfo {author} {\bibfnamefont {A.}~\bibnamefont {Nitzan}},\ and\ \bibinfo {author} {\bibfnamefont {J.~E.}\ \bibnamefont {Subotnik}},\ }\bibfield  {title} {\bibinfo {title} {{Polariton Relaxation under Vibrational Strong Coupling: Comparing Cavity Molecular Dynamics Simulations against Fermi's Golden Rule Rate}},\ }\href {https://doi.org/10.1063/5.0079784} {\bibfield  {journal} {\bibinfo  {journal} {J. Chem. Phys.}\ }\textbf {\bibinfo {volume} {156}},\ \bibinfo {pages} {134106} (\bibinfo {year} {2022}{\natexlab{c}})}\BibitemShut {NoStop}%
\bibitem [{\citenamefont {Galego}\ \emph {et~al.}(2019)\citenamefont {Galego}, \citenamefont {Climent}, \citenamefont {Garcia-Vidal},\ and\ \citenamefont {Feist}}]{Galego2019}%
  \BibitemOpen
  \bibfield  {author} {\bibinfo {author} {\bibfnamefont {J.}~\bibnamefont {Galego}}, \bibinfo {author} {\bibfnamefont {C.}~\bibnamefont {Climent}}, \bibinfo {author} {\bibfnamefont {F.~J.}\ \bibnamefont {Garcia-Vidal}},\ and\ \bibinfo {author} {\bibfnamefont {J.}~\bibnamefont {Feist}},\ }\bibfield  {title} {\bibinfo {title} {{Cavity Casimir-Polder Forces and Their Effects in Ground-State Chemical Reactivity}},\ }\href {https://doi.org/10.1103/PhysRevX.9.021057} {\bibfield  {journal} {\bibinfo  {journal} {Phys. Rev. X}\ }\textbf {\bibinfo {volume} {9}},\ \bibinfo {pages} {021057} (\bibinfo {year} {2019})}\BibitemShut {NoStop}%
\bibitem [{\citenamefont {Hern{\'{a}}ndez}\ and\ \citenamefont {Herrera}(2019)}]{Hernandez2019}%
  \BibitemOpen
  \bibfield  {author} {\bibinfo {author} {\bibfnamefont {F.~J.}\ \bibnamefont {Hern{\'{a}}ndez}}\ and\ \bibinfo {author} {\bibfnamefont {F.}~\bibnamefont {Herrera}},\ }\bibfield  {title} {\bibinfo {title} {{Multi-level Quantum Rabi Model for Anharmonic Vibrational Polaritons}},\ }\href {https://doi.org/10.1063/1.5121426} {\bibfield  {journal} {\bibinfo  {journal} {J. Chem. Phys.}\ }\textbf {\bibinfo {volume} {151}},\ \bibinfo {pages} {144116} (\bibinfo {year} {2019})}\BibitemShut {NoStop}%
\bibitem [{\citenamefont {Campos-Gonzalez-Angulo}\ \emph {et~al.}(2019)\citenamefont {Campos-Gonzalez-Angulo}, \citenamefont {Ribeiro},\ and\ \citenamefont {Yuen-Zhou}}]{Campos-Gonzalez-Angulo2019}%
  \BibitemOpen
  \bibfield  {author} {\bibinfo {author} {\bibfnamefont {J.~A.}\ \bibnamefont {Campos-Gonzalez-Angulo}}, \bibinfo {author} {\bibfnamefont {R.~F.}\ \bibnamefont {Ribeiro}},\ and\ \bibinfo {author} {\bibfnamefont {J.}~\bibnamefont {Yuen-Zhou}},\ }\bibfield  {title} {\bibinfo {title} {{Resonant Catalysis of Thermally Activated Chemical Reactions with Vibrational Polaritons}},\ }\href {https://doi.org/10.1038/s41467-019-12636-1} {\bibfield  {journal} {\bibinfo  {journal} {Nat. Commun.}\ }\textbf {\bibinfo {volume} {10}},\ \bibinfo {pages} {4685} (\bibinfo {year} {2019})}\BibitemShut {NoStop}%
\bibitem [{\citenamefont {Hoffmann}\ \emph {et~al.}(2020)\citenamefont {Hoffmann}, \citenamefont {Lacombe}, \citenamefont {Rubio},\ and\ \citenamefont {Maitra}}]{Hoffmann2020}%
  \BibitemOpen
  \bibfield  {author} {\bibinfo {author} {\bibfnamefont {N.~M.}\ \bibnamefont {Hoffmann}}, \bibinfo {author} {\bibfnamefont {L.}~\bibnamefont {Lacombe}}, \bibinfo {author} {\bibfnamefont {A.}~\bibnamefont {Rubio}},\ and\ \bibinfo {author} {\bibfnamefont {N.~T.}\ \bibnamefont {Maitra}},\ }\bibfield  {title} {\bibinfo {title} {{Effect of Many Modes on Self-Polarization and Photochemical Suppression in Cavities}},\ }\href {https://doi.org/10.1063/5.0012723} {\bibfield  {journal} {\bibinfo  {journal} {J. Chem. Phys.}\ }\textbf {\bibinfo {volume} {153}},\ \bibinfo {pages} {104103} (\bibinfo {year} {2020})}\BibitemShut {NoStop}%
\bibitem [{\citenamefont {Botzung}\ \emph {et~al.}(2020)\citenamefont {Botzung}, \citenamefont {Hagenm{\"{u}}ller}, \citenamefont {Sch{\"{u}}tz}, \citenamefont {Dubail}, \citenamefont {Pupillo},\ and\ \citenamefont {Schachenmayer}}]{Botzung2020}%
  \BibitemOpen
  \bibfield  {author} {\bibinfo {author} {\bibfnamefont {T.}~\bibnamefont {Botzung}}, \bibinfo {author} {\bibfnamefont {D.}~\bibnamefont {Hagenm{\"{u}}ller}}, \bibinfo {author} {\bibfnamefont {S.}~\bibnamefont {Sch{\"{u}}tz}}, \bibinfo {author} {\bibfnamefont {J.}~\bibnamefont {Dubail}}, \bibinfo {author} {\bibfnamefont {G.}~\bibnamefont {Pupillo}},\ and\ \bibinfo {author} {\bibfnamefont {J.}~\bibnamefont {Schachenmayer}},\ }\bibfield  {title} {\bibinfo {title} {{Dark state semilocalization of quantum emitters in a cavity}},\ }\href {https://doi.org/10.1103/PhysRevB.102.144202} {\bibfield  {journal} {\bibinfo  {journal} {Phys. Rev. B}\ }\textbf {\bibinfo {volume} {102}},\ \bibinfo {pages} {144202} (\bibinfo {year} {2020})}\BibitemShut {NoStop}%
\bibitem [{\citenamefont {Li}\ \emph {et~al.}(2021{\natexlab{b}})\citenamefont {Li}, \citenamefont {Mandal},\ and\ \citenamefont {Huo}}]{LiHuo2021}%
  \BibitemOpen
  \bibfield  {author} {\bibinfo {author} {\bibfnamefont {X.}~\bibnamefont {Li}}, \bibinfo {author} {\bibfnamefont {A.}~\bibnamefont {Mandal}},\ and\ \bibinfo {author} {\bibfnamefont {P.}~\bibnamefont {Huo}},\ }\bibfield  {title} {\bibinfo {title} {{Cavity Frequency-Dependent Theory for Vibrational Polariton Chemistry}},\ }\href {https://doi.org/10.1038/s41467-021-21610-9} {\bibfield  {journal} {\bibinfo  {journal} {Nat. Commun.}\ }\textbf {\bibinfo {volume} {12}},\ \bibinfo {pages} {1315} (\bibinfo {year} {2021}{\natexlab{b}})}\BibitemShut {NoStop}%
\bibitem [{\citenamefont {Fischer}\ and\ \citenamefont {Saalfrank}(2021)}]{Fischer2021}%
  \BibitemOpen
  \bibfield  {author} {\bibinfo {author} {\bibfnamefont {E.~W.}\ \bibnamefont {Fischer}}\ and\ \bibinfo {author} {\bibfnamefont {P.}~\bibnamefont {Saalfrank}},\ }\bibfield  {title} {\bibinfo {title} {{Ground State Properties and Infrared Spectra of Anharmonic Vibrational Polaritons of Small Molecules in Cavities}},\ }\href {https://doi.org/10.1063/5.0040853} {\bibfield  {journal} {\bibinfo  {journal} {J. Chem. Phys.}\ }\textbf {\bibinfo {volume} {154}},\ \bibinfo {pages} {104311} (\bibinfo {year} {2021})}\BibitemShut {NoStop}%
\bibitem [{\citenamefont {Yang}\ and\ \citenamefont {Cao}(2021)}]{YangCao2021}%
  \BibitemOpen
  \bibfield  {author} {\bibinfo {author} {\bibfnamefont {P.~Y.}\ \bibnamefont {Yang}}\ and\ \bibinfo {author} {\bibfnamefont {J.}~\bibnamefont {Cao}},\ }\bibfield  {title} {\bibinfo {title} {{Quantum Effects in Chemical Reactions under Polaritonic Vibrational Strong Coupling}},\ }\href {https://doi.org/10.1021/ACS.JPCLETT.1C02210/SUPPL_FILE/JZ1C02210_SI_001.PDF} {\bibfield  {journal} {\bibinfo  {journal} {J. Phys. Chem. Lett.}\ }\textbf {\bibinfo {volume} {12}},\ \bibinfo {pages} {9531} (\bibinfo {year} {2021})}\BibitemShut {NoStop}%
\bibitem [{\citenamefont {Wang}\ \emph {et~al.}(2022)\citenamefont {Wang}, \citenamefont {Neuman}, \citenamefont {Yelin},\ and\ \citenamefont {Flick}}]{Wang2022JPCL}%
  \BibitemOpen
  \bibfield  {author} {\bibinfo {author} {\bibfnamefont {D.~S.}\ \bibnamefont {Wang}}, \bibinfo {author} {\bibfnamefont {T.}~\bibnamefont {Neuman}}, \bibinfo {author} {\bibfnamefont {S.~F.}\ \bibnamefont {Yelin}},\ and\ \bibinfo {author} {\bibfnamefont {J.}~\bibnamefont {Flick}},\ }\bibfield  {title} {\bibinfo {title} {{Cavity-Modified Unimolecular Dissociation Reactions via Intramolecular Vibrational Energy Redistribution}},\ }\href {https://doi.org/10.1021/acs.jpclett.2c00558} {\bibfield  {journal} {\bibinfo  {journal} {J. Phys. Chem. Lett}\ }\textbf {\bibinfo {volume} {13}},\ \bibinfo {pages} {3317} (\bibinfo {year} {2022})}\BibitemShut {NoStop}%
\bibitem [{\citenamefont {Flick}\ \emph {et~al.}(2017)\citenamefont {Flick}, \citenamefont {Ruggenthaler}, \citenamefont {Appel},\ and\ \citenamefont {Rubio}}]{Flick2017}%
  \BibitemOpen
  \bibfield  {author} {\bibinfo {author} {\bibfnamefont {J.}~\bibnamefont {Flick}}, \bibinfo {author} {\bibfnamefont {M.}~\bibnamefont {Ruggenthaler}}, \bibinfo {author} {\bibfnamefont {H.}~\bibnamefont {Appel}},\ and\ \bibinfo {author} {\bibfnamefont {A.}~\bibnamefont {Rubio}},\ }\bibfield  {title} {\bibinfo {title} {{Atoms and Molecules in Cavities, from Weak to Strong Coupling in Quantum-Electrodynamics (QED) Chemistry}},\ }\href {https://doi.org/10.1073/pnas.1615509114} {\bibfield  {journal} {\bibinfo  {journal} {Proc. Natl. Acad. Sci.}\ }\textbf {\bibinfo {volume} {114}},\ \bibinfo {pages} {3026} (\bibinfo {year} {2017})}\BibitemShut {NoStop}%
\bibitem [{\citenamefont {Riso}\ \emph {et~al.}(2022)\citenamefont {Riso}, \citenamefont {Haugland}, \citenamefont {Ronca},\ and\ \citenamefont {Koch}}]{Riso2022}%
  \BibitemOpen
  \bibfield  {author} {\bibinfo {author} {\bibfnamefont {R.~R.}\ \bibnamefont {Riso}}, \bibinfo {author} {\bibfnamefont {T.~S.}\ \bibnamefont {Haugland}}, \bibinfo {author} {\bibfnamefont {E.}~\bibnamefont {Ronca}},\ and\ \bibinfo {author} {\bibfnamefont {H.}~\bibnamefont {Koch}},\ }\bibfield  {title} {\bibinfo {title} {{Molecular Orbital Theory in Cavity QED Environments}},\ }\href {https://doi.org/10.1038/s41467-022-29003-2} {\bibfield  {journal} {\bibinfo  {journal} {Nat. Commun.}\ }\textbf {\bibinfo {volume} {13}},\ \bibinfo {pages} {1368} (\bibinfo {year} {2022})}\BibitemShut {NoStop}%
\bibitem [{\citenamefont {Sch{\"{a}}fer}\ \emph {et~al.}(2022)\citenamefont {Sch{\"{a}}fer}, \citenamefont {Flick}, \citenamefont {Ronca}, \citenamefont {Narang},\ and\ \citenamefont {Rubio}}]{Schafer2021}%
  \BibitemOpen
  \bibfield  {author} {\bibinfo {author} {\bibfnamefont {C.}~\bibnamefont {Sch{\"{a}}fer}}, \bibinfo {author} {\bibfnamefont {J.}~\bibnamefont {Flick}}, \bibinfo {author} {\bibfnamefont {E.}~\bibnamefont {Ronca}}, \bibinfo {author} {\bibfnamefont {P.}~\bibnamefont {Narang}},\ and\ \bibinfo {author} {\bibfnamefont {A.}~\bibnamefont {Rubio}},\ }\bibfield  {title} {\bibinfo {title} {{Shining Light on the Microscopic Resonant Mechanism Responsible for Cavity-Mediated Chemical Reactivity}},\ }\href {https://doi.org/10.1038/s41467-022-35363-6} {\bibfield  {journal} {\bibinfo  {journal} {Nat. Commun.}\ }\textbf {\bibinfo {volume} {13}},\ \bibinfo {pages} {7817} (\bibinfo {year} {2022})}\BibitemShut {NoStop}%
\bibitem [{\citenamefont {Bonini}\ and\ \citenamefont {Flick}(2021)}]{Bonini2021}%
  \BibitemOpen
  \bibfield  {author} {\bibinfo {author} {\bibfnamefont {J.}~\bibnamefont {Bonini}}\ and\ \bibinfo {author} {\bibfnamefont {J.}~\bibnamefont {Flick}},\ }\bibfield  {title} {\bibinfo {title} {{Ab Initio Linear-Response Approach to Vibro-polaritons in the Cavity Born-Oppenheimer Approximation}},\ }\href {https://doi.org/10.1021/ACS.JCTC.1C01035/ASSET/IMAGES/LARGE/CT1C01035_0005.JPEG} {\bibfield  {journal} {\bibinfo  {journal} {J. Chem. Theory Comput.}\ }\textbf {\bibinfo {volume} {18}},\ \bibinfo {pages} {2764} (\bibinfo {year} {2021})}\BibitemShut {NoStop}%
\bibitem [{\citenamefont {Yang}\ \emph {et~al.}(2021)\citenamefont {Yang}, \citenamefont {Ou}, \citenamefont {Pei}, \citenamefont {Wang}, \citenamefont {Weng}, \citenamefont {Shuai}, \citenamefont {Mullen},\ and\ \citenamefont {Shao}}]{Yang2021}%
  \BibitemOpen
  \bibfield  {author} {\bibinfo {author} {\bibfnamefont {J.}~\bibnamefont {Yang}}, \bibinfo {author} {\bibfnamefont {Q.}~\bibnamefont {Ou}}, \bibinfo {author} {\bibfnamefont {Z.}~\bibnamefont {Pei}}, \bibinfo {author} {\bibfnamefont {H.}~\bibnamefont {Wang}}, \bibinfo {author} {\bibfnamefont {B.}~\bibnamefont {Weng}}, \bibinfo {author} {\bibfnamefont {Z.}~\bibnamefont {Shuai}}, \bibinfo {author} {\bibfnamefont {K.}~\bibnamefont {Mullen}},\ and\ \bibinfo {author} {\bibfnamefont {Y.}~\bibnamefont {Shao}},\ }\bibfield  {title} {\bibinfo {title} {{Quantum-Electrodynamical Time-Dependent Density Functional Theory within Gaussian Atomic Basis}},\ }\href {https://doi.org/10.1063/5.0057542} {\bibfield  {journal} {\bibinfo  {journal} {J. Chem. Phys.}\ }\textbf {\bibinfo {volume} {155}},\ \bibinfo {pages} {064107} (\bibinfo {year} {2021})}\BibitemShut {NoStop}%
\bibitem [{\citenamefont {Rosenzweig}\ \emph {et~al.}(2022)\citenamefont {Rosenzweig}, \citenamefont {Hoffmann}, \citenamefont {Lacombe},\ and\ \citenamefont {Maitra}}]{Rosenzweig2022}%
  \BibitemOpen
  \bibfield  {author} {\bibinfo {author} {\bibfnamefont {B.}~\bibnamefont {Rosenzweig}}, \bibinfo {author} {\bibfnamefont {N.~M.}\ \bibnamefont {Hoffmann}}, \bibinfo {author} {\bibfnamefont {L.}~\bibnamefont {Lacombe}},\ and\ \bibinfo {author} {\bibfnamefont {N.~T.}\ \bibnamefont {Maitra}},\ }\bibfield  {title} {\bibinfo {title} {{Analysis of the Classical Trajectory Treatment of Photon Dynamics for Polaritonic Phenomena}},\ }\href {https://doi.org/10.1063/5.0079379/5.0079379.MM.ORIGINAL.V2.MP4} {\bibfield  {journal} {\bibinfo  {journal} {J. Chem. Phys.}\ }\textbf {\bibinfo {volume} {156}},\ \bibinfo {pages} {054101} (\bibinfo {year} {2022})}\BibitemShut {NoStop}%
\bibitem [{\citenamefont {Triana}\ \emph {et~al.}(2020)\citenamefont {Triana}, \citenamefont {Hern{\'{a}}ndez},\ and\ \citenamefont {Herrera}}]{Triana2020Shape}%
  \BibitemOpen
  \bibfield  {author} {\bibinfo {author} {\bibfnamefont {J.~F.}\ \bibnamefont {Triana}}, \bibinfo {author} {\bibfnamefont {F.~J.}\ \bibnamefont {Hern{\'{a}}ndez}},\ and\ \bibinfo {author} {\bibfnamefont {F.}~\bibnamefont {Herrera}},\ }\bibfield  {title} {\bibinfo {title} {{The Shape of the Electric Dipole Function Determines the Sub-picosecond Dynamics of Anharmonic Vibrational Polaritons}},\ }\href {https://doi.org/10.1063/5.0009869} {\bibfield  {journal} {\bibinfo  {journal} {J. Chem. Phys.}\ }\textbf {\bibinfo {volume} {152}},\ \bibinfo {pages} {234111} (\bibinfo {year} {2020})}\BibitemShut {NoStop}%
\bibitem [{\citenamefont {Haugland}\ \emph {et~al.}(2020)\citenamefont {Haugland}, \citenamefont {Ronca}, \citenamefont {Kj{\o}nstad}, \citenamefont {Rubio},\ and\ \citenamefont {Koch}}]{Haugland2020}%
  \BibitemOpen
  \bibfield  {author} {\bibinfo {author} {\bibfnamefont {T.~S.}\ \bibnamefont {Haugland}}, \bibinfo {author} {\bibfnamefont {E.}~\bibnamefont {Ronca}}, \bibinfo {author} {\bibfnamefont {E.~F.}\ \bibnamefont {Kj{\o}nstad}}, \bibinfo {author} {\bibfnamefont {A.}~\bibnamefont {Rubio}},\ and\ \bibinfo {author} {\bibfnamefont {H.}~\bibnamefont {Koch}},\ }\bibfield  {title} {\bibinfo {title} {{Coupled Cluster Theory for Molecular Polaritons: Changing Ground and Excited States}},\ }\href {https://doi.org/10.1103/PhysRevX.10.041043} {\bibfield  {journal} {\bibinfo  {journal} {Phys. Rev. X}\ }\textbf {\bibinfo {volume} {10}},\ \bibinfo {pages} {041043} (\bibinfo {year} {2020})}\BibitemShut {NoStop}%
\bibitem [{\citenamefont {Philbin}\ \emph {et~al.}(2023)\citenamefont {Philbin}, \citenamefont {Haugland}, \citenamefont {Ghosh}, \citenamefont {Ronca}, \citenamefont {Chen}, \citenamefont {Narang},\ and\ \citenamefont {Koch}}]{Philbin2022}%
  \BibitemOpen
  \bibfield  {author} {\bibinfo {author} {\bibfnamefont {J.~P.}\ \bibnamefont {Philbin}}, \bibinfo {author} {\bibfnamefont {T.~S.}\ \bibnamefont {Haugland}}, \bibinfo {author} {\bibfnamefont {T.~K.}\ \bibnamefont {Ghosh}}, \bibinfo {author} {\bibfnamefont {E.}~\bibnamefont {Ronca}}, \bibinfo {author} {\bibfnamefont {M.}~\bibnamefont {Chen}}, \bibinfo {author} {\bibfnamefont {P.}~\bibnamefont {Narang}},\ and\ \bibinfo {author} {\bibfnamefont {H.}~\bibnamefont {Koch}},\ }\bibfield  {title} {\bibinfo {title} {{Molecular van der Waals Fluids in Cavity Quantum Electrodynamics}},\ }\href {https://doi.org/10.1021/acs.jpclett.3c01790} {\bibfield  {journal} {\bibinfo  {journal} {J. Phys. Chem. Lett.}\ }\textbf {\bibinfo {volume} {14}},\ \bibinfo {pages} {8988} (\bibinfo {year} {2023})},\ \Eprint {https://arxiv.org/abs/2209.07956} {arXiv:2209.07956} \BibitemShut {NoStop}%
\bibitem [{\citenamefont {Poh}\ \emph {et~al.}(2023)\citenamefont {Poh}, \citenamefont {Pannir-Sivajothi},\ and\ \citenamefont {Yuen-Zhou}}]{Poh2023}%
  \BibitemOpen
  \bibfield  {author} {\bibinfo {author} {\bibfnamefont {Y.~R.}\ \bibnamefont {Poh}}, \bibinfo {author} {\bibfnamefont {S.}~\bibnamefont {Pannir-Sivajothi}},\ and\ \bibinfo {author} {\bibfnamefont {J.}~\bibnamefont {Yuen-Zhou}},\ }\bibfield  {title} {\bibinfo {title} {{Understanding the Energy Gap Law under Vibrational Strong Coupling}},\ }\href {https://doi.org/10.1021/acs.jpcc.2c07047} {\bibfield  {journal} {\bibinfo  {journal} {J. Phys. Chem. C}\ }\textbf {\bibinfo {volume} {127}},\ \bibinfo {pages} {5491} (\bibinfo {year} {2023})},\ \Eprint {https://arxiv.org/abs/2210.04986} {arXiv:2210.04986} \BibitemShut {NoStop}%
\bibitem [{\citenamefont {Suyabatmaz}\ and\ \citenamefont {Ribeiro}(2023)}]{Suyabatmaz2023}%
  \BibitemOpen
  \bibfield  {author} {\bibinfo {author} {\bibfnamefont {E.}~\bibnamefont {Suyabatmaz}}\ and\ \bibinfo {author} {\bibfnamefont {R.~F.}\ \bibnamefont {Ribeiro}},\ }\bibfield  {title} {\bibinfo {title} {{Vibrational Polariton Transport in Disordered Media}},\ }\href {https://doi.org/10.1063/5.0156008/2902632} {\bibfield  {journal} {\bibinfo  {journal} {J. Chem. Phys.}\ }\textbf {\bibinfo {volume} {159}},\ \bibinfo {pages} {034701} (\bibinfo {year} {2023})}\BibitemShut {NoStop}%
\bibitem [{\citenamefont {Yu}\ and\ \citenamefont {Bowman}(2024)}]{Yu2024}%
  \BibitemOpen
  \bibfield  {author} {\bibinfo {author} {\bibfnamefont {Q.}~\bibnamefont {Yu}}\ and\ \bibinfo {author} {\bibfnamefont {J.~M.}\ \bibnamefont {Bowman}},\ }\bibfield  {title} {\bibinfo {title} {{Fully Quantum Simulation of Polaritonic Vibrational Spectra of Large Cavity-Molecule System}},\ }\href {https://doi.org/10.1021/acs.jctc.4c00129} {\bibfield  {journal} {\bibinfo  {journal} {J. Chem. Theory Comput.}\ }\textbf {\bibinfo {volume} {20}},\ \bibinfo {pages} {4278} (\bibinfo {year} {2024})}\BibitemShut {NoStop}%
\bibitem [{\citenamefont {Xiang}\ \emph {et~al.}(2019)\citenamefont {Xiang}, \citenamefont {Ribeiro}, \citenamefont {Chen}, \citenamefont {Wang}, \citenamefont {Du}, \citenamefont {Yuen-Zhou},\ and\ \citenamefont {Xiong}}]{Xiang2019State}%
  \BibitemOpen
  \bibfield  {author} {\bibinfo {author} {\bibfnamefont {B.}~\bibnamefont {Xiang}}, \bibinfo {author} {\bibfnamefont {R.~F.}\ \bibnamefont {Ribeiro}}, \bibinfo {author} {\bibfnamefont {L.}~\bibnamefont {Chen}}, \bibinfo {author} {\bibfnamefont {J.}~\bibnamefont {Wang}}, \bibinfo {author} {\bibfnamefont {M.}~\bibnamefont {Du}}, \bibinfo {author} {\bibfnamefont {J.}~\bibnamefont {Yuen-Zhou}},\ and\ \bibinfo {author} {\bibfnamefont {W.}~\bibnamefont {Xiong}},\ }\bibfield  {title} {\bibinfo {title} {{State-Selective Polariton to Dark State Relaxation Dynamics}},\ }\href {https://doi.org/10.1021/acs.jpca.9b04601} {\bibfield  {journal} {\bibinfo  {journal} {J. Phys. Chem. A}\ }\textbf {\bibinfo {volume} {123}},\ \bibinfo {pages} {5918} (\bibinfo {year} {2019})}\BibitemShut {NoStop}%
\bibitem [{\citenamefont {Ribeiro}\ \emph {et~al.}(2021)\citenamefont {Ribeiro}, \citenamefont {Campos-Gonzalez-Angulo}, \citenamefont {Giebink}, \citenamefont {Xiong},\ and\ \citenamefont {Yuen-Zhou}}]{Ribeiro2020}%
  \BibitemOpen
  \bibfield  {author} {\bibinfo {author} {\bibfnamefont {R.~F.}\ \bibnamefont {Ribeiro}}, \bibinfo {author} {\bibfnamefont {J.~A.}\ \bibnamefont {Campos-Gonzalez-Angulo}}, \bibinfo {author} {\bibfnamefont {N.~C.}\ \bibnamefont {Giebink}}, \bibinfo {author} {\bibfnamefont {W.}~\bibnamefont {Xiong}},\ and\ \bibinfo {author} {\bibfnamefont {J.}~\bibnamefont {Yuen-Zhou}},\ }\bibfield  {title} {\bibinfo {title} {{Enhanced Optical Nonlinearities under Collective Strong Light-Matter Coupling}},\ }\href {https://doi.org/10.1103/PhysRevA.103.063111} {\bibfield  {journal} {\bibinfo  {journal} {Phys. Rev. A}\ }\textbf {\bibinfo {volume} {103}},\ \bibinfo {pages} {063111} (\bibinfo {year} {2021})}\BibitemShut {NoStop}%
\bibitem [{\citenamefont {Wright}\ \emph {et~al.}(2023)\citenamefont {Wright}, \citenamefont {Nelson},\ and\ \citenamefont {Weichman}}]{Wright2023}%
  \BibitemOpen
  \bibfield  {author} {\bibinfo {author} {\bibfnamefont {A.~D.}\ \bibnamefont {Wright}}, \bibinfo {author} {\bibfnamefont {J.~C.}\ \bibnamefont {Nelson}},\ and\ \bibinfo {author} {\bibfnamefont {M.~L.}\ \bibnamefont {Weichman}},\ }\bibfield  {title} {\bibinfo {title} {{Rovibrational Polaritons in Gas-Phase Methane}},\ }\href {https://doi.org/10.1021/jacs.3c00126} {\bibfield  {journal} {\bibinfo  {journal} {J. Am. Chem. Soc.}\ }\textbf {\bibinfo {volume} {145}},\ \bibinfo {pages} {5982} (\bibinfo {year} {2023})}\BibitemShut {NoStop}%
\bibitem [{\citenamefont {Crawford}\ \emph {et~al.}(1952)\citenamefont {Crawford}, \citenamefont {Welsh},\ and\ \citenamefont {Harrold}}]{Crawford1952}%
  \BibitemOpen
  \bibfield  {author} {\bibinfo {author} {\bibfnamefont {M.~F.}\ \bibnamefont {Crawford}}, \bibinfo {author} {\bibfnamefont {H.~L.}\ \bibnamefont {Welsh}},\ and\ \bibinfo {author} {\bibfnamefont {J.~H.}\ \bibnamefont {Harrold}},\ }\bibfield  {title} {\bibinfo {title} {{Rotational Wings Of Raman Bands And Free Rotation In Liquid Oxygen, Nitrogen, And Methane}},\ }\href {https://doi.org/10.1139/p52-008} {\bibfield  {journal} {\bibinfo  {journal} {Can. J. Phys.}\ }\textbf {\bibinfo {volume} {30}},\ \bibinfo {pages} {81} (\bibinfo {year} {1952})}\BibitemShut {NoStop}%
\bibitem [{\citenamefont {Chapados}\ and\ \citenamefont {Cabana}(1972)}]{Chapados1972}%
  \BibitemOpen
  \bibfield  {author} {\bibinfo {author} {\bibfnamefont {C.}~\bibnamefont {Chapados}}\ and\ \bibinfo {author} {\bibfnamefont {A.}~\bibnamefont {Cabana}},\ }\bibfield  {title} {\bibinfo {title} {{Infrared Spectra and Structures of Solid CH4 and CD4 in Phases I and II}},\ }\href {https://doi.org/10.1139/v72-566} {\bibfield  {journal} {\bibinfo  {journal} {Can. J. Phys.}\ }\textbf {\bibinfo {volume} {50}},\ \bibinfo {pages} {3521} (\bibinfo {year} {1972})}\BibitemShut {NoStop}%
\bibitem [{\citenamefont {Sun}(1998)}]{Sun1998}%
  \BibitemOpen
  \bibfield  {author} {\bibinfo {author} {\bibfnamefont {H.}~\bibnamefont {Sun}},\ }\bibfield  {title} {\bibinfo {title} {{COMPASS: An ab Initio Force-Field Optimized for Condensed-Phase Applications: Overview with Details on Alkane and Benzene Compounds}},\ }\href {https://doi.org/10.1021/jp980939v} {\bibfield  {journal} {\bibinfo  {journal} {J. Phys. Chem. B}\ }\textbf {\bibinfo {volume} {102}},\ \bibinfo {pages} {7338} (\bibinfo {year} {1998})}\BibitemShut {NoStop}%
\bibitem [{\citenamefont {Veit}\ \emph {et~al.}(2019)\citenamefont {Veit}, \citenamefont {Jain}, \citenamefont {Bonakala}, \citenamefont {Rudra}, \citenamefont {Hohl},\ and\ \citenamefont {Cs{\'{a}}nyi}}]{Veit2019}%
  \BibitemOpen
  \bibfield  {author} {\bibinfo {author} {\bibfnamefont {M.}~\bibnamefont {Veit}}, \bibinfo {author} {\bibfnamefont {S.~K.}\ \bibnamefont {Jain}}, \bibinfo {author} {\bibfnamefont {S.}~\bibnamefont {Bonakala}}, \bibinfo {author} {\bibfnamefont {I.}~\bibnamefont {Rudra}}, \bibinfo {author} {\bibfnamefont {D.}~\bibnamefont {Hohl}},\ and\ \bibinfo {author} {\bibfnamefont {G.}~\bibnamefont {Cs{\'{a}}nyi}},\ }\bibfield  {title} {\bibinfo {title} {{Equation of State of Fluid Methane from First Principles with Machine Learning Potentials}},\ }\href {https://doi.org/10.1021/acs.jctc.8b01242} {\bibfield  {journal} {\bibinfo  {journal} {J. Chem. Theory Comput.}\ }\textbf {\bibinfo {volume} {15}},\ \bibinfo {pages} {2574} (\bibinfo {year} {2019})}\BibitemShut {NoStop}%
\bibitem [{\citenamefont {Deringer}\ \emph {et~al.}(2021)\citenamefont {Deringer}, \citenamefont {Bart{\'{o}}k}, \citenamefont {Bernstein}, \citenamefont {Wilkins}, \citenamefont {Ceriotti},\ and\ \citenamefont {Cs{\'{a}}nyi}}]{Deringer2021}%
  \BibitemOpen
  \bibfield  {author} {\bibinfo {author} {\bibfnamefont {V.~L.}\ \bibnamefont {Deringer}}, \bibinfo {author} {\bibfnamefont {A.~P.}\ \bibnamefont {Bart{\'{o}}k}}, \bibinfo {author} {\bibfnamefont {N.}~\bibnamefont {Bernstein}}, \bibinfo {author} {\bibfnamefont {D.~M.}\ \bibnamefont {Wilkins}}, \bibinfo {author} {\bibfnamefont {M.}~\bibnamefont {Ceriotti}},\ and\ \bibinfo {author} {\bibfnamefont {G.}~\bibnamefont {Cs{\'{a}}nyi}},\ }\bibfield  {title} {\bibinfo {title} {{Gaussian Process Regression for Materials and Molecules}},\ }\href {https://doi.org/10.1021/acs.chemrev.1c00022} {\bibfield  {journal} {\bibinfo  {journal} {Chem. Rev.}\ }\textbf {\bibinfo {volume} {121}},\ \bibinfo {pages} {10073} (\bibinfo {year} {2021})}\BibitemShut {NoStop}%
\bibitem [{\citenamefont {Thompson}\ \emph {et~al.}(2022)\citenamefont {Thompson}, \citenamefont {Aktulga}, \citenamefont {Berger}, \citenamefont {Bolintineanu}, \citenamefont {Brown}, \citenamefont {Crozier}, \citenamefont {{in 't Veld}}, \citenamefont {Kohlmeyer}, \citenamefont {Moore}, \citenamefont {Nguyen}, \citenamefont {Shan}, \citenamefont {Stevens}, \citenamefont {Tranchida}, \citenamefont {Trott},\ and\ \citenamefont {Plimpton}}]{Thompson2022}%
  \BibitemOpen
  \bibfield  {author} {\bibinfo {author} {\bibfnamefont {A.~P.}\ \bibnamefont {Thompson}}, \bibinfo {author} {\bibfnamefont {H.~M.}\ \bibnamefont {Aktulga}}, \bibinfo {author} {\bibfnamefont {R.}~\bibnamefont {Berger}}, \bibinfo {author} {\bibfnamefont {D.~S.}\ \bibnamefont {Bolintineanu}}, \bibinfo {author} {\bibfnamefont {W.~M.}\ \bibnamefont {Brown}}, \bibinfo {author} {\bibfnamefont {P.~S.}\ \bibnamefont {Crozier}}, \bibinfo {author} {\bibfnamefont {P.~J.}\ \bibnamefont {{in 't Veld}}}, \bibinfo {author} {\bibfnamefont {A.}~\bibnamefont {Kohlmeyer}}, \bibinfo {author} {\bibfnamefont {S.~G.}\ \bibnamefont {Moore}}, \bibinfo {author} {\bibfnamefont {T.~D.}\ \bibnamefont {Nguyen}}, \bibinfo {author} {\bibfnamefont {R.}~\bibnamefont {Shan}}, \bibinfo {author} {\bibfnamefont {M.~J.}\ \bibnamefont {Stevens}}, \bibinfo {author} {\bibfnamefont {J.}~\bibnamefont {Tranchida}}, \bibinfo {author} {\bibfnamefont {C.}~\bibnamefont {Trott}},\ and\ \bibinfo {author} {\bibfnamefont {S.~J.}\ \bibnamefont {Plimpton}},\
  }\bibfield  {title} {\bibinfo {title} {{LAMMPS - A Flexible Simulation Tool for Particle-based Materials Modeling at the Atomic, Neso, and Continuum Scales}},\ }\href {https://doi.org/10.1016/j.cpc.2021.108171} {\bibfield  {journal} {\bibinfo  {journal} {Comput. Phys. Commun.}\ }\textbf {\bibinfo {volume} {271}},\ \bibinfo {pages} {108171} (\bibinfo {year} {2022})}\BibitemShut {NoStop}%
\bibitem [{\citenamefont {Litman}\ \emph {et~al.}(2024)\citenamefont {Litman}, \citenamefont {Kapil}, \citenamefont {Feldman}, \citenamefont {Tisi}, \citenamefont {Begu{\v{s}}i{\'{c}}}, \citenamefont {Fidanyan}, \citenamefont {Fraux}, \citenamefont {Higer}, \citenamefont {Kellner}, \citenamefont {Li}, \citenamefont {P{\'{o}}s}, \citenamefont {Stocco}, \citenamefont {Trenins}, \citenamefont {Hirshberg}, \citenamefont {Rossi},\ and\ \citenamefont {Ceriotti}}]{Litman2024}%
  \BibitemOpen
  \bibfield  {author} {\bibinfo {author} {\bibfnamefont {Y.}~\bibnamefont {Litman}}, \bibinfo {author} {\bibfnamefont {V.}~\bibnamefont {Kapil}}, \bibinfo {author} {\bibfnamefont {Y.~M.}\ \bibnamefont {Feldman}}, \bibinfo {author} {\bibfnamefont {D.}~\bibnamefont {Tisi}}, \bibinfo {author} {\bibfnamefont {T.}~\bibnamefont {Begu{\v{s}}i{\'{c}}}}, \bibinfo {author} {\bibfnamefont {K.}~\bibnamefont {Fidanyan}}, \bibinfo {author} {\bibfnamefont {G.}~\bibnamefont {Fraux}}, \bibinfo {author} {\bibfnamefont {J.}~\bibnamefont {Higer}}, \bibinfo {author} {\bibfnamefont {M.}~\bibnamefont {Kellner}}, \bibinfo {author} {\bibfnamefont {T.~E.}\ \bibnamefont {Li}}, \bibinfo {author} {\bibfnamefont {E.~S.}\ \bibnamefont {P{\'{o}}s}}, \bibinfo {author} {\bibfnamefont {E.}~\bibnamefont {Stocco}}, \bibinfo {author} {\bibfnamefont {G.}~\bibnamefont {Trenins}}, \bibinfo {author} {\bibfnamefont {B.}~\bibnamefont {Hirshberg}}, \bibinfo {author} {\bibfnamefont {M.}~\bibnamefont {Rossi}},\ and\ \bibinfo {author} {\bibfnamefont
  {M.}~\bibnamefont {Ceriotti}},\ }\bibfield  {title} {\bibinfo {title} {{i-PI 3.0: A Flexible and Efficient Framework for Advanced Atomistic Simulations}},\ }\href {https://doi.org/10.1063/5.0215869} {\bibfield  {journal} {\bibinfo  {journal} {J. Chem. Phys.}\ }\textbf {\bibinfo {volume} {161}},\ \bibinfo {pages} {062504} (\bibinfo {year} {2024})}\BibitemShut {NoStop}%
\bibitem [{\citenamefont {McQuarrie}(1976)}]{McQuarrie1976}%
  \BibitemOpen
  \bibfield  {author} {\bibinfo {author} {\bibfnamefont {D.~A.}\ \bibnamefont {McQuarrie}},\ }\href@noop {} {\emph {\bibinfo {title} {{Statistical Mechanics}}}}\ (\bibinfo  {publisher} {Harper-Collins Publish- ers},\ \bibinfo {address} {New York},\ \bibinfo {year} {1976})\BibitemShut {NoStop}%
\bibitem [{\citenamefont {Gaigeot}\ and\ \citenamefont {Sprik}(2003)}]{Gaigeot2003}%
  \BibitemOpen
  \bibfield  {author} {\bibinfo {author} {\bibfnamefont {M.-P.}\ \bibnamefont {Gaigeot}}\ and\ \bibinfo {author} {\bibfnamefont {M.}~\bibnamefont {Sprik}},\ }\bibfield  {title} {\bibinfo {title} {{Ab Initio Molecular Dynamics Computation of the Infrared Spectrum of Aqueous Uracil}},\ }\href {https://doi.org/10.1021/jp034788u} {\bibfield  {journal} {\bibinfo  {journal} {J. Phys. Chem. B}\ }\textbf {\bibinfo {volume} {107}},\ \bibinfo {pages} {10344} (\bibinfo {year} {2003})}\BibitemShut {NoStop}%
\bibitem [{\citenamefont {Habershon}\ \emph {et~al.}(2008)\citenamefont {Habershon}, \citenamefont {Fanourgakis},\ and\ \citenamefont {Manolopoulos}}]{Habershon2008}%
  \BibitemOpen
  \bibfield  {author} {\bibinfo {author} {\bibfnamefont {S.}~\bibnamefont {Habershon}}, \bibinfo {author} {\bibfnamefont {G.~S.}\ \bibnamefont {Fanourgakis}},\ and\ \bibinfo {author} {\bibfnamefont {D.~E.}\ \bibnamefont {Manolopoulos}},\ }\bibfield  {title} {\bibinfo {title} {{Comparison of Path Integral Molecular Dynamics Methods for the Infrared Absorption Spectrum of Liquid Water}},\ }\href {https://doi.org/10.1063/1.2968555} {\bibfield  {journal} {\bibinfo  {journal} {J. Chem. Phys.}\ }\textbf {\bibinfo {volume} {129}},\ \bibinfo {pages} {074501} (\bibinfo {year} {2008})}\BibitemShut {NoStop}%
\bibitem [{\citenamefont {Childs}\ and\ \citenamefont {Ahn}(1939)}]{Childs1939}%
  \BibitemOpen
  \bibfield  {author} {\bibinfo {author} {\bibfnamefont {W.~H.~J.}\ \bibnamefont {Childs}}\ and\ \bibinfo {author} {\bibfnamefont {H.~A.~J.}\ \bibnamefont {Ahn}},\ }\bibfield  {title} {\bibinfo {title} {{A New Coriolis Perturbation in the Methane Spectrum III. Intensities and Optical Spectrum}},\ }\href {https://doi.org/10.1098/rspa.1939.0009} {\bibfield  {journal} {\bibinfo  {journal} {Proc. R. soc. Lond. Ser. A}\ }\textbf {\bibinfo {volume} {169}},\ \bibinfo {pages} {451} (\bibinfo {year} {1939})}\BibitemShut {NoStop}%
\bibitem [{\citenamefont {Robiette}\ and\ \citenamefont {Mills}(1979)}]{Robiette1979}%
  \BibitemOpen
  \bibfield  {author} {\bibinfo {author} {\bibfnamefont {A.}~\bibnamefont {Robiette}}\ and\ \bibinfo {author} {\bibfnamefont {I.}~\bibnamefont {Mills}},\ }\bibfield  {title} {\bibinfo {title} {{Intensity Perturbations due to $\nu$3/$\nu$4 Coriolis Interaction in Methane}},\ }\href {https://doi.org/10.1016/0022-2852(79)90195-4} {\bibfield  {journal} {\bibinfo  {journal} {J. Mol. Spectrosc.}\ }\textbf {\bibinfo {volume} {77}},\ \bibinfo {pages} {48} (\bibinfo {year} {1979})}\BibitemShut {NoStop}%
\bibitem [{\citenamefont {Tipping}\ \emph {et~al.}(2001)\citenamefont {Tipping}, \citenamefont {Brown}, \citenamefont {Ma}, \citenamefont {Hartmann}, \citenamefont {Boulet},\ and\ \citenamefont {Li{\'{e}}vin}}]{Tipping2001}%
  \BibitemOpen
  \bibfield  {author} {\bibinfo {author} {\bibfnamefont {R.~H.}\ \bibnamefont {Tipping}}, \bibinfo {author} {\bibfnamefont {A.}~\bibnamefont {Brown}}, \bibinfo {author} {\bibfnamefont {Q.}~\bibnamefont {Ma}}, \bibinfo {author} {\bibfnamefont {J.~M.}\ \bibnamefont {Hartmann}}, \bibinfo {author} {\bibfnamefont {C.}~\bibnamefont {Boulet}},\ and\ \bibinfo {author} {\bibfnamefont {J.}~\bibnamefont {Li{\'{e}}vin}},\ }\bibfield  {title} {\bibinfo {title} {{Collision-induced Absorption in the $\nu$2 Fundamental Band of CH4. I. Determination of the Quadrupole Transition Moment}},\ }\href {https://doi.org/10.1063/1.1408915} {\bibfield  {journal} {\bibinfo  {journal} {J. Chem. Phys.}\ }\textbf {\bibinfo {volume} {115}},\ \bibinfo {pages} {8852} (\bibinfo {year} {2001})}\BibitemShut {NoStop}%
\bibitem [{\citenamefont {Ying}\ \emph {et~al.}(2024)\citenamefont {Ying}, \citenamefont {Mondal},\ and\ \citenamefont {Huo}}]{ying2024theory}%
  \BibitemOpen
  \bibfield  {author} {\bibinfo {author} {\bibfnamefont {W.}~\bibnamefont {Ying}}, \bibinfo {author} {\bibfnamefont {M.~E.}\ \bibnamefont {Mondal}},\ and\ \bibinfo {author} {\bibfnamefont {P.}~\bibnamefont {Huo}},\ }\bibfield  {title} {\bibinfo {title} {Theory and quantum dynamics simulations of exciton-polariton motional narrowing},\ }\href@noop {} {\bibfield  {journal} {\bibinfo  {journal} {J. Chem. Phys.}\ }\textbf {\bibinfo {volume} {161}},\ \bibinfo {pages} {064105} (\bibinfo {year} {2024})}\BibitemShut {NoStop}%
\bibitem [{\citenamefont {Groenhof}\ \emph {et~al.}(2019)\citenamefont {Groenhof}, \citenamefont {Climent}, \citenamefont {Feist}, \citenamefont {Morozov},\ and\ \citenamefont {Toppari}}]{Groenhof2019}%
  \BibitemOpen
  \bibfield  {author} {\bibinfo {author} {\bibfnamefont {G.}~\bibnamefont {Groenhof}}, \bibinfo {author} {\bibfnamefont {C.}~\bibnamefont {Climent}}, \bibinfo {author} {\bibfnamefont {J.}~\bibnamefont {Feist}}, \bibinfo {author} {\bibfnamefont {D.}~\bibnamefont {Morozov}},\ and\ \bibinfo {author} {\bibfnamefont {J.~J.}\ \bibnamefont {Toppari}},\ }\bibfield  {title} {\bibinfo {title} {{Tracking Polariton Relaxation with Multiscale Molecular Dynamics Simulations}},\ }\href {https://doi.org/10.1021/acs.jpclett.9b02192} {\bibfield  {journal} {\bibinfo  {journal} {J. Phys. Chem. Lett.}\ }\textbf {\bibinfo {volume} {10}},\ \bibinfo {pages} {5476} (\bibinfo {year} {2019})}\BibitemShut {NoStop}%
\bibitem [{\citenamefont {Chng}\ \emph {et~al.}(2024)\citenamefont {Chng}, \citenamefont {Ying}, \citenamefont {Lai}, \citenamefont {Vamivakas}, \citenamefont {Cundiff}, \citenamefont {Krauss},\ and\ \citenamefont {Huo}}]{Chng2024}%
  \BibitemOpen
  \bibfield  {author} {\bibinfo {author} {\bibfnamefont {B.~X.~K.}\ \bibnamefont {Chng}}, \bibinfo {author} {\bibfnamefont {W.}~\bibnamefont {Ying}}, \bibinfo {author} {\bibfnamefont {Y.}~\bibnamefont {Lai}}, \bibinfo {author} {\bibfnamefont {A.~N.}\ \bibnamefont {Vamivakas}}, \bibinfo {author} {\bibfnamefont {S.~T.}\ \bibnamefont {Cundiff}}, \bibinfo {author} {\bibfnamefont {T.~D.}\ \bibnamefont {Krauss}},\ and\ \bibinfo {author} {\bibfnamefont {P.}~\bibnamefont {Huo}},\ }\bibfield  {title} {\bibinfo {title} {{Mechanism of Molecular Polariton Decoherence in the Collective Light–Matter Couplings Regime}},\ }\href {https://doi.org/10.1021/acs.jpclett.4c03049} {\bibfield  {journal} {\bibinfo  {journal} {J. Phys. Chem. Lett.}\ }\textbf {\bibinfo {volume} {15}},\ \bibinfo {pages} {11773} (\bibinfo {year} {2024})}\BibitemShut {NoStop}%
\bibitem [{\citenamefont {Lazzeretti}\ \emph {et~al.}(1987)\citenamefont {Lazzeretti}, \citenamefont {Zanasi}, \citenamefont {Sadlej},\ and\ \citenamefont {Raynes}}]{Lazzeretti1987}%
  \BibitemOpen
  \bibfield  {author} {\bibinfo {author} {\bibfnamefont {P.}~\bibnamefont {Lazzeretti}}, \bibinfo {author} {\bibfnamefont {R.}~\bibnamefont {Zanasi}}, \bibinfo {author} {\bibfnamefont {A.}~\bibnamefont {Sadlej}},\ and\ \bibinfo {author} {\bibfnamefont {W.}~\bibnamefont {Raynes}},\ }\bibfield  {title} {\bibinfo {title} {{Magnetizability and Carbon-13 Shielding Surfaces for the Methane Molecule}},\ }\href {https://doi.org/10.1080/00268978700102431} {\bibfield  {journal} {\bibinfo  {journal} {Mol. Phys.}\ }\textbf {\bibinfo {volume} {62}},\ \bibinfo {pages} {605} (\bibinfo {year} {1987})}\BibitemShut {NoStop}%
\bibitem [{\citenamefont {Wang}\ and\ \citenamefont {Carrington}(2003)}]{Wang2003}%
  \BibitemOpen
  \bibfield  {author} {\bibinfo {author} {\bibfnamefont {X.-G.}\ \bibnamefont {Wang}}\ and\ \bibinfo {author} {\bibfnamefont {T.}~\bibnamefont {Carrington}},\ }\bibfield  {title} {\bibinfo {title} {{Deficiencies of the Bend Symmetry Coordinates Used for Methane}},\ }\href {https://doi.org/10.1063/1.1557455} {\bibfield  {journal} {\bibinfo  {journal} {J. Chem. Phys.}\ }\textbf {\bibinfo {volume} {118}},\ \bibinfo {pages} {6260} (\bibinfo {year} {2003})}\BibitemShut {NoStop}%
\bibitem [{\citenamefont {Carusotto}\ and\ \citenamefont {Ciuti}(2013)}]{Carusotto2013}%
  \BibitemOpen
  \bibfield  {author} {\bibinfo {author} {\bibfnamefont {I.}~\bibnamefont {Carusotto}}\ and\ \bibinfo {author} {\bibfnamefont {C.}~\bibnamefont {Ciuti}},\ }\bibfield  {title} {\bibinfo {title} {{Quantum Fluids of Light}},\ }\href {https://doi.org/10.1103/RevModPhys.85.299} {\bibfield  {journal} {\bibinfo  {journal} {Rev. Mod. Phys.}\ }\textbf {\bibinfo {volume} {85}},\ \bibinfo {pages} {299} (\bibinfo {year} {2013})}\BibitemShut {NoStop}%
\bibitem [{\citenamefont {Bhuyan}\ \emph {et~al.}(2023)\citenamefont {Bhuyan}, \citenamefont {Mony}, \citenamefont {Kotov}, \citenamefont {Castellanos}, \citenamefont {{G{\'{o}}mez Rivas}}, \citenamefont {Shegai},\ and\ \citenamefont {B{\"{o}}rjesson}}]{Bhuyan2023}%
  \BibitemOpen
  \bibfield  {author} {\bibinfo {author} {\bibfnamefont {R.}~\bibnamefont {Bhuyan}}, \bibinfo {author} {\bibfnamefont {J.}~\bibnamefont {Mony}}, \bibinfo {author} {\bibfnamefont {O.}~\bibnamefont {Kotov}}, \bibinfo {author} {\bibfnamefont {G.~W.}\ \bibnamefont {Castellanos}}, \bibinfo {author} {\bibfnamefont {J.}~\bibnamefont {{G{\'{o}}mez Rivas}}}, \bibinfo {author} {\bibfnamefont {T.~O.}\ \bibnamefont {Shegai}},\ and\ \bibinfo {author} {\bibfnamefont {K.}~\bibnamefont {B{\"{o}}rjesson}},\ }\bibfield  {title} {\bibinfo {title} {{The Rise and Current Status of Polaritonic Photochemistry and Photophysics}},\ }\href {https://doi.org/10.1021/ACS.CHEMREV.2C00895/ASSET/IMAGES/LARGE/CR2C00895_0015.JPEG} {\bibfield  {journal} {\bibinfo  {journal} {Chem. Rev.}\ }\textbf {\bibinfo {volume} {123}},\ \bibinfo {pages} {10877} (\bibinfo {year} {2023})}\BibitemShut {NoStop}%
\end{thebibliography}


\clearpage


\section*{Supplementary material}

# Supplementary Information

# Selective Excitation of IR-Inactive Vibrations via Vibrational Polaritons: Insights from Atomistic Simulations

Xinwei Ji[1] and Tao E. Li[1,*]

[1]*Department of Physics and Astronomy,*
*University of Delaware, Newark, Delaware 19716, USA*

# CONTENTS

---

* taoeli@udel.edu



# I. ANALYTICAL THEORY OF UP$_{v_4}$ DECAY RATES

In this section, we present the analytical derivation of the UP$_{v_4}$ decay rate discussed in the main text. Following Ref. [S1], we first describe the light-matter system, including the polariton states, within the harmonic limit. Then, considering the weak couplings between the polariton states and various molecular dark states induced by molecular interactions, we derive the UP$_{v_4}$ decay rate into these dark states using the Fermi's golden rule.

We consider the following Tavis–Cummings Hamiltonian to describe vibrational strong coupling (VSC) between a single cavity photon mode and $N$ IR-active $v_4$ vibrational transitions:

$$\hat{H} = \hbar\omega_c \hat{a}^\dagger \hat{a} + \hbar\omega_0 \sum_{n=1}^{N} \hat{b}_n^\dagger \hat{b}_n + \hbar g_0 \sum_{n=1}^{N} \left( \hat{a}^\dagger \hat{b}_n + \hat{a} \hat{b}_n^\dagger \right). \tag{S1}$$

Here, $\omega_c$ and $\omega_0$ denote the frequencies of the cavity and IR-active $v_4$ vibrational mode, respectively; $\hat{a}^\dagger$ ($\hat{b}_n^\dagger$) and $\hat{a}$ ($\hat{b}_n$) represent the creation and annihilation operators of the cavity mode ($n$-th $v_4$ vibrational mode), respectively; and $g_0$ is the coupling strength between the cavity mode and each $v_4$ vibrational mode. Unlike the standard Tavis–Cummings Hamiltonian [S2, S3], in which molecular excitations are described using two-level systems, here, quantum harmonic oscillators are employed to better represent molecular vibrations. As the $v_4$ vibrational transitions are triply degenerate in each CH$_4$ molecule, the inclusion of $N$ $v_4$ vibrational modes in Eq. (S1) implies that the Hamiltonian describes a system containing $N/3$ CH$_4$ molecules.

Following the standard protocol, we define the bright-mode creation and annihilation operators for $N$ IR-active $v_4$ vibrational transitions as the symmetric combinations of local molecular operators:

$$\hat{B}^\dagger = \frac{1}{\sqrt{N}} \sum_{n=1}^{N} \hat{b}_n^\dagger, \tag{S2a}$$

$$\hat{B} = \frac{1}{\sqrt{N}} \sum_{n=1}^{N} \hat{b}_n. \tag{S2b}$$

The remaining asymmetric $N-1$ linear combinations of $v_4$ operators form the dark-state

manifold. The creation and annihilation operators of these $N-1$ dark states are given by:

$$\hat{D}_\mu^\dagger = \frac{1}{\sqrt{N}} \sum_{n=1}^{N} e^{i2\pi n\mu/N} \hat{b}_n^\dagger, \tag{S3a}$$

$$\hat{D}_\mu = \frac{1}{\sqrt{N}} \sum_{n=1}^{N} e^{i2\pi n\mu/N} \hat{b}_n, \tag{S3b}$$

where $\mu = 1, 2, \ldots, N-1$ indexes the dark modes.

Using the definitions of the bright- and dark-state operators, we can rewrite the Tavis–Cummings Hamiltonian in Eq. (S1) as:

$$\hat{H} = \hbar\omega_c \hat{a}^\dagger \hat{a} + \hbar\omega_0 \hat{B}^\dagger \hat{B} + \frac{1}{2}\hbar\Omega_N \left(\hat{a}^\dagger \hat{B} + \hat{a}\hat{B}^\dagger\right) + \hat{H}_D. \tag{S4}$$

Here, $\Omega_N = 2g_0\sqrt{N}$ represents the collective Rabi splitting, and the dark-state Hamiltonian $\hat{H}_D$ reads:

$$\hat{H}_D = \sum_{\mu=1}^{N-1} \hbar\omega_0 \hat{D}_\mu^\dagger \hat{D}_\mu. \tag{S5}$$

Clearly, these dark states are decoupled from the cavity mode.

Eq. (S4) can be further diagonalized, leading to the standard polariton Hamiltonian:

$$\hat{H} = \hbar\omega_+ \hat{P}_+^\dagger \hat{P}_+ + \hbar\omega_- \hat{P}_-^\dagger \hat{P}_- + \hat{H}_D. \tag{S6}$$

Here, $\omega_+$ and $\omega_-$ represent the frequencies of the upper and the lower polariton (UP and LP), respectively:

$$\omega_\pm = \frac{1}{2}\left[\omega_0 + \omega_c \pm \sqrt{\Omega_N^2 + (\omega_0 - \omega_c)^2}\right]. \tag{S7}$$

In Eq. (S6), the polariton creation and annihilation operators ($\hat{P}_\pm^\dagger$ and $\hat{P}_\pm$) are represented as:

$$\hat{P}_\pm^\dagger = X_\pm^{(B)} \hat{B}^\dagger + X_\pm^{(c)} \hat{a}^\dagger, \tag{S8a}$$

$$\hat{P}_\pm = X_\pm^{(B)} \hat{B} + X_\pm^{(c)} \hat{a}. \tag{S8b}$$

Here, the Hopfield coefficients of the polariton states are defined as $\hat{X}_+^{(B)} = -\hat{X}_-^{(c)} = -\sin\theta$ and $\hat{X}_-^{(B)} = \hat{X}_+^{(c)} = \cos\theta$, where the mixing angle $\theta$ quantifies the hybridization between the molecular and photonic states:

$$\theta = \frac{1}{2}\arctan\left(\frac{\Omega_N}{\omega_c - \omega_0}\right). \tag{S9}$$

Maximal light-matter hybridization occurs at resonance when $\omega_c = \omega_0$. In this limit, for a finite Rabi splitting $\Omega_N$, the ratio $\frac{\Omega_N}{\omega_c - \omega_0}$ approaches $\pm\infty$, leading to $\theta \to \pm\frac{\pi}{4}$ and $|\hat{X}_\pm^{(B)}| = |\hat{X}_\pm^{(c)}| = \frac{1}{\sqrt{2}}$. In other words, at resonance strong coupling, the molecular weight $|\hat{X}_\pm^{(B)}|^2$ and the photonic weight $|\hat{X}_\pm^{(c)}|^2$ in each polariton state both become $1/2$.

### A. The rate of $UP_{v_4}$ dephasing into $v_4$ dark modes

After obtaining the harmonic polariton Hamiltonian in Eq. (S6), we now derive the relaxation rate from the UP to the dark-mode manifold of IR-active $v_4$ vibrations. A similar rate was derived in Ref. [S1] for VSC in a liquid $CO_2$ system. In that molecular system, intermolecular dipole-dipole interactions serve as the only pathway for polariton dephasing into dark modes. In contrast, for the liquid $CH_4$ system studied in this manuscript, the polaritons formed by the $v_4$ transitions can dephase into the $v_4$ dark modes through two distinct pathways: (i) intermolecular dipole-dipole interactions between $v_4$ vibrations, and (ii) intramolecular anharmonic interactions within the triply degenerate $v_4$ transitions. We consider these two pathways separately.

#### 1. Contribution from intermolecular dipole-dipole interactions

Following Ref. [S1], we express the intermolecular dipole-dipole coupling between $v_4$ transitions in neighboring $CH_4$ molecules using the following tight-binding form:

$$\hat{V}_{\mathrm{dd}} = \sum_{n=1}^{N} \hbar \Delta_n \left[ \sum_{M_n=1}^{N_{\mathrm{nn}}} \left( \hat{b}_n^\dagger \hat{b}_{M_n} + \hat{b}_n \hat{b}_{M_n}^\dagger \right) \right]. \quad (S10)$$

Here, $M_n$ denotes all the possible nearest neighbors of the $n$-th $v_4$ vibrational transition; $N_{\mathrm{nn}}$ represents the total number of nearest neighbors for each $v_4$ transition; and $\Delta_n$ is the intermolecular dipole-dipole coupling strength between neighboring $v_4$ vibrational transitions in different molecules.

According to Eq. (S2), each molecular $v_4$ transition contains a small bright-state contribution. In other words, $\hat{b}_{M_n} = \hat{B}/\sqrt{N} + \cdots$, where $\cdots$ represents a linear combination of the dark-state operators $\hat{D}_\mu$ for $\mu = 1, 2, \cdots, N-1$. Utilizing $\hat{b}_{M_n} = \hat{B}/\sqrt{N} + \cdots$, we can

rewrite the local dipole-dipole coupling in Eq. (S10) as

$$\hat{V}_{\text{dd}} = \frac{\hbar N_{\text{nn}}}{\sqrt{N}} \sum_{n=1}^{N} \Delta_n \left( \hat{b}_n^\dagger \hat{B} + \hat{b}_n \hat{B}^\dagger \right) + \cdots \tag{S11}$$

In the large $N$ limit, the local $v_4$ operators $\hat{b}_n^\dagger$ and $\hat{b}_n$ are predominately contributed by the $v_4$ dark modes defined in Eq. (S3). Hence, Eq. (S11) provides the interactions between the bright and dark modes of $v_4$ transitions.

According to the Fermi's golden rule, the energy transfer rate from the $\text{UP}_{v_4}$ to the $v_4$ dark modes can be calculated using

$$\gamma_{\text{UP} \to \text{D}_{v_4}} = \sum_f \frac{2\pi}{\hbar^2} |V_{fi}|^2 \delta(\omega - \omega_f), \tag{S12}$$

where $f$ and $i$ index the final and initial states, respectively, $\delta(\omega - \omega_f)$ denotes the density of states for the state $f$, and $V_{fi} = \langle i | \hat{V} | f \rangle$ represents the transition matrix element.

When the intermolecular dipole-dipole coupling defined in Eq. (S11) is taken into account, the initial state $|i\rangle$ corresponds to the $\text{UP}_{v_4}$ state, given by $|i\rangle = \hat{P}_+^\dagger |0\rangle$, where $|0\rangle$ denotes the vacuum state; the final states are the $v_4$ dark modes, expressed as $|f\rangle = \hat{D}_\mu^\dagger |0\rangle$ for $\mu = 1, 2, \cdots, N-1$. In the large $N$ limit, since each local molecular transition is predominately composed of the dark modes, we can approximately write $|f\rangle \approx \hat{b}_f^\dagger |0\rangle$ for $f = 1, 2, \cdots, N$. With this approximation,

$$V_{fi} \approx \left\langle 0 \left| \hat{P}_+ \hat{V}_{\text{dd}} \hat{b}_f^\dagger \right| 0 \right\rangle = X_+^{(\text{B})} \left\langle 0 \left| \hat{B} \hat{V}_{\text{dd}} \hat{b}_f^\dagger \right| 0 \right\rangle, \tag{S13}$$

where the definition of $\hat{P}_+$ from Eq. (S8) has been applied.

By substituting Eq. (S11) into Eq. (S13), we further obtain

$$V_{fi} = X_+^{(\text{B})} \frac{\hbar N_{\text{nn}} \Delta_f}{\sqrt{N}}. \tag{S14}$$

According to Eq. (S14), the golden rule decay rate in Eq. (S12) can be rewritten as

$$\begin{aligned}\gamma_{\text{UP} \to \text{D}_{v_4}} &= \sum_{f=1}^{N} \frac{2\pi}{\hbar^2} \left( X_+^{(\text{B})} \frac{\hbar N_{\text{nn}} \Delta_f}{\sqrt{N}} \right)^2 \delta(\omega - \omega_f), \\ &= 2\pi |X_+^{(\text{B})}|^2 \Delta_{\text{dd}}^2 \rho_{v_4}(\omega).\end{aligned} \tag{S15}$$

Here, $\Delta_{\text{dd}}^2 \equiv \sum_{f=1}^{N} N_{\text{nn}}^2 \Delta_f^2 \delta(\omega - \omega_f) / \sum_{f=1}^{N} \delta(\omega - \omega_f)$ represents the average intermolecular dipole-dipole coupling between $v_4$ transitions, and $\rho_{v_4}(\omega) \equiv \frac{1}{N} \sum_{f=1}^{N} \delta(\omega - \omega_f)$ denotes the vibrational density of states per $v_4$ transition.

Notably, this polariton dephasing rate does not exhibit an explicit $N$ dependence, as the number of final states ($N$) cancels the $1/N$-scaled $|V_{fi}|^2$ term. This cancellation suggests that polariton relaxation dynamics in Fabry–Pérot cavities can be accurately simulated using only a finite number of molecules.

In experiments, since the initial polariton state has also a finite linewidth, we may include a phenomenological density of state for the UP, $\rho_+(\omega)$, with the normalization condition $\int_0^{+\infty} d\omega \rho_+(\omega) = 1$. Including this density of states smooths out the UP dephasing rate in Eq. (S15), leading to the final expression:

$$\gamma_{\text{UP} \to \text{D}_{v_4}} = 2\pi |X_+^{(\text{B})}|^2 \Delta_{\text{dd}}^2 J_{\text{UP},v_4}. \tag{S16a}$$

Here, the spectral overlap is defined as

$$J_{\text{UP},v_4} = \int_0^{+\infty} d\omega \rho_+(\omega) \rho_{v_4}(\omega). \tag{S16b}$$

2. *Contribution from intramolecular $v_4$ anharmonic interactions*

The above derivation demonstrates that intermolecular dipole-dipole interactions between local $v_4$ transitions can induce the UP$_{v_4}$ dephasing into $v_4$ dark modes. Similarly, because each CH$_4$ molecule contains three degenerate $v_4$ transitions, the anharmonic coupling among these triply degenerate $v_4$ transitions within the same molecule can also induce the UP$_{v_4}$ dephasing into $v_4$ dark modes.

Following the derivation above, we express the intramolecular $v_4$ anharmonic interactions as

$$\hat{V}_{44} = \sum_{n=1}^{N} \hbar \xi_{44} \left[ \sum_{M_n=1}^{N_{\text{nn}}} \left( \hat{b}_n^\dagger \hat{b}_n^\dagger \hat{b}_n \hat{b}_{M_n} + \hat{b}_n \hat{b}_n \hat{b}_n^\dagger \hat{b}_{M_n}^\dagger \right) \right]. \tag{S17}$$

Here, $\xi_{44}$ quantifies the magnitude of the intramolecular anharmonic coupling among the triply degenerate $v_4$ transitions, and $N_{\text{nn}} = 3$ denotes the total number of nearest neighbors within the triply degenerate $v_4$ transitions. The anharmonicity within the same vibrational transition (i.e., $M_n = n$) is also included.

Using $\hat{b}_{M_n} = \hat{B}/\sqrt{N} + \cdots$, where $\cdots$ represents a linear combination of the dark-state operators $\hat{D}_\mu$ for $\mu = 1, 2, \cdots, N-1$, we can rewrite Eq. (S17) as

$$\hat{V}_{44} = \sum_{n=1}^{N} \frac{\hbar \xi_{44}}{\sqrt{N}} \left[ \sum_{M_n=1}^{N_{\text{nn}}} \left( \hat{b}_n^\dagger \hat{b}_n^\dagger \hat{b}_n \hat{B} + \hat{b}_n \hat{b}_n \hat{b}_n^\dagger \hat{B}^\dagger \right) \right] + \cdots \tag{S18}$$

Following the derivations from Eq. (S12) to Eq. (S16), we obtain the anharmonicity-induced $\text{UP}_{v_4}$ dephasing rate into $v_4$ dark modes as

$$\gamma_{\text{UP}\to \text{D}_{v_4}} = 2\pi |X_+^{(\text{B})}|^2 \Xi_{44}^2 J_{\text{UP},v_4}, \quad (\text{S19})$$

where $\Xi_{44}^2 \equiv \sum_{f=1}^{N} 4N_{\text{nn}}^2 \xi_{44}^2 \delta(\omega - \omega_f)/\sum_{f=1}^{N} \delta(\omega - \omega_f)$ represents the average intramolecular anharmonic coupling among $v_4$ transitions.

Combining Eqs. (S16) and (S19), we obtain the overall polariton dephasing rate into $v_4$ dark modes:

$$\gamma_{\text{UP}\to \text{D}_{v_4}} = 2\pi |X_+^{(\text{B})}|^2 (\Delta_{\text{dd}}^2 + \Xi_{44}^2) J_{\text{UP},v_4}. \quad (\text{S20})$$

### B. The rate of $\text{UP}_{v_4}$ energy transfer to $v_2$ IR-inactive modes

Apart from the $v_4$ dark modes, the $\text{UP}_{v_4}$ can also transfer energy to the IR-inactive $v_2$ states. However, because $v_2$ vibrations have zero net transition dipole moments, intermolecular dipole-dipole interactions between $v_4$ and $v_2$ are absent. As a result, the $\text{UP}_{v_4}$ cannot transfer energy to $v_2$ states via this pathway. Instead, only intramolecular molecular interactions can induce this polariton energy transfer. In the following section, We consider possible $v_4$-$v_2$ intramolecular molecular interactions to calculate the rate of $\text{UP}_{v_4}$ energy transfer to IR-inactive $v_2$ modes.

#### 1. Contribution from direct intramolecular $v_4$-$v_2$ anharmonic interactions

Analogous to Eq. (S17), we may express the direct intramolecular $v_4$-$v_2$ anharmonic interactions as

$$\hat{V}_{24} = \sum_{n=1}^{2N/3} \hbar \xi_{24} \left[ \sum_{M_n=1}^{N_{\text{nn}}} \left( \hat{c}_n^\dagger \hat{c}_n^\dagger \hat{c}_n \hat{b}_{M_n} + \hat{c}_n \hat{c}_n \hat{c}_n^\dagger \hat{b}_{M_n}^\dagger \right) \right], \quad (\text{S21a})$$

$$= \sum_{n=1}^{2N/3} \frac{\hbar \xi_{24}}{\sqrt{N}} \left[ \sum_{M_n=1}^{N_{\text{nn}}} \left( \hat{c}_n^\dagger \hat{c}_n^\dagger \hat{c}_n \hat{B} + \hat{c}_n \hat{c}_n \hat{c}_n^\dagger \hat{B}^\dagger \right) \right] + \cdots. \quad (\text{S21b})$$

In Eq. (S21a), $\xi_{24}$ represents the magnitude of intramolecular $v_4$-$v_2$ anharmonic coupling; $\hat{c}_n^\dagger$ and $\hat{c}_n$ denote the creation and annihilation operators of each $v_2$ transition, respectively; and $N_{\text{nn}} = 3$ denotes the number of $v_4$ transitions anharmonically coupled to each $v_2$ state. The summation index $n$ runs up to $2N/3$ to account for the total number of the doubly

degenerate $v_2$ transitions. In Eq. (S21b), we have again applied $\hat{b}_{M_n} = \hat{B}/\sqrt{N} + \cdots$, where $\cdots$ represents a linear combination of the dark-state operators $\hat{D}_\mu$ for $\mu = 1, 2, \cdots, N-1$.

For the evaluation of the transition matrix element $V_{fi}$, the initial state corresponds to the $\mathrm{UP}_{v_4}$, and the final states are the IR-inactive $v_2$ transitions. Thus, $V_{fi}$ can be computed as

$$V_{fi} = \left\langle 0 \middle| \hat{P}_+ \hat{V}_{24} \hat{c}_f^\dagger \middle| 0 \right\rangle, \tag{S22a}$$

$$= X_+^{(\mathrm{B})} \left\langle 0 \middle| \hat{B} \hat{V}_{24} \hat{c}_f^\dagger \middle| 0 \right\rangle, \tag{S22b}$$

$$= X_+^{(\mathrm{B})} \frac{\hbar \xi_{24} N_{\mathrm{nn}}}{\sqrt{N}} \left\langle 0 \middle| \hat{B} \hat{c}_f \hat{c}_f \hat{c}_f^\dagger \hat{B}^\dagger \hat{c}_f^\dagger \middle| 0 \right\rangle, \tag{S22c}$$

$$= X_+^{(\mathrm{B})} \frac{2\hbar \xi_{24} N_{\mathrm{nn}}}{\sqrt{N}}. \tag{S22d}$$

Applying the Fermi's golden rule, we express the $\mathrm{UP}_{v_4}$ energy transfer rate to IR-inactive $v_2$ modes as

$$\gamma_{\mathrm{UP} \to v_2} = \sum_{f=1}^{2N/3} \frac{2\pi}{\hbar^2} \left( X_+^{(\mathrm{B})} \frac{2\hbar N_{\mathrm{nn}} \xi_{24}}{\sqrt{N}} \right)^2 \delta(\omega - \omega_f),$$
$$= 2\pi |X_+^{(\mathrm{B})}|^2 \Xi_{24}^2 \rho_{v_2}(\omega). \tag{S23}$$

Here, $\Xi_{24}^2 \equiv \sum_{f=1}^{2N/3} \frac{8}{3} N_{\mathrm{nn}}^2 \xi_{24}^2 \delta(\omega - \omega_f) / \sum_{f=1}^{2N/3} \delta(\omega - \omega_f)$ represents the average intramolecular anharmonic coupling between the $v_4$ and $v_2$ transitions; $\rho_{v_2}(\omega) \equiv \frac{3}{2N} \sum_{f=1}^{2N/3} \delta(\omega - \omega_f)$ denotes the vibrational density of states per $v_2$ transition.

By further accounting for the finite linewidth of the $\mathrm{UP}_{v_4}$, we obtain the final expression for the $\mathrm{UP}_{v_4}$ energy transfer rate to IR-inactive $v_2$ transitions:

$$\gamma_{\mathrm{UP} \to v_2} = 2\pi |X_+^{(\mathrm{B})}|^2 \Xi_{24}^2 J_{\mathrm{UP}, v_2}, \tag{S24a}$$

where the spectral overlap $J_{\mathrm{UP}, v_2}$ is defined as

$$J_{\mathrm{UP}, v_2} = \int_0^{+\infty} d\omega \rho_+(\omega) \rho_{v_2}(\omega). \tag{S24b}$$

2. *Contribution from Coriolis interactions between $v_2$ and $v_4$ transitions*

Apart from the direct intramolecular anharmonic coupling between the $v_2$ and $v_4$ transitions, Coriolis interactions can also induce the rovibrational coupling between the $v_2$ and $v_4$

transitions [S4–S6]:

$$\hat{V}'_{24} = \sum_{n=1}^{2N/3} \hbar\zeta_{24} \left[\sum_\alpha \sum_{M_n=1}^{N_{\rm nn}} J_\alpha \left(\hat{c}_n^\dagger \hat{b}_{M_n} + \hat{c}_n \hat{b}_{M_n}^\dagger\right)\right]. \tag{S25}$$

Here, $\zeta_{24}$ represents the magnitude of $v_2$-$v_4$ intramolecular Coriolis coupling; the index $\alpha$ runs over the three cyclic permutations of $\alpha$, $\beta$, $\gamma$ on $x$, $y$, $z$; $J_\alpha$ represents the angular momentum operator of each molecule along the $\alpha$ direction; and $N_{\rm nn} = 3$ represents the number of $v_4$ transitions coupled to each $v_2$ transition. In the high-temperature limit, we treat $J_\alpha$ as a classical variable to simplify the calculation.

Because $\hat{b}_{M_n} = \hat{B}/\sqrt{N} + \cdots$, where $\cdots$ represents a linear combination of the dark-state operators $\hat{D}_\mu$ for $\mu = 1, 2, \cdots, N-1$, we can rewrite Eq. (S25) as

$$\hat{V}'_{24} = \sum_{n=1}^{2N/3} \frac{\hbar\zeta_{24}}{\sqrt{N}} \left[\sum_\alpha \sum_{M_n=1}^{N_{\rm nn}} J_\alpha \left(\hat{c}_n^\dagger \hat{B} + \hat{c}_n \hat{B}^\dagger\right)\right] + \cdots. \tag{S26}$$

Following the procedure from Eq. (S22) to Eq. (S24), we obtain the Coriolis-interaction-induced $\text{UP}_{v_4}$ energy transfer rate to $v_2$ transitions as

$$\gamma_{\text{UP}\to v_2} = 2\pi|X_+^{(\text{B})}|^2 Z_{24}^2 \rho_{v_2}(\omega). \tag{S27}$$

Here, $Z_{24}^2 \equiv \sum_{f=1}^{2N/3} \frac{2}{3} N_{\rm nn}^2 \zeta_{24}^2 (\sum_\alpha J_\alpha)^2 \delta(\omega - \omega_f) / \sum_{f=1}^{2N/3} \delta(\omega - \omega_f)$ represents the average intramolecular Coriolis coupling between $v_4$ and $v_2$ transitions.

Combining Eq. (S24) and Eq. (S27), we obtain the overall $\text{UP}_{v_4}$ energy transfer rate to the IR-inactive $v_2$ vibrations:

$$\gamma_{\text{UP}\to v_2} = 2\pi|X_+^{(\text{B})}|^2 (\Xi_{24}^2 + Z_{24}^2) \rho_{v_2}(\omega). \tag{S28}$$

## II. BRIEF REVIEW OF CAVMD

Within the framework of CavMD, the equations of motion for the coupled photon–nuclear system are

$$M_{nj}\ddot{\mathbf{R}}_{nj} = \mathbf{F}_{nj}^{(0)} + \mathbf{F}_{nj}^{\text{cav}}, \tag{S29a}$$

$$m_{k,\lambda}\ddot{\tilde{q}}_{k,\lambda} = -m_{k,\lambda}\omega_{k,\lambda}^2 \tilde{q}_{k,\lambda} - \widetilde{\varepsilon}_{k,\lambda} \sum_{n=1}^{N_{\rm simu}} d_{ng,\lambda}. \tag{S29b}$$

In Eq. (S29a), $M_{nj}$, $R_{nj}$, and $\mathbf{F}_{nj}^{(0)}$ represent the mass, position, and nuclear force outside the cavity for the nucleus indexed by $nj$, where $nj$ denotes the $j$-th nucleus of the $n$-th molecule. The term $\mathbf{F}_{nj}^{\text{cav}}$ represents the cavity-induced contribution to the nuclear force, the explicit form of which will be introduced in Eq. (S30) below.

In Eq. (S29b), $m_{k,\lambda}$, $\tilde{\tilde{q}}_{k,\lambda}$, and $\omega_{k,\lambda}$ represent the auxiliary mass, position, and frequency of the cavity photon mode characterized by the wave vector $\mathbf{k}$ and the polarization direction $\boldsymbol{\xi}_\lambda$. The parameter $\tilde{\varepsilon}_{k,\lambda}$ represents the effective light-matter coupling strength for the cavity photon mode indexed by $k, \lambda$. Each cavity photon mode interacts with the total dipole moment of the explicitly simulated molecular system, given by $\sum_{n=1}^{N_{\text{simu}}} d_{ng,\lambda}$, where $N_{\text{simu}}$ denotes the total number of molecules explicitly simulated, and $d_{ng,\lambda}$ represents the electronic ground-state dipole moment of the $n$-th molecule projected along the cavity polarization direction $\boldsymbol{\xi}_\lambda$. In previous CavMD studies [S7, S8], $N_{\text{simu}}$ was also referred to as $N_{\text{sub}}$.

In Eq. (S29a), the cavity contribution of the nuclear force is defined as

$$\mathbf{F}_{nj}^{\text{cav}} = -\sum_{k,\lambda}\left(\tilde{\varepsilon}_{k,\lambda}\tilde{\tilde{q}}_{k,\lambda} + \frac{\tilde{\varepsilon}_{k,\lambda}^2}{m_{k,\lambda}\omega_{k,\lambda}^2}\sum_{l=1}^{N_{\text{simu}}} d_{lg,\lambda}\right)\frac{\partial d_{ng,\lambda}}{\partial \mathbf{R}_{nj}}. \tag{S30}$$

In the framework of CavMD, the coupled cavity-molecular system can interact with a time-dependent external driving electric field, $\mathbf{E}_{\text{ext}}(t)$. In our simulations, this external field is assumed to interact entirely with either the molecular or the cavity subsystem. When the external field interacts with only the molecular subsystem, the nuclear equation of motion in Eq. (S29a) is modified to

$$M_{nj}\ddot{R}_{nj} = \mathbf{F}_{nj}^{(0)} + \mathbf{F}_{nj}^{\text{cav}} + \mathbf{F}_{nj}^{\text{ext}}(t). \tag{S31}$$

Here, the external driving force $\mathbf{F}_{nj}^{\text{ext}}(t) = -Q_{nj}\mathbf{E}_{\text{ext}}(t)$ is included in the dynamics, where $Q_{nj}$ denotes the partial charge of each nucleus.

When the external field is assumed to interact with only the cavity, each cavity photon mode is coupled to the external field $\mathbf{E}_{\text{ext}}(t)$ via an effective dipole moment $\mu_{k,\lambda} \equiv Q_{k,\lambda}\tilde{\tilde{q}}_{k,\lambda}$, where $Q_{k,\lambda}$ denotes the partial charge of the cavity photon mode. Consequently, the photonic equation of motion in Eq. (S29b) is modified to

$$m_{k,\lambda}\ddot{\tilde{\tilde{q}}}_{k,\lambda} = -m_{k,\lambda}\omega_{k,\lambda}^2\tilde{\tilde{q}}_{k,\lambda} - \tilde{\varepsilon}_{k,\lambda}\sum_{n=1}^{N_{\text{simu}}} d_{ng,\lambda} + \mathbf{F}_{k,\lambda}^{\text{ext}}(t). \tag{S32}$$

Here, the external driving force acting on each photon mode, $\mathbf{F}_{k,\lambda}^{\text{ext}}(t) = -Q_{k,\lambda}\mathbf{E}_{\text{ext}}(t)$, is included in the dynamics. For further details on the fundamentals of CavMD, see also Refs. S7–S9.

## III. SIMULATION DETAILS

For the CavMD simulations of liquid $CH_4$ under VSC, a schematic of the cavity structure is shown in Fig. 1b in the main text. The cavity was assumed to be placed along the $z$-axis, and only a single cavity photon mode was considered in the calculations. This cavity photon mode was polarized along both the $x$- and $y$-directions. The effective light-matter coupling strength per molecule, $\widetilde{\varepsilon}$, was varied between zero (corresponding to the outside-cavity condition) and $5 \times 10^{-4}$ a.u.

### A. Equilibrium simulations

For the liquid $CH_4$ system, $N_{\text{simu}} = 400$ molecules were explicitly simulated in a cubic box with a length of 29.14 Å under periodic boundary conditions. This simulation system corresponded to a molecular number density of $\rho = 16.16$ nm$^{-3}$, or a molecular density of 0.43 g/cm$^3$, matching the experimental liquid $CH_4$ density at 110 K. The standard COMPASS force field [S10] was used to describe the methane system. Intermolecular Coulomb interactions were computed using the Edward summation method. The CavMD simulations were performed using a modified version of the i-PI package [S7, S11], with nuclear forces outside the cavity evaluated via calls to the LAMMPS package [S12].

The initial molecular geometry was generated using the PACKMOL package [S13]. After energy minimization, this geometry was equilibrated through an NVT simulation for 150 ps at 110 K. For the NVT simulation, a Langevin thermostat with a relaxation lifetime of 100 fs was attached to both the nuclei and the cavity photon mode. Starting from the thermally equilibrated geometry, 40 consecutive 20-ps NVE trajectories were simulated. The initial geometry of each NVE simulation was set as the final configuration of the preceding NVE simulation, while the initial velocities of all particles were resampled following a Maxwell-Boltzmann distribution under 110 K. The time step for the molecular dynamics simulations was set to 0.5 fs, and snapshots of the NVE simulation trajectories were stored every 2 fs.

## B. Polariton pumping simulations

After the equilibrium simulations, additional nonequilibrium simulations were conducted to study the polariton relaxation and energy transfer dynamics under the NVE ensemble. A total of 40 nonequilibrium simulations were performed to obtained thermally averaged results. The initial geometry of each nonequilibrium simulation was selected from the starting geometry of the corresponding equilibrium NVE simulation. During the nonequilibrium simulations, an external pulse was applied to the coupled cavity-molecular system to excite the polariton, and each nonequilibrium simulation was run for 20 ps. Here, the NVE simulations implied that the cavity loss was set to zero.

In the *Results* section of the main text, the external pulse was assumed to interact solely with the molecular subsystem. In this case, a cw pulse was used, defined as

$$\mathbf{E}_{\text{ext}}(t) = E_0 \cos{(\omega t + \phi)}\mathbf{e}_{\text{x}}. \tag{S33}$$

Here, $E_0$ and $\omega$ denote the pulse amplitude and center frequency, respectively. The phase $\phi \in [0, 2\pi)$ was assigned as a random value, and $\mathbf{e}_x$ represents a unit vector along the $x$-axis. This pulse was applied within the time window $0.1 < t < 0.6$ ps. Three different pumping fluences were used in simulations: strong pumping $E_0 = 3.084 \times 10^7$ V/m ($6 \times 10^{-3}$ a.u.), medium pumping $E_0 = 1.542 \times 10^7$ V/m ($3 \times 10^{-3}$ a.u.), and weak pumping $E_0 = 3.084 \times 10^6$ V/m ($6 \times 10^{-4}$ a.u.). The corresponding pulse fluences, given by $F = \frac{1}{2}\epsilon_0 c E_0^2 (t_{\text{start}} - t_{\text{end}})$, for the three different pulses were $F = 632$ mJ/cm$^2$, $158$ mJ/cm$^2$, and $6.32$ mJ/cm$^2$, respectively.

In the *Discussion* section of the main text, more realistic simulations were performed. In both equilibrium and nonequilibrium simulations, the cavity loss was incorporated by attaching a Langevin thermostat exclusively to the cavity photon mode. The relaxation lifetime of the Langevin thermostat, representing the cavity lifetime, was set to 0.75 ps. For nonequilibrium simulations, the following Gaussian pulse was applied to excite the cavity photon mode:

$$\mathbf{E}_{\text{ext}}(t) = E_0 \exp\left[-2\ln 2 \frac{(t - t_0 - 4\tau)^2}{\tau^2}\right] \sin{(\omega t + \phi)}\mathbf{e}_{\text{x}}. \tag{S34}$$

Here, $E_0$ and $\omega$ denote the pulse amplitude and center frequency, respectively, while the phase $\phi \in [0, 2\pi)$ was assigned as a random value. The remaining parameters were set as $t_0 = 10$ fs, $\tau = 500$ fs, and $E_0 = 3.084 \times 10^7$ V/m ($6 \times 10^{-3}$ a.u.). As described below Eq.

(S32), the external field interacted with the cavity photon mode via $\mathbf{F}_{k,\lambda}^{\text{ext}}(t) = -Q_{k,\lambda}\mathbf{E}_{\text{ext}}(t)$. The partial charge of the cavity photon mode, $Q_c$, was chosen as 0.028 a.u. This value properly balanced with the cavity lifetime (0.75 ps), in consistent with the input-output theory [S14]; see also the SI of Ref. [S9] for detailed derivations.

### C. Linear-response polariton spectra

With equilibrium NVE trajectories, the linear-response polariton spectrum was computed by evaluating the Fourier transform of the dipole autocorrelation function of the molecular system: [S8, S15–S17]:

$$I(\omega) \propto \omega^2 \int_{-\infty}^{\infty} dt\, e^{-i\omega t} \langle \boldsymbol{\mu}(0) \cdot \boldsymbol{\mu}(t) \rangle. \tag{S35}$$

Here, $\boldsymbol{\mu}(t)$ represents the total dipole moment vector of the molecular system at time $t$.

### D. Time-resolved CH$_4$ bending spectra

At each snapshot of the time-resolved CH$_4$ bending spectra shown in Fig. 2 of the main text, the corresponding spectrum was calculated by evaluating the following Fourier transform:

$$A(\omega) \propto \frac{\omega^2}{6N_{\text{simu}}} \sum_{j=1}^{6} \sum_{k=1}^{N_{\text{simu}}} \int_{-\infty}^{\infty} dt\, e^{-i\omega t} \langle \alpha_{jk}(0)\alpha_{jk}(t) \rangle. \tag{S36}$$

Here, $\alpha_{jk}$ represents the $j$-th H−C−H bending angle in the $k$-th simulated molecule. The time-resolved bending spectrum at time $T_i$ was obtained by Fourier transforming the nonequilibrium angle trajectory, $\alpha_{jk}(t)$, over the time window $[T_i, T_i + \Delta T]$, where $\Delta T$ = 5 ps.

### E. CH$_4$ Symmetry coordinates

A single CH$_4$ molecule has nine distinct vibrational normal modes. Due to the $T_d$ symmetry of CH$_4$, these nine normal modes exhibit degeneracy and can be classified into only four unique vibrational frequencies. One approach to characterizing these normal modes is

to use symmetry coordinates [S18, S19]:

$$\begin{aligned}
v_1 &= \frac{1}{2}(r_1 + r_2 + r_3 + r_4), \\
v_{2a} &= \frac{1}{\sqrt{12}}(2\alpha_{12} - \alpha_{13} - \alpha_{14} - \alpha_{23} - \alpha_{24} + 2\alpha_{34}), \\
v_{2b} &= \frac{1}{2}(\alpha_{13} - \alpha_{14} - \alpha_{23} + \alpha_{24}), \\
v_{3x} &= \frac{1}{2}(r_1 - r_2 + r_3 - r_4), \\
v_{3y} &= \frac{1}{2}(r_1 - r_2 - r_3 + r_4), \\
v_{3z} &= \frac{1}{2}(r_1 + r_2 - r_3 - r_4), \\
v_{4x} &= \frac{1}{\sqrt{2}}(\alpha_{24} - \alpha_{13}), \\
v_{4y} &= \frac{1}{\sqrt{2}}(\alpha_{23} - \alpha_{14}), \\
v_{4z} &= \frac{1}{\sqrt{2}}(\alpha_{34} - \alpha_{12}).
\end{aligned} \quad (S37)$$

Here, $r_1, r_2, r_3, r_4$ represent the four C−H bond length displacements relative to the equilibrium geometry, while $\alpha_{12}, \alpha_{13}, \alpha_{14}, \alpha_{23}, \alpha_{24}, \alpha_{34}$ denote the six H−C−H bending angle displacements relative to the equilibrium geometry.

Due to the $T_d$ symmetry of methane, the symmetry coordinates correspond to specific vibrational modes as follows:

- $v_1$ corresponds to the non-degenerate symmetric stretching mode.

- $v_{2a}$ and $v_{2b}$ correspond to the doubly degenerate symmetric bending mode.

- $v_{3x}$, $v_{3y}$ and $v_{3z}$ correspond to the triply degenerate asymmetric stretching mode.

- $v_{4x}$, $v_{4y}$ and $v_{4z}$ correspond to the triply degenerate asymmetric bending mode.

Given the symmetry coordinates defined above, the average vibrational energy in each symmetry coordinate at time $t$, according to the harmonic approximation, is proportional to

$$[v_{\sigma\lambda}(t)]^2 = \frac{1}{N_{\text{simu}}} \sum_{n=1}^{N_{\text{simu}}} \left[v_{\sigma\lambda}^{(n)}(t)\right]^2, \quad (S38)$$

where $\sigma = 1, 2, 3, 4$ represents the four unique vibrational modes in methane and $\lambda$ indexes the degeneracy.

At thermal equilibrium, the average vibrational energy per symmetry coordinate is $k_\mathrm{B}T$. Therefore, the excess vibrational energy in each vibrational mode during polariton pumping can also be expressed in units of $k_\mathrm{B}T$:

$$E_\sigma(t) = \eta_\sigma \left( \frac{\sum_\lambda [v_{\sigma\lambda}^\mathrm{noneq}(t)]^2}{\sum_\lambda \langle [v_{\sigma\lambda}^\mathrm{eq}(t)]^2 \rangle} - 1 \right) \qquad (\sigma = 1, 2, 3, 4). \tag{S39}$$

Here, $[v_{\sigma\lambda}^\mathrm{eq}(t)]^2$ and $[v_{\sigma\lambda}^\mathrm{noneq}(t)]^2$ represent the values of $[v_{\sigma\lambda}(t)]^2$ during equilibrium and nonequilibrium (with polariton pumping) simulations at time $t$, respectively. The notation $\langle \cdots \rangle$ denotes the time average, and $\eta_\sigma = 1, 2, 3, 3$ corresponds to the degeneracies of the $v_1$-$v_4$ vibrational modes, respectively. Since $E_\sigma(t)$ is defined as a ratio between nonequilibrium and equilibrium values of $[v_{\sigma\lambda}(t)]^2$, $E_\sigma(t)$ has units of $k_\mathrm{B}T$.

Eq. (S39) was used to evaluate the time-resolved vibrational energy dynamics in the $v_1$-$v_4$ normal modes, as shown in Fig. 2k-o of the main text.

### F. Estimating the maximal efficiency of polariton energy transfer

In Fig. 4f of the main text, at the fixed UP frequency, the cavity frequency that maximizes $E_{v_2} = E_0^2 |X_+^{(c)}|^4 |X_+^{(B)}|^2$ [Eq. (2) of the main text] can be determined as follows. Since $|X_+^{(c)}|^2 + |X_+^{(B)}|^2 = 1$, the maximum value of $|X_+^{(c)}|^4 |X_+^{(B)}|^2$ occurs when the photonic weight is $|X_+^{(c)}|^2 = 2/3$. According to Eq. (S9), setting $|X_+^{(c)}|^2 = \cos^2\theta = 2/3$ leads to the relation $\Omega_N^2 = 8(\omega_c - \omega_0)^2$. Substituting this relation into Eq. (S7), we obtain $\omega_c = (\omega_+ + \omega_0)/2$ when $\omega_c > \omega_0$.

Given the molecular frequency $\omega_0 = 1311$ cm$^{-1}$ and the UP frequency $\omega_+ = 1619$ cm$^{-1}$, the optimal cavity frequency is calculated as

$$\omega_c = \frac{\omega_+ + \omega_0}{2} = 1466 \text{ cm}^{-1}. \tag{S40}$$

## IV. SUPPLEMENTARY SIMULATION DATA

In this section, supplementary simulation data are given for facilitating the understanding of the main text.

### A. Liquid CH$_4$ $v_1$-$v_4$ vibrational frequencies from symmetry coordinates

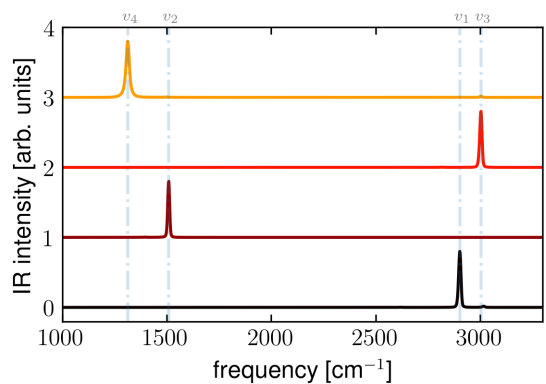

FIG. S1. Vibrational frequencies of the four CH$_4$ $v_1$-$v_4$ symmetry coordinates. The spectrum of each symmetry coordinate is obtained by computing the Fourier transform of the autocorrelation function of the symmetry coordinate $v_{\sigma\lambda}$ in Eq. (S37) from equilibrium molecular dynamics trajectories outside the cavity.

## B. Photonic dynamics corresponding to Fig. 2 in the main text

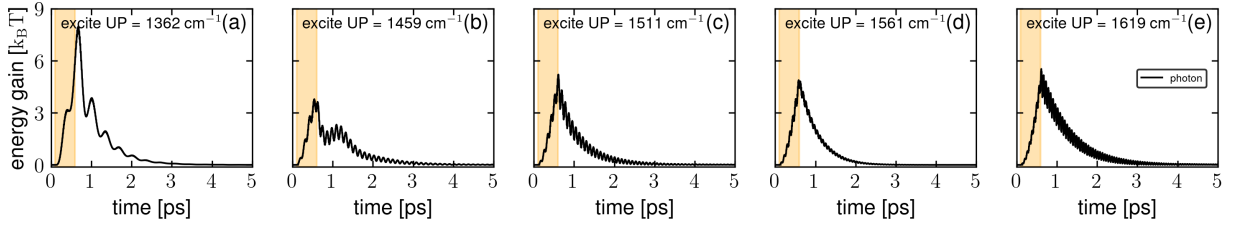

FIG. S2. Photonic energy dynamics under the cw pulse excitation with a pulse fluence of $F = 632$ mJ/cm$^2$. The simulation conditions are identical to those in Fig. 2k-o of the main text. The UP decay rates presented in Fig. 1d are obtained by fitting the photonic energy dynamics after the pulse pumping ($t > 0.6$ ps) with an exponential function $Ae^{-kt}$, where $k$ represents the UP decay rate.

## C. Symmetry coordinate dynamics under different pulse fluences

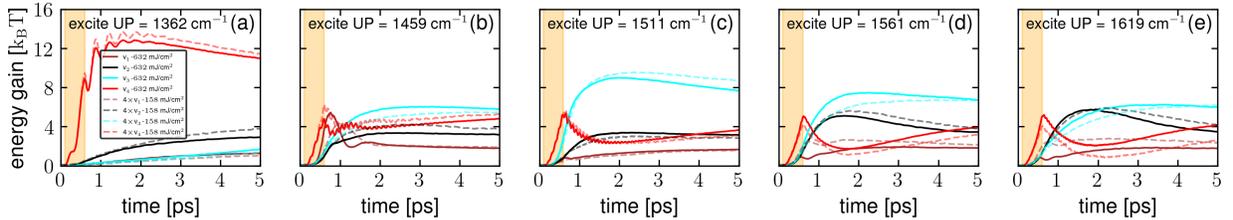

FIG. S3. Average vibrational energy dynamics per molecule during the cw pulse excitation, analogous to Fig. 2k-o in the main text. Vibrational energy trajectories are shown for pulse fluences of $F = 632$ mJ/cm$^2$ (solid lines, identical to Fig. 2k-o in the main text) and $F = 158$ mJ/cm$^2$ (dashed lines). To facilitate the assessment of potential nonlinear effects, the vibrational energy signals for $F = 158$ mJ/cm$^2$ have been scaled by a factor of four. Overall, the good overlap between the signals at different pulse fluences suggests that the nonlinear effects in UP energy transfer are weak. However, part (b) indicates the presence of a weak nonlinear UP energy transfer pathway: $\text{UP}_{v_4} \to 2v_1$.

## D. Symmetry coordinate dynamics corresponding to Fig. 3a in the main text

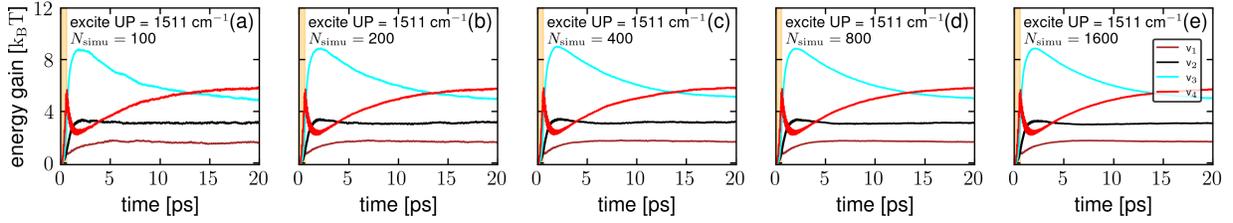

FIG. S4. Average vibrational energy dynamics per molecule during the cw pulse excitation, corresponding to Fig. 3a in the main text for the UP frequency of 1511 cm$^{-1}$. From left to right, the Rabi splitting is fixed, while the number of explicitly simulated molecules is varied as $N_{\text{simu}} = 100, 200, 400, 800, 1600$, respectively.

## E. Symmetry coordinate dynamics corresponding to Fig. 3b in the main text

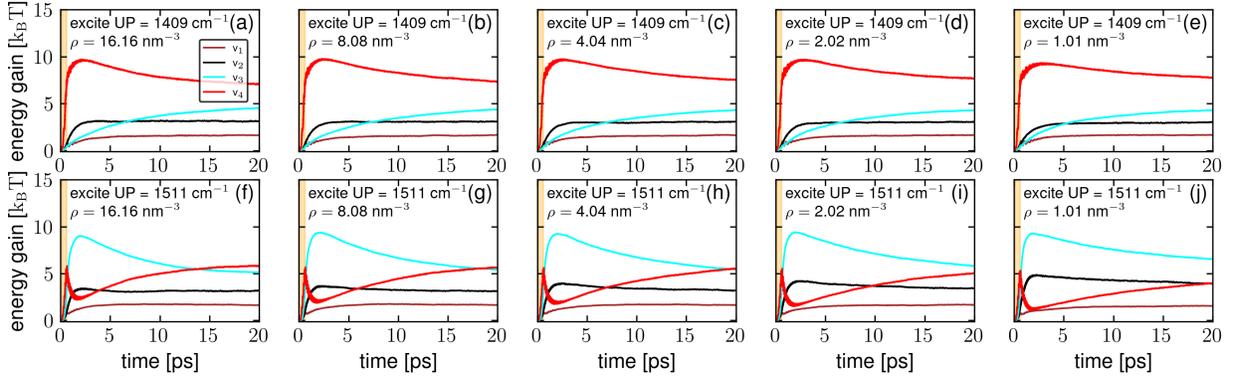

FIG. S5. Average vibrational energy dynamics per molecule during the cw pulse excitation, corresponding to Fig. 3b in the main text. The dynamics for the UP frequency of 1409 cm$^{-1}$ (upper panel) and 1511 cm$^{-1}$ (lower panel) are shown. The number of simulated molecules is fixed, while the molecular number density is varied as $\rho =$ 16.16 nm$^{-3}$, 8.08 nm$^{-3}$, 4.04 nm$^{-3}$, 2.02 nm$^{-3}$, 1.01 nm$^{-3}$ (from left to right), respectively.

## F. Symmetry coordinate and photonic dynamics under Gaussian pulse excitation of a lossy cavity mode with $\omega_c = 1311$ cm$^{-1}$

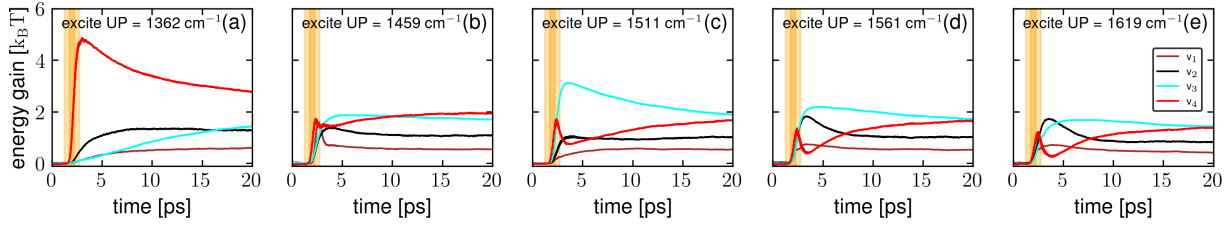

FIG. S6. Average vibrational energy dynamics per molecule under the same conditions as Fig. 2k-o in the main text, except with a Gaussian pulse used to excite a lossy cavity mode with $\tau_c = 0.75$ ps. Part (c) here is also plotted as Fig. 4a in the main text.

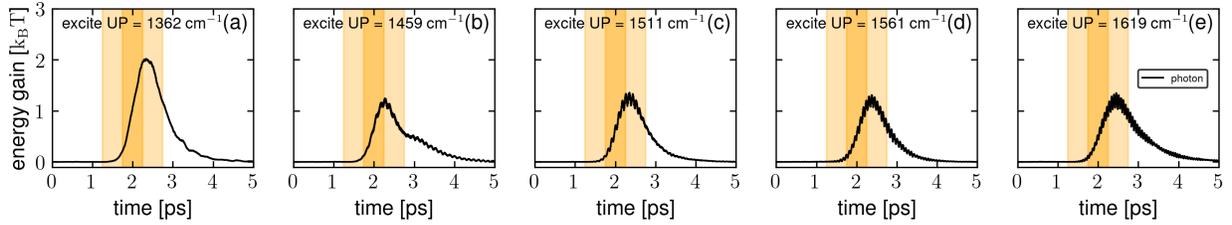

FIG. S7. Photonic energy dynamics corresponding to Fig. S6.

## G. Symmetry coordinate and photonic dynamics under Gaussian pulse excitation of a lossy cavity mode with $\omega_c = 1500$ cm$^{-1}$

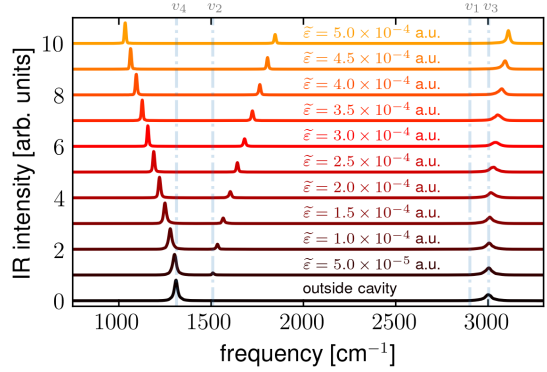

FIG. S8. Equilibrium IR spectra analogous to Fig. 1c in the main text, except with the cavity frequency blueshifted to $\omega_c = 1500$ cm$^{-1}$.

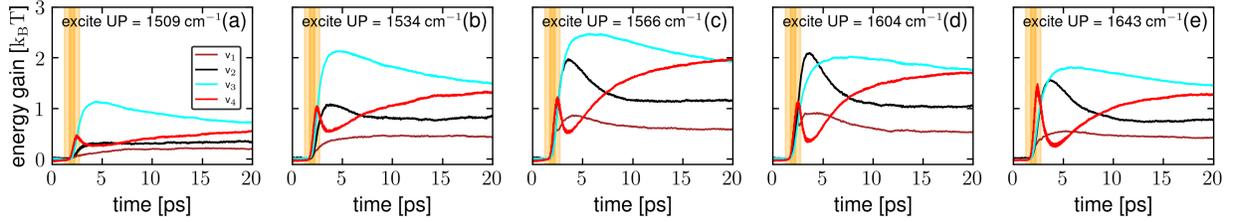

FIG. S9. Average vibrational energy dynamics per molecule when a Gaussian pulse is used to excite the UP, with the cavity mode frequency set to $\omega_c = 1500$ cm$^{-1}$. Cavity loss is also included. Part (a) here is also plotted as Fig. 4b in the main text.

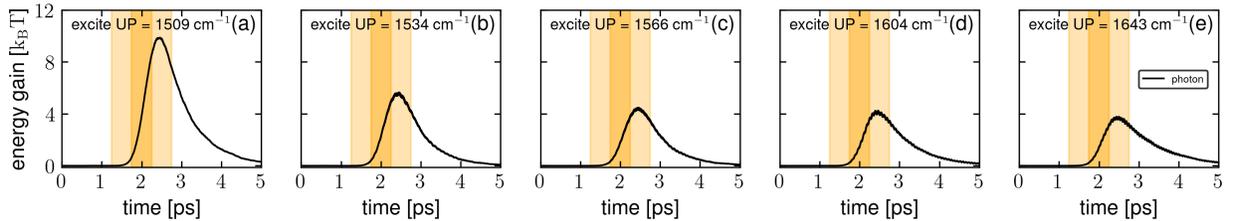

FIG. S10. Photonic energy dynamics corresponding to Fig. S9.

S22

## H. Symmetry coordinate and photonic dynamics corresponding to Figs. 4d-f in the main text

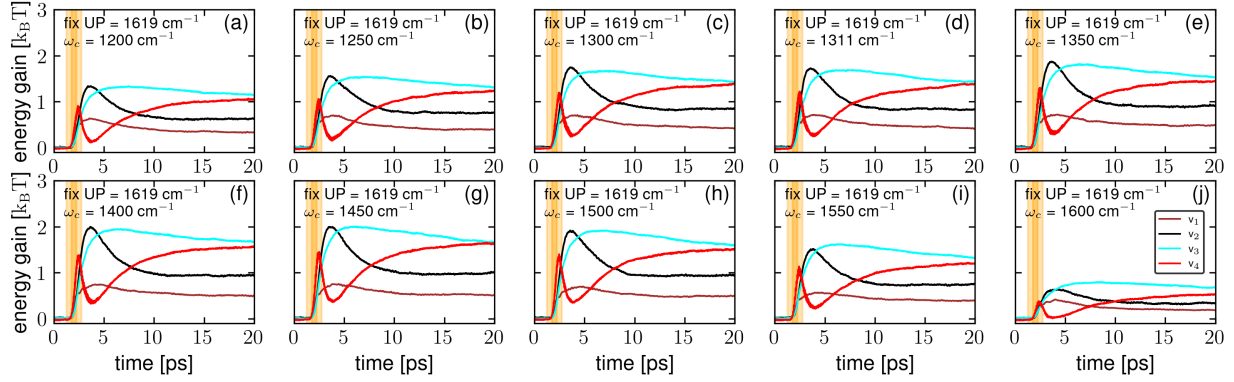

FIG. S11. Average vibrational energy dynamics per molecule when a Gaussian pulse is used to excite the UP, corresponding to Fig. 4d in the main text. In each part, a different combination of $\{\omega_c, \widetilde{\varepsilon}\}$ is used to maintain the UP frequency at 1619 cm$^{-1}$. Here, the maximum excitation energy of $v_2$ and $v_2 + v_3$ signals in each part is used to plot Fig. 4f of the main text.

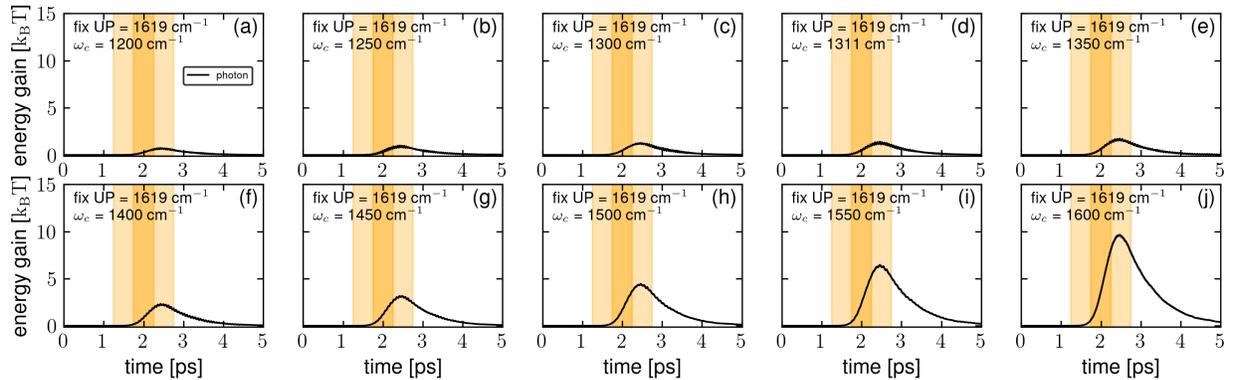

FIG. S12. Photonic energy dynamics corresponding to Fig. S11. Here, the maximum photonic energy in each part is used to plot Fig. 4e of the main text.

S23

## I. Liquid CD$_4$ simulation results

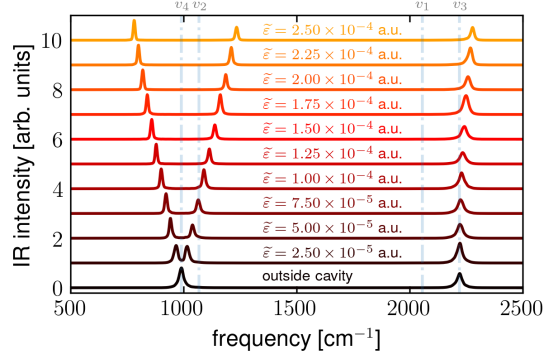

FIG. S13. Equilibrium IR spectra for liquid CD$_4$ under VSC. The cavity frequency is set to $\omega_c = 990.9$ cm$^{-1}$, at resonance with the $v_4$ mode of CD$_4$. All other simulation details are identical to those in Fig. 1c of the main text.

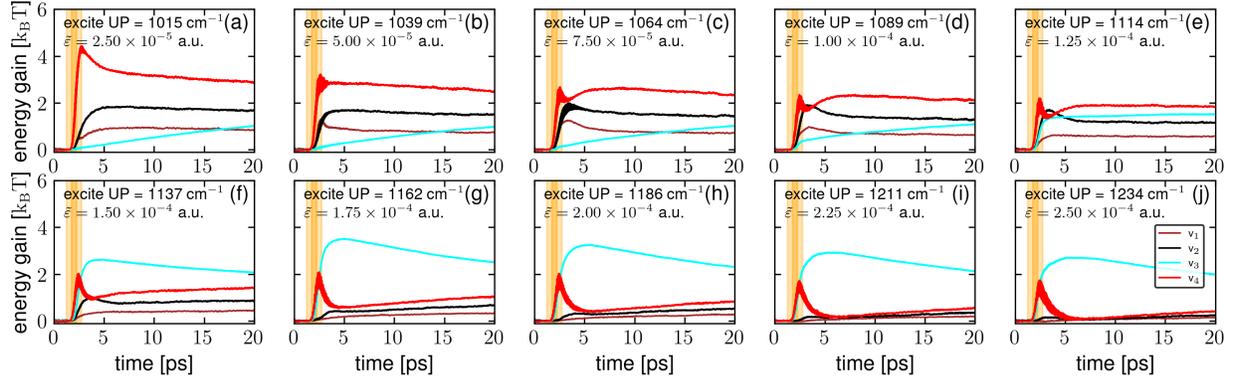

FIG. S14. Corresponding average vibrational energy dynamics per molecule when a Guassian pulse is used to excite each UP shown in Fig. S13. From left to right, the effective light-matter coupling strength $\widetilde{\varepsilon}$ (labeled in each part) increases from $2.50 \times 10^{-5}$ to $2.50 \times 10^{-4}$ a.u.

[S1] T. E. Li, A. Nitzan, and J. E. Subotnik, Polariton Relaxation under Vibrational Strong Coupling: Comparing Cavity Molecular Dynamics Simulations against Fermi's Golden Rule Rate, J. Chem. Phys. **156**, 134106 (2022).

[S2] M. Tavis and F. W. Cummings, Exact Solution for an N-Molecule—Radiation-Field Hamiltonian, Phys. Rev. **170**, 379 (1968).

[S3] M. Tavis and F. W. Cummings, Approximate Solutions for an N-Molecule-Radiation-Field Hamiltonian, Phys. Rev. **188**, 692 (1969).

[S4] W. H. J. Childs and H. A. J. Ahn, A New Coriolis Perturbation in the Methane Spectrum III. Intensities and Optical Spectrum, Proc. R. soc. Lond. Ser. A **169**, 451 (1939).

[S5] A. Robiette and I. Mills, Intensity Perturbations due to $\nu 3/\nu 4$ Coriolis Interaction in Methane, J. Mol. Spectrosc. **77**, 48 (1979).

[S6] R. H. Tipping, A. Brown, Q. Ma, J. M. Hartmann, C. Boulet, and J. Liévin, Collision-induced Absorption in the $\nu 2$ Fundamental Band of CH4. I. Determination of the Quadrupole Transition Moment, J. Chem. Phys. **115**, 8852 (2001).

[S7] T. E. Li, J. E. Subotnik, and A. Nitzan, Cavity Molecular Dynamics Simulations of Liquid Water under Vibrational Ultrastrong Coupling, Proc. Natl. Acad. Sci. **117**, 18324 (2020).

[S8] T. E. Li, A. Nitzan, and J. E. Subotnik, Cavity Molecular Dynamics Simulations of Vibrational Polariton-Enhanced Molecular Nonlinear Absorption, J. Chem. Phys. **154**, 094124 (2021).

[S9] T. E. Li, A. Nitzan, and J. E. Subotnik, Energy-Efficient Pathway for Selectively Exciting Solute Molecules to High Vibrational States via Solvent Vibration-Polariton Pumping, Nat. Commun. **13**, 4203 (2022).

[S10] H. Sun, COMPASS: An ab Initio Force-Field Optimized for Condensed-Phase Applications: Overview with Details on Alkane and Benzene Compounds, J. Phys. Chem. B **102**, 7338 (1998).

[S11] Y. Litman, V. Kapil, Y. M. Feldman, D. Tisi, T. Begušić, K. Fidanyan, G. Fraux, J. Higer, M. Kellner, T. E. Li, E. S. Pós, E. Stocco, G. Trenins, B. Hirshberg, M. Rossi, and M. Ceriotti, i-PI 3.0: A Flexible and Efficient Framework for Advanced Atomistic Simulations, J. Chem. Phys. **161**, 10.1063/5.0215869 (2024).

[S12] A. P. Thompson, H. M. Aktulga, R. Berger, D. S. Bolintineanu, W. M. Brown, P. S. Crozier, P. J. in 't Veld, A. Kohlmeyer, S. G. Moore, T. D. Nguyen, R. Shan, M. J. Stevens, J. Tranchida, C. Trott, and S. J. Plimpton, LAMMPS - A Flexible Simulation Tool for Particle-based Materials Modeling at the Atomic, Neso, and Continuum Scales, Comput. Phys. Commun. **271**, 108171 (2022).

[S13] L. Martínez, R. Andrade, E. G. Birgin, and J. M. Martínez, PACKMOL: A Package for Building Initial Configurations for Molecular Dynamics Simulations, J. Comput. Chem. **30**, 2157 (2009).

[S14] I. Carusotto and C. Ciuti, Quantum Fluids of Light, Rev. Mod.Phys. **85**, 299 (2013).

[S15] D. A. McQuarrie, *Statistical Mechanics* (Harper-Collins Publish- ers, New York, 1976).

[S16] M.-P. Gaigeot and M. Sprik, Ab Initio Molecular Dynamics Computation of the Infrared Spectrum of Aqueous Uracil, J. Phys. Chem. B **107**, 10344 (2003).

[S17] S. Habershon, G. S. Fanourgakis, and D. E. Manolopoulos, Comparison of Path Integral Molecular Dynamics Methods for the Infrared Absorption Spectrum of Liquid Water, J. Chem. Phys. **129**, 074501 (2008).

[S18] P. Lazzeretti, R. Zanasi, A. Sadlej, and W. Raynes, Magnetizability and Carbon-13 Shielding Surfaces for the Methane Molecule, Mol. Phys. **62**, 605 (1987).

[S19] X.-G. Wang and T. Carrington, Deficiencies of the Bend Symmetry Coordinates Used for Methane, J. Chem. Phys. **118**, 6260 (2003).


    \end{document}